\documentclass[a4paper,11pt]{article}
\pdfoutput=1 

\usepackage{jheppub} 
\usepackage{slashed}
\usepackage[T1]{fontenc} 
\usepackage{caption}
\usepackage{subcaption}

\title{\boldmath Low and moderate $x$ gluon contribution to exclusive Compton scattering processes}

\author[a]{R. Boussarie}
\author[b,c]{Y. Mehtar-Tani}

\affiliation[a]{CPHT, CNRS, Ecole Polytechnique, Institut Polytechnique de Paris, 91128 Palaiseau, France}
\affiliation[b]{Physics Department, Brookhaven National Laboratory, Upton, NY 11973, USA}
\affiliation[c]{RIKEN BNL Research Center, Brookhaven National Laboratory, Upton, NY 11973, USA}

\emailAdd{renaud.boussarie@polytechnique.edu}
\emailAdd{mehtartani@bnl.gov}

\abstract{We revisit the high energy semi-classical description of the exclusive processes DVCS, TCS, and Double DVCS by explicitly keeping track of the Feynman $x$ dependence in both the hard and the hadronic matrix elements. This is achieved by a modification of the standard shock wave approximation to derive the effective Feynman rules, which leads to a generic expression on which we then perform a partial twist expansion to get rid of quantities suppressed by the proper physical scales. We obtain a compact factorized master formula that can be used to investigate the Bjorken limit at leading twist. In particular, we recover the full one-loop result in the collinear limit for pure gluon exchange with the target. Finally, we discuss the subtleties in taking the simultaneous collinear and small $x$ limit.}

\begin{document} 
\maketitle
\flushbottom

\section{Introduction}
Exclusive processes such as Deeply Virtual Compton Scattering (DVCS)
$\gamma^{\ast}P\rightarrow\gamma P$ play a crucial role in providing precise information about the partonic structure of hadrons. These experiments can be viewed as performing a 3-dimensional tomography of hadrons through the requirement that undisturbed final state hadrons are measured. Such a requirement provides valuable insights into non-diagonal parton distributions. By applying Fourier transformations with respect to the momentum transfer between non-diagonal states, researchers can access not only the longitudinal momentum fraction $x$ of partons inside hadrons but also their spatial positions.

Alongside DVCS, Timelike Compton Scattering (TCS) $\gamma P\rightarrow\gamma^{\ast}P$
and Double DVCS (DDVCS) $\gamma^{\ast}P\rightarrow\gamma^{\ast}P$
can be described theoretically in very similar fashions. Although
the cross section for DDVCS is expected to be challenging to measure
experimentally, it is the most valuable case study for the theoretical
purposes of this article.

The physics of exclusive Compton scattering amplitudes, like many other processes in Quantum Chromodynamics (QCD), exhibits distinct characteristics in two primary kinematical regimes: the Bjorken limit and the Regge limit.

In the Bjorken limit, the hard scale parameter $Q$ and the center-of-mass energy $\sqrt{s}$ are approximately of the same order of magnitude. Here, the amplitude elegantly factorizes to all orders within the framework of QCD factorization~\cite{Collins:1998be}. This factorization involves an expansion in powers of the hard scale, with a resummation of logarithmic terms. Within this framework, the factorized amplitude is described by a Generalized Parton Distribution (GPD)~\cite{Muller:1994ses, Ji:1996nm, Radyushkin:1997ki, Diehl:2003ny, Belitsky:2005qn}, which characterizes the 3D distribution of pointlike parton pairs inside the hadron.

On the other hand, in the Regge limit, where $Q$ and $\sqrt{s}$ are widely separated scales, expanding in powers of the hard scale is not suitable. Instead, in the most efficient frameworks for the Regge limit, a factorized form is constructed order-by-order in perturbation theory, relying on a semi-classical effective approach~\cite{McLerran:1993ni, McLerran:1993ka, McLerran:1994vd, Balitsky:1995ub, Iancu:2003xm, Gelis:2010nm}. In this limit, Generalized Parton Distributions (GPDs) are substituted by non-diagonal matrix elements of infinite Wilson lines. This transformation gives rise to an intriguing scenario: rather than dealing with pointlike partons, we encounter a large number of gluons carrying transverse momentum and undergoing recombination processes with each other.

It was recently shown that all QCD processes in their Regge limit
can actually be rewritten in a form that explicitly involves parton
distributions~\cite{Altinoluk:2019fui, Altinoluk:2019wyu, Boussarie:2020vzf}. For inclusive and semi-inclusive processes one finds Transverse Momentum
Dependent (TMD) distributions which are 3-dimensional momentum space
generalizations of Parton Distribution Functions (PDFs). For exclusive
processes one finds Generalized TMD (GTMD) distributions, the 5-dimensional
generalization of GPDs. This result generalizes
the case of DVCS studied in~\cite{Hatta:2017cte}. An unfortunate feature
of this rewriting is the fact that all distributions are evaluated
in the strict $x=0$ limit. For inclusive processes, this property
results in inconsistencies in the collinear corner of phase space~\cite{Boussarie:2020fpb, Boussarie:2021wkn}, confirming observations made prior to the rewriting of small $x$ physics in terms of actual parton distributions~\cite{Bialas:2000xs, Kutak:2004ym}.

In~\cite{Boussarie:2020fpb, Boussarie:2021wkn}, we proposed a consistent effective description
in the case of an inclusive process where both limits are taken into
account in their entirety. This led to a form where instead of a TMD
at $x=0$, a new type of unintegrated distribution with $x\neq0$
and with transverse momenta, arose. In this article, we will test
our effective framework on the more complicated case of DDVCS. Indeed
as we will see below, this process involves three different longitudinal
variables so it is more challenging in principle to take into account
in a framework which is consistent with the Regge limit where all
these variables are small. Such an exclusive process also has the
potentially interesting feature that the presence of the additional
longitudinal variable $\xi$ in the distribution might regularize
the issues at $x=0$.

The purpose of this article is thus to apply the effective framework put forward
in~\cite{Boussarie:2020fpb, Boussarie:2021wkn} for the case of inclusive Deep Inelastic Scattering (DIS) to DVCS, TCS and DDVCS and to recover all
the expected limits. In particular, we will thoroughly discuss the
interplay between the leading power of the Regge expressions and the
high energy limit of the Bjorken expression, as definite proof that
having $x=0$ in the distributions is unavoidable in semi-classical
Regge physics. Indeed, we will write the following unified form for the amplitude:
\begin{align}
    {\cal A}\sim  \int {\rm d} x \int {\rm d} k_\perp G(x,k_\perp) H(x,k_\perp),
\end{align}
where $G$ is an unintegrated distribution with dependence on intrinsic transverse momentum $k_\perp$ and longitudinal momentum fraction $x$, and $H$ is a hard subamplitude. The Bjorken limit will be recovered by neglecting $k_\perp$ in the hard subamplitude:
\begin{align}
     {\cal A}_{\rm Bjorken}\sim \int {\rm d} x \left( \int {\rm d} k_\perp G(x,k_\perp)\right)  H(x,k_\perp=0_\perp),
\end{align}
and the Regge limit will be recovered by setting $x=0$ in the distribution:
\begin{align}
     {\cal A}_{\rm Regge}\sim \int {\rm d} k_\perp  G(x=0,k_\perp) \left( \int {\rm d} x H(x,k_\perp) \right).
\end{align}
This article is structured as follows. In Section~\ref{sec:gen}, In Section~\ref{sec:gen}, we provide a brief overview of the effective framework and apply it to the $\gamma^{(\ast)}P\rightarrow\gamma^{(\ast)}P$ amplitude in full generality. Moving on to Section~\ref{sec:PTE}, we employ the Partial Twist Expansion (PTE) method as introduced in~\cite{Boussarie:2020fpb, Boussarie:2021wkn} to derive the final interpolating formula for this amplitude. In Section~\ref{sec:Bjorken}, we delve into the Bjorken limit and show that we recover the well-known results for the one-loop DDVCS amplitude. In Section~\ref{sec:Regge} we show how to recover the usual semi-classical description in the Regge limits.
Finally, in Section~\ref{sec:Double Limit}, we perform a comparison between the leading power limit of the Regge expressions and the high-energy limit of the Bjorken expressions.

\section{Generic amplitude}\label{sec:gen}

The observable we want to discuss is the exclusive amplitude for $\gamma^{(\ast)}(q)P(p)\rightarrow\gamma^{(\ast)}(q^{\prime})P(p^{\prime})$~, where at least one photon has a perturbatively large virtuality:
$-q^{2}\equiv Q^{2}\gg\Lambda_{{\rm QCD}}^{2}$ (DVCS case) or $q^{\prime2}\equiv Q^{\prime2}\gg\Lambda_{{\rm QCD}}^{2}$
(TCS case) or both (DDVCS case). We shall not address the Bethe-Heitler contributions to our amplitude, though it does contribute to final observables: our only focus is the perturbative QCD part of the amplitude. 

Although quarks would also contribute in the Bjorken regime, we will restrain our analysis for the time being to the gluon-mediated amplitude that is expected to dominate at high energy so we can interpolate smoothly with the gluon-exclusive Regge limit. Because we deal with high density effects in the target, our approach goes beyond a plain perturbative expansion: we will use the effective Feynman rules for the scattering off an external gluon field $A^\mu(x)= A^-(x^+,0^-,{\boldsymbol x})$ produced by a target moving near the light-cone along $-z$ direction. Indeed, in the Regge limit, only the $A^-$ term dominates in power counting of center-of-mass energy $s$. For an observable such as exclusive Compton scattering which factorizes with collinear distributions (as opposed to TMD-factorized observables), although transverse gluon fields contribute it is possible to perform the computation with pure $A^-$ fields if we are only interested in the leading twist result~\cite{Boussarie:2020fpb, Boussarie:2021wkn}. We shall resum all orders in $g A^\mu$ rather than relying on a naive expansion in powers of $g$ \footnote{Throughout this article, we consistently employ light-cone coordinates defined as $x^\pm=(t\pm z)/\sqrt{2}$}. Indeed, dense target effects can result in $gA$ being of order~$1$. Here we choose the convention in which the light cone $+$ (resp. $-$) direction is that of the photon (resp. target hadron). These rules are discussed in~\cite{Boussarie:2020fpb, Boussarie:2021wkn}.

Because only the $\gamma^+ A^-$ interaction term contributes, the quark spin is conserved along its trajectory and the Dirac structure factors out. The corresponding effective fermionic propagator $D_F$ in $D$ dimensions reads in momentum space:
\begin{equation}
D_{F}(k_2,k_1)=\frac{i\gamma^{+}}{2k_1^{+}}(2\pi)^{D}\delta^{D}(k_2-k_1)+i\frac{\slashed{k_2}\gamma^{+}\slashed{k}_1}{2k_1^{+}}G_{{\rm scal}}(k_2,k_1),\label{eq:fermion-prop-exp}
\end{equation}
where $G_{\rm scal}$ is the effective scalar propagator. In position space, it is the solution of the equations
\begin{equation}
\left[-\square_{x}+2igA^{-}(x)\frac{\partial}{\partial x^{-}}\right]G_{\mathrm{scal}}(x,x_{0})=\delta^{D}(x-x_{0}).\label{eq:KG}
\end{equation}
and
\begin{equation}
G_{\mathrm{scal}}(x,x_{0})\left[-\overleftarrow{\square}_{x_{0}}-2ig\frac{\overleftarrow{\partial}}{\partial x_{0}^{-}}A^{-}(x_{0})\right]=\delta^{D}(x-x_{0}),\label{eq:KGx0}
\end{equation}

In terms of these objects, we will write the amplitude in dimensional regularization, with dim reg parameter $\mu$. Since the amplitude we are computing is effectively a one-loop quantity, we know that the renormalization of $g$ will require us to introduce $\mu^{2-d}$, where $d=D-2$ is the dimension of the transverse space, so we included it right away with this power and we will leave $g$ untouched in the following. 

\begin{figure} 
\centering
\includegraphics[width=8cm]{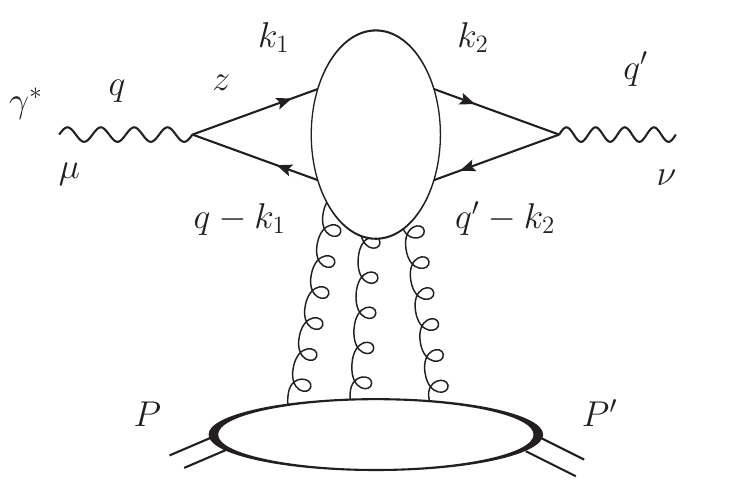}
\caption{Diagrammatic representation of the amplitude $\cal A$ for the process $\gamma^{(\ast)}(q)P(p)\rightarrow\gamma^{(\ast)}(q^{\prime})P(p^{\prime})$ where only gluons are exchanged with the target in the $t$-channel. The incoming photon splits into a quark and antiquark pair which then propagates in the background field of the target. 
\label{fig:figure1}}
\end{figure}

The gluon-mediated exclusive amplitude for $\gamma^{(\ast)}(q)P(p)\rightarrow\gamma^{(\ast)}(q^{\prime})P(p^{\prime})$ to all orders in the background target field reads
\begin{align}
{\cal A} & =-\sum_{f}(-ieq_{f})^{2}\mu^{2-d}\int\!\frac{{\rm d}^{D}k_1}{(2\pi)^{D}}\frac{{\rm d}^{D}k_2}{(2\pi)^{D}}\left\langle p^{\prime}\left|{\rm tr}\slashed{\varepsilon}^{\prime\ast}D_{F}(k_2,k_1)\slashed{\varepsilon}D_{F}(k_1-q,k_2-q^{\prime})\right|p\right\rangle .\label{eq:amp-ini}
\end{align}
This corresponds to the amplitude for a photon splitting into a quark-antiquark pair with momenta $k_1$ snd $q-k_1$, respectively, and subsequently propagating through the target before merging to form the final state photon as depicted by Figure~\ref{fig:figure1}. 
Here, $q_{f}$ is the charge of the quark of flavor $f$, and $D_{F}$ are effective fermionic propagators.

QED gauge invariance can be proven with similar steps as described in the appendices of~\cite{Boussarie:2020fpb, Boussarie:2021wkn}. Owing to this property that ensures that $q_\mu {\rm tr}\gamma^\mu D_F \gamma^\nu D_F=0$, we may shift the polarization vectors by terms proportional to $q$ and $q'$ is the amplitude:
\begin{align}
{\cal A} & =-4\pi\alpha_{{\rm em}}\sum_{f}q_{f}^{2}\mu^{2-d}\left(\varepsilon_{\mu}-\frac{\varepsilon^{+}}{q^{+}}q_{\mu}\right)\left(\varepsilon_{\nu}^{\prime\ast}-\frac{\varepsilon^{\prime\ast+}}{q^{\prime+}}q_{\nu}^{\prime}\right)\int\!\frac{{\rm d}^{D}k_1}{(2\pi)^{D}}\frac{{\rm d}^{D}k_2}{(2\pi)^{D}}\nonumber \\
 & \times{\rm tr}\frac{\gamma^{\nu}\slashed{k}_2\gamma^{+}\slashed{k}_1\gamma^{\mu}(\slashed{k}_1-\slashed{q})\gamma^{+}(\slashed{k}_2-\slashed{q}^{\prime})}{4k_1^{+}(k_1^{+}-q^{+})}\label{eq:amp-gscal} \left\langle p^{\prime}\left|{\rm tr}G_{{\rm scal}}(k_2,k_1)G_{{\rm scal}}(k_1-q,k_2-q^{\prime})\right|p\right\rangle . 
\end{align}
In Landau gauge, it is possible to parametrize the photon polarizations
as follows:
\begin{equation}
\varepsilon_{L}=\frac{q}{Q}+ \frac{Q}{q^{+}}n_{2},\quad\varepsilon_{\pm}=e_{\pm\perp}+\frac{\boldsymbol{e}_{\pm}\cdot\boldsymbol{q}}{q^{+}}n_{2},\label{eq:pola-def}
\end{equation}
where $e_{\pm}$ are purely transverse vectors which verify
\begin{equation}
\sum_{h=\pm}e_{h}^{\mu}e_{h}^{\nu\ast}=-g_{\perp}^{\mu\nu},\quad\boldsymbol{e}_{h}\cdot\boldsymbol{e}_{h^{\prime}}^\ast=\delta_{hh^{\prime}}.\label{eq:tpola-def}
\end{equation}
With this parametrization, we have $\varepsilon_{L}^{2}=+1,\varepsilon_{h}^{2}=-1,$
and the completeness relation
\begin{equation}
\varepsilon_{L}^{\mu}\varepsilon_{L}^{\nu}-\sum_{h}\varepsilon_{h}^{\mu}\varepsilon_{h}^{\nu\ast} = g^{\mu\nu}-\frac{q^{\mu}q^{\nu}}{q^{2}}.\label{eq:pola-completeness}
\end{equation}
 Similarly, we can write
\begin{equation}
\varepsilon_{L}^{\prime\ast}=\frac{q^{\prime}}{Q^{\prime}}-\frac{Q^{\prime}}{q^{\prime+}}n_{2},\quad\varepsilon_{h=\pm}^{\prime\ast}=e_{h=\pm\perp}^{\prime\ast}+\frac{\boldsymbol{e}_{h=\pm}^{\prime\ast}\cdot\boldsymbol{q}^{\prime}}{q^{\prime+}}n_{2}.\label{eq:pola-def-1}
\end{equation}
We shall now reduce the Dirac structure of the fermion loop that reads
\begin{align}
 & A^{\mu\nu}\equiv{\rm tr}[\gamma^{\nu}\slashed{k}_2\gamma^{+}\slashed{k}_1\gamma^{\mu}(\slashed{k}_1-\slashed{q})\gamma^{+}(\slashed{k}_2-\slashed{q}^{\prime})].
\end{align}
Owing to the fact that $\gamma^+ = \frac{1}{2}\gamma^+\gamma^-\gamma^+$, we can rewrite it as
\begin{equation}
A^{\mu\nu}= {\rm tr}[\gamma^{-}K_2^{\nu}\gamma^{-}K_1^{\mu}],
\end{equation}
with 
\begin{align}
K^{\nu}_2 & =\frac{1}{2}\gamma^{+}(\slashed{k}_2-\slashed{q}^{\prime})\gamma^{\nu}\slashed{k}_2\gamma^{+}\\
 & =2k_2^{+}\left(k_2^{+}-q^{\prime+}\right)n_{1}^{\nu}\gamma^{+}-n_{2}^{\nu}\left(\slashed{k}_{2\perp}-\slashed{q}_{\perp}^{\prime}\right)\slashed{k}_{2\perp}\gamma^{+}\nonumber \\
 & +\left(k_2^{+}-q^{\prime+}\right)\gamma_{\perp}^{\nu}\slashed{k}_{2\perp}\gamma^{+}+k_2^{+}\left(\slashed{k}_{2\perp}-\slashed{q}_{\perp}^{\prime}\right)\gamma_{\perp}^{\nu}\gamma^{+},
\end{align}
and
\begin{align}
K_1^{\mu} & =\frac{1}{2}\gamma^{+}\slashed{k}_1\gamma^{\mu}(\slashed{k}_1-\slashed{q})\gamma^{+}\\
 & =2k_1^{+}\left(k_1^{+}-q^{+}\right)n_{1}^{\mu}\gamma^{+}-n_{2}^{\mu}\slashed{k}_{1\perp}\left(\slashed{k}_{1\perp}-\slashed{q}_{\perp}\right)\gamma^{+}\nonumber \\
 & +k_1^{+}\gamma_{\perp}^{\mu}\left(\slashed{k}_{1\perp}-\slashed{q}_{\perp}\right)\gamma^{+}+\left(k_1^{+}-q^{+}\right)\gamma^{+}\slashed{k}_{1\perp}\gamma_{\perp}^{\mu}.
\end{align}
To get the final explicitly transverse boost invariant formula, it is convenient to reconstruct $q^{\prime \nu}$ as follows:
\begin{align}
2k^{+}_2\left(k_2^{+}-q^{\prime+}\right)n_{1}^{\nu}\gamma^{+} & =2\frac{k_2^{+}\left(k_2^{+}-q^{\prime+}\right)}{q^{\prime+}}q^{\prime\nu}\gamma^{+}-2\frac{k_2^{+}\left(k_2^{+}-q^{\prime+}\right)}{q^{\prime+}}q^{\prime-}n_{2}^{\nu}\gamma^{+}\nonumber \\
 & -\frac{k_2^{+}\left(k_2^{+}-q^{\prime+}\right)}{q^{\prime+}}\left(\gamma_{\perp}^{\nu}\slashed{q}_{\perp}^{\prime}+\slashed{q}_{\perp}^{\prime}\gamma_{\perp}^{\nu}\right)\gamma^{+}
\end{align}
and to combine the last three terms with the remaining similar ones in $K^\nu_2$. This way, we have
\begin{align}
K_2^{\nu} & =2\frac{k_2^{+}\left(k_2^{+}-q^{\prime+}\right)}{q^{\prime+}}q^{\prime\nu}\gamma^{+} +\left(k_{2\perp\beta}-\frac{k_2^{+}}{q^{\prime+}}q_{\perp\beta}^{\prime}\right)\left[k_2^{+}\gamma_{\perp}^{\beta}\gamma_{\perp}^{\nu}+\left(k_2^{+}-q^{\prime+}\right)\gamma_{\perp}^{\nu}\gamma_{\perp}^{\beta}\right]\gamma^{+} \nonumber\\
 & -n_{2}^{\nu}\left[2\frac{k_2^{+}\left(k_2^{+}-q^{\prime+}\right)}{q^{\prime+}}q^{\prime-}+\left(\slashed{k}_{2\perp}-\slashed{q}_{\perp}^{\prime}\right)\slashed{k}_{2\perp}\right]\gamma^ {+} \,,
\end{align}
and similarly,
\begin{align}
K_1^{\mu} & =2\frac{k_1^{+}\left(k_1^{+}-q^{+}\right)}{q^{+}}q^{\mu}\gamma^{+} +\left(k_{1\perp\alpha}-\frac{k_1^{+}}{q^{+}}q_{\perp\alpha}\right)\left[\gamma_{\perp}^{\alpha}\left(k_1^{+}-q^{+}\right)\gamma_{\perp}^{\mu}+k_1^{+}\gamma_{\perp}^{\alpha}\gamma_{\perp}^{\mu}\right]\gamma^{+}\nonumber \\
 & -n_{2}^{\mu}\left[2\frac{k_1^{+}\left(k_1^{+}-q^{+}\right)}{q^{+}}q^{-}+\slashed{k}_{1\perp}\left(\slashed{k}_{1\perp}-\slashed{q}_{\perp}\right)\right]\gamma^{+}.
\end{align} 
Note that terms that are proportional to $n_2^\mu$ or $n_2^\nu$ do not contribute to the amplitude because of the overall factor $(\varepsilon_{\mu}-\frac{\varepsilon^{+}}{q^{+}}q_{\mu})(\varepsilon_{L\nu}^{\prime\ast}-\frac{\varepsilon_{L}^{\prime\ast+}}{q^{\prime+}}q_{\nu}^{\prime})$ which vanishes upon contraction with either of these 4-vectors. We can thus neglect them to obtain for the following Dirac structure
\begin{align}
A^{\mu\nu} & =\left\{ 2\frac{k_{1}^{+}\left(k_{1}^{+}-q^{+}\right)}{q^{+}}q^{\mu}+\left(k_{1\perp\alpha}-\frac{k_{1}^{+}}{q^{+}}q_{\perp\alpha}\right)\left[\gamma_{\perp}^{\alpha}\left(k_{1}^{+}-q^{+}\right)\gamma_{\perp}^{\mu}+k_{1}^{+}\gamma_{\perp}^{\alpha}\gamma_{\perp}^{\mu}\right]\right\} \label{eq:dirac-alg-1}\\
 & \times\left\{2\frac{k_{2}^{+}\left(k_{2}^{+}-q^{\prime+}\right)}{q^{\prime+}}q^{\prime\nu}+\left(k_{2\perp\beta}-\frac{k_{2}^{+}}{q^{\prime+}}q_{\perp\beta}^{\prime}\right)\left[k_{2}^{+}\gamma_{\perp}^{\beta}\gamma_{\perp}^{\nu}+\left(k_{2}^{+}-q^{\prime+}\right)\gamma_{\perp}^{\nu}\gamma_{\perp}^{\beta}\right]\right\}\,. \nonumber 
\end{align}
Let us define the longitudinal momentum fractions $z_{1}\equiv k_1^{+}/q^{+}\equiv 1-\overline{z}_{1}$
and $z_{2}\equiv k_2^{+}/q^{\prime+}\equiv 1-\overline{z}_{2}$, and the following photon wave function numerators:
\begin{equation}
\psi_{L}(z_{1},\boldsymbol{k}_1-z_{1}\boldsymbol{q})\equiv-2z_{1}\overline{z}_{1}q^{\mu}\left(\varepsilon_{L\mu}-\frac{\varepsilon_{L}^{+}}{q^{+}}q_{\mu}\right),\label{eq:psiL-ini-def}
\end{equation}
\begin{equation}
\psi_{L}^{\prime\ast}(z_{2},\boldsymbol{k}_2-z_{2}\boldsymbol{q}^{\prime})\equiv-2z_{2}\overline{z}_{2}q^{\prime\nu}\left(\varepsilon_{L\nu}^{\prime\ast}-\frac{\varepsilon_{L}^{\prime\ast+}}{q^{\prime+}}q_{\nu}^{\prime}\right),\label{eq:psiL-fin-def}
\end{equation}
for the longitudinal ones, and 
\begin{equation}
\psi_{h}(z_{1},\boldsymbol{k}_1-z_{1}\boldsymbol{q})\equiv\left(\varepsilon_{h\perp\rho}-\frac{\varepsilon_{h}^{+}}{q^{+}}q_{\perp\rho}\right)(k_{1\perp\alpha}-z_{1}q_{\perp\alpha})(z_{1}\gamma_{\perp}^{\rho}\gamma_{\perp}^{\alpha}-\overline{z}_{1}\gamma_{\perp}^{\alpha}\gamma_{\perp}^{\rho}),\label{eq:psiT-ini-def}
\end{equation}

\begin{equation}
\psi_{h^{\prime}}^{\prime\ast}(z_{2},\boldsymbol{k}_2-z_{2}\boldsymbol{q}^{\prime})\equiv\left(\varepsilon_{h^{\prime}\perp\sigma}^{\prime\ast}-\frac{\varepsilon_{h^{\prime}}^{\prime\ast+}}{q^{\prime+}}q_{\perp\sigma}^{\prime}\right)(k_{2\perp\beta}-z_{2}q_{\perp\beta}^{\prime})(z_{2}\gamma_{\perp}^{\beta}\gamma_{\perp}^{\sigma}-\overline{z}_{2}\gamma_{\perp}^{\sigma}\gamma_{\perp}^{\beta})\,,\label{eq:psiT-fin-def}
\end{equation}
for the transverse ones. Note that in this case the $q^\mu$ and $q'^\mu$ terms do not contribute due to the transversity condition $\varepsilon_h\cdot q=0$ which is valid within our choice of gauge and parametrization of polarizations. With explicit photon polarization vectors
\begin{align}
& \left(\varepsilon_{L\mu}-\frac{\varepsilon_{L}^{+}}{q^{+}}q_{\mu}\right) = \frac{Q}{q^+}n_{2\mu}\\
& \left(\varepsilon_{h\rho}-\frac{\varepsilon^+_h}{q^{+}}q_{\rho}\right)=e_{\perp,h\rho} -\frac{\boldsymbol{e}_{h}\cdot\boldsymbol{q}}{q^+}n_{2\rho},
\end{align} 
we have
\begin{equation}
\psi_{L}(z_{1},\boldsymbol{k}_1-z_{1}\boldsymbol{q})\equiv-2z_{1}\overline{z}_{1}Q,\label{eq:psiL-ini-exp}
\end{equation}

\begin{equation}
\psi_{L}^{\prime\ast}(z_{2},\boldsymbol{k}_2-z_{2}\boldsymbol{q}^{\prime})\equiv2z_{2}\overline{z}_{2}Q^{\prime},\label{eq:psiL-fin-exp}\,
\end{equation}
for the projection on the longitudinal polarization vectors. Note the difference in sign between $\psi_{L}$ and $\psi'_{L}$ that is due to the fact that the incoming photon is spacelike whereas the outgoing one is timelike. Namely, we have $q^2=-Q^2<0$ and $q'^2=Q'^2>0$. 

The projections on the transverse polarizations are also straightforward and read
\begin{equation}
\psi_{h}(z_{1},\boldsymbol{k}_1-z_{1}\boldsymbol{q})\equiv e_{h\perp\rho}(k_{1\perp\alpha}-z_{1}q_{\perp\alpha})(z_{1}\gamma_{\perp}^{\rho}\gamma_{\perp}^{\alpha}-\overline{z}_{1}\gamma_{\perp}^{\alpha}\gamma_{\perp}^{\rho}) ,\label{eq:psiT-ini-exp}
\end{equation}
and
\begin{equation}
\psi_{h^{\prime}}^{\prime\ast}(z_{2},\boldsymbol{k}_2-z_{2}\boldsymbol{q}^{\prime})\equiv e_{h^{\prime}\perp\sigma}^{\prime\ast}(k_{2\perp\beta}-z_{2}q_{\perp\beta}^{\prime})(z_{2}\gamma_{\perp}^{\beta}\gamma_{\perp}^{\sigma}-\overline{z}_{2}\gamma_{\perp}^{\sigma}\gamma_{\perp}^{\beta})\,.\label{eq:psiT-fin-xp}
\end{equation}

Contracting the Dirac structure from Eq.~(\ref{eq:dirac-alg-1})
with all 4 possible photon polarization transitions (LL, LT, TL, TT),
we obtain for the LL transition
\begin{equation}
  \left(\varepsilon_{L\mu}-\frac{\varepsilon_{L}^{+}}{q^{+}}q_{\mu}\right)\left(\varepsilon_{L\nu}^{\prime\ast}-\frac{\varepsilon_{L}^{\prime\ast+}}{q^{\prime+}}q_{\nu}^{\prime}\right)A^{\mu \nu}\label{eq:phiLL}
  =2q^{+}q^{\prime+}{\rm tr}[\psi_{L}(z_{1},\boldsymbol{k}_1-z_{1}\boldsymbol{q})\psi_{L}^{\prime\ast}(z_{2},\boldsymbol{k}_2-z_{2}\boldsymbol{q}^{\prime})] ,
\end{equation}
the LT transition,
\begin{equation}
 \left(\varepsilon_{L\mu}-\frac{\varepsilon_{L}^{+}}{q^{+}}q_{\mu}\right)\left(\varepsilon_{h^{\prime}\nu}^{\prime\ast}-\frac{\varepsilon_{h^{\prime}}^{\prime\ast+}}{q^{\prime+}}q_{\nu}^{\prime}\right)A^{\mu \nu}
 =2q^{+}q^{\prime+}{\rm tr}[\psi_{L}(z_{1},\boldsymbol{k}_1-z_{1}\boldsymbol{q})\psi_{h^{\prime}}^{\prime\ast}(z_{2},\boldsymbol{k}_2-z_{2}\boldsymbol{q}^{\prime})],\label{eq:phiLT}
\end{equation}
the TL transition
\begin{equation}
 \left(\varepsilon_{h\mu}-\frac{\varepsilon_{h}^{+}}{q^{+}}q_{\mu}\right)\left(\varepsilon_{L\nu}^{\prime\ast}-\frac{\varepsilon_{L}^{\prime\ast+}}{q^{\prime+}}q_{\nu}^{\prime}\right)A^{\mu \nu}
  =2q^{+}q^{\prime+}{\rm tr}[\psi_{h}(z_{1},\boldsymbol{k}_1-z_{1}\boldsymbol{q})\psi_{L}^{\prime\ast}(z_{2},\boldsymbol{k}_2-z_{2}\boldsymbol{q}^{\prime})],\label{eq:phiTL}
\end{equation}
and the TT transition:
\begin{equation}
 \left(\varepsilon_{h\mu}-\frac{\varepsilon_{h}^{+}}{q^{+}}q_{\mu}\right)\left(\varepsilon_{h^{\prime}\nu}^{\prime\ast}-\frac{\varepsilon_{h^{\prime}}^{\prime\ast+}}{q^{\prime+}}q_{\nu}^{\prime}\right)A^{\mu \nu}
 =2q^{+}q^{\prime+}{\rm tr}[\psi_{h}(z_{1},\boldsymbol{k}_1-z_{1}\boldsymbol{q})\psi_{h^{\prime}}^{\prime\ast}(z_{2},\boldsymbol{k}_2-z_{2}\boldsymbol{q}^{\prime})].\label{eq:phiTT}
\end{equation}
 This way, defining $\lambda^{(\prime)}\in\left\{ L,h^{(\prime)}\right\} $,
we can consistently write the amplitude as follows:
\begin{align}
{\cal A}_{\lambda\lambda^{\prime}} & =4\pi\alpha_{{\rm em}}\sum_{f}q_{f}^{2}\mu^{2-d}(q^{+})^{2}\int\!\frac{{\rm d}z_{1}}{2\pi}\int\!\frac{{\rm d}z_{2}}{2\pi}\int\!\frac{{\rm d}^{d}\boldsymbol{k}_1}{(2\pi)^{d}}\frac{{\rm d}^{d}\boldsymbol{k}_2}{(2\pi)^{d}}\nonumber \\
 & \times\frac{q^{\prime +}}{2z_1\overline{z}_1 q^+}{\rm tr}[\psi_{\lambda}(z_1,\boldsymbol{k}_1-z_1\boldsymbol{q})\psi_{\lambda^{\prime}}^{\prime\ast}(z_2,\boldsymbol{k}_2-z_2\boldsymbol{q}^{\prime})]\\
 & \times\int\!\frac{{\rm d}k_2^{-}}{2\pi}\frac{{\rm d}k_1^{-}}{2\pi}\left\langle p^{\prime}\left|{\rm tr}G_{{\rm scal}}(k_2,k_1)G_{{\rm scal}}(k_1-q,k_2-q^{\prime})\right|p\right\rangle .\nonumber 
\end{align}

We will now follow closely the procedure outlined in~\cite{Boussarie:2020fpb, Boussarie:2021wkn} that  consists of factorizing the photon wave functions from the target operator. This is achieved by extracting the first and last interaction of the quark-antiquark pair with the target gauge field $A^-$. As a result, we obtain four terms corresponding to the various gauge field insertions: 

\begin{align}
 & \int\!\frac{{\rm d}k_2^{-}}{2\pi}\frac{{\rm d}k_1^{-}}{2\pi}\left\langle p^{\prime}\left|{\rm tr}G_{{\rm scal}}(k_2,k_1)G_{{\rm scal}}(k_1-q,k_2-q^{\prime})\right|p\right\rangle \nonumber \\
 & =-g^{2}\int\!{\rm d}^{D}x_{1}\,{\rm d}^{D}x_{2}\,{\rm d}^{D}y_{1}\,{\rm d}^{D}y_{2}\,\delta(x_{1}^{+}-y_{1}^{+})\,\delta(x_{2}^{+}-y_{2}^{+})\nonumber \\
 & \times{\rm e}^{-iq^{-}x_{1}^{+}-ik_1^{+}x_{1}^{-}+i(k_1^{+}-q^{+})y_{1}^{-}+i\boldsymbol{k}_1\cdot(\boldsymbol{x}_{1}-\boldsymbol{y}_{1})+i\boldsymbol{q}\cdot\boldsymbol{y}_{1}}\nonumber \\
 & \times{\rm e}^{iq^{\prime-}x_{2}^{+}+ik_2^{+}x_{2}^{-}-i(k_2^{+}-q^{\prime+})y_{2}^{-}-i\boldsymbol{k}_2\cdot(\boldsymbol{x}_{2}-\boldsymbol{y}_{2})-i\boldsymbol{q}^{\prime}\cdot\boldsymbol{y}_{2}}\nonumber \\
 & \times\frac{\theta(k_1^{+})\theta(q^{+}-k_1^{+})\theta(k_2^{+})\theta(q^{\prime+}-k_2^{+})}{\left(q^{-}-\frac{\boldsymbol{k}_1^{2}}{2k_1^{+}}-\frac{(\boldsymbol{q}-\boldsymbol{k}_1)^{2}}{2(q^{+}-k_1^{+})}+i0\right)\left(q^{\prime-}-\frac{\boldsymbol{k}_2^{2}}{2k_2^{+}}-\frac{(\boldsymbol{q}^{\prime}-\boldsymbol{k}_2)^{2}}{2(q^{\prime+}-k_2^{+})}+i0\right)}\label{eq:GGtoGAGA}\\
 & \times\left\langle p^{\prime}\left|{\rm tr}\left\{ [A^{-}(y_{1})-A^{-}(x_{1})]G_{{\rm scal}}(y_{1},y_{2})[A^{-}(y_{2})-A^{-}(x_{2})]G_{{\rm scal}}(x_{2},x_{1})\right\} \right|p\right\rangle .\nonumber 
\end{align}
This allows us to generate the so-called energy denominators which are necessary to recover the photon wave functions: 
\begin{equation}
q^{-}-\frac{\boldsymbol{k}_1^{2}}{2k_1^{+}}-\frac{(\boldsymbol{q}-\boldsymbol{k}_1)^{2}}{2(q^{+}-k_1^{+})}+i0=-\frac{(\boldsymbol{k}_1-z_1\boldsymbol{q})^{2}+z_1\overline{z}_1 Q^{2}-i0}{2z_1\overline{z}_1 q^{+}}\label{eq:ed1}
\end{equation}
and
\begin{equation}
q^{\prime-}-\frac{\boldsymbol{k}_2^{2}}{2k_2^{+}}-\frac{(\boldsymbol{q}^{\prime}-\boldsymbol{k}_2)^{2}}{(q^{\prime+}-k_2^{+})}+i0=-\frac{(\boldsymbol{k}_2-z_2\boldsymbol{q}^{\prime})^{2}-z_2\overline{z}_2Q^{\prime2}-i0}{2z_2\overline{z}_2q^{\prime+}}.\label{eq:ed2}
\end{equation}
 We can then complete our wave functions into
\begin{equation}
\Psi_{\lambda}(z_{1},\boldsymbol{k}_1-z_{1}\boldsymbol{q})\equiv\frac{\psi_{\lambda}(z_{1},\boldsymbol{k}_1-z_{1}\boldsymbol{q})}{(\boldsymbol{k}_1-z_{1}\boldsymbol{q})^{2}+z_{1}\overline{z}_{1}Q^{2}-i0}\label{eq:Psi-ini-def}
\end{equation}
and
\begin{equation}
\Psi_{\lambda^{\prime}}^{\prime\ast}(z_{2},\boldsymbol{k}_2-z_{2}\boldsymbol{q}^{\prime})\equiv\frac{\psi_{\lambda^{\prime}}^{\prime\ast}(z_{2},\boldsymbol{k}_2-z_{2}\boldsymbol{q}^{\prime})}{(\boldsymbol{k}_2-z_{2}\boldsymbol{q}^{\prime})^{2}-z_{2}\overline{z}_{2}Q^{\prime2}-i0}\label{eq:Psi-fin-def}
\end{equation}
in the amplitude, to obtain:
\begin{align}
{\cal A}_{\lambda\lambda^{\prime}} & =-4\pi\alpha_{{\rm em}}\sum_{f}q_{f}^{2}\mu^{2-d}g^{2}(q^{+})^{2}\int_{0}^{1}\!\frac{{\rm d}z_{2}}{2\pi}\int_{0}^{1}\!\frac{{\rm d}z_{1}}{2\pi}\int\!\frac{{\rm d}^{d}\boldsymbol{k}_1}{(2\pi)^{d}}\frac{{\rm d}^{d}\boldsymbol{k}_2}{(2\pi)^{d}}\nonumber \\
 & \times2z_2\overline{z}_2(q^{+})^{2}{\rm tr}[\Psi_{\lambda}(z_{1},\boldsymbol{k}_1-z_{1}\boldsymbol{q})\Psi_{\lambda^{\prime}}^{\prime\ast}(z_{2},\boldsymbol{k}_2-z_{2}\boldsymbol{q}^{\prime})]\nonumber \\
 & \times\int\!{\rm d}^{D}x_{1}\,{\rm d}^{D}x_{2}\,{\rm d}^{D}y_{1}\,{\rm d}^{D}y_{2}\,\,\delta(x_{1}^{+}-y_{1}^{+})\,\delta(x_{2}^{+}-y_{2}^{+})\nonumber \\
 & \times{\rm e}^{-iq^{-}x_{1}^{+}-iz_1q^+x_{1}^{-}-i\bar{z}_1q^+y_{1}^{-}+i\boldsymbol{k}_1\cdot(\boldsymbol{x}_{1}-\boldsymbol{y}_{1})+i\boldsymbol{q}\cdot\boldsymbol{y}_{1}}\label{eq:amp-with-den}\\
 & \times{\rm e}^{iq^{\prime-}x_{2}^{+}+iz_2 q^{\prime +}x_{2}^{-}+i\bar{z}_2q^{\prime +}y_{2}^{-}+i(\boldsymbol{k}_2-\boldsymbol{q}^{\prime})\cdot\boldsymbol{y}_{2}-i\boldsymbol{k}_2\cdot\boldsymbol{x}_{2}}\nonumber\\
 & \times\left\langle p^{\prime}\left|{\rm tr}\left\{ [A^{-}(y_{1})-A^{-}(x_{1})]G_{{\rm scal}}(y_{1},y_{2})[A^{-}(y_{2})-A^{-}(x_{2})]G_{{\rm scal}}(x_{2},x_{1})\right\} \right|p\right\rangle .\nonumber
\end{align}
It is possible to leverage the independence of the gluon fields on the $-$ coordinates and its direct consequence on the dependence of the effective propagators to simplify our expression.  To do so, we define $(d+1)$-dimensional propagators ${\cal G}$ via \cite{Blaizot:2015lma,Blaizot:2012fh}
\begin{equation}
G_{{\rm scal}}(x_{2},x_{1})=\int\frac{{\rm d}p^{+}}{2\pi}\frac{{\rm e}^{-ip^{+}(x_{2}^{-}-x_{1}^{-})}}{2ip^{+}} ( \boldsymbol{x}_{2} | {\cal G}_{p^{+}}(x_{2}^{+},x_{1}^{+}) | \boldsymbol{x}_{1}).\label{eq:calG-def}
\end{equation}
This allows us to perform a few more integrals. Notably, it will set $z_1=z_2$ and we will then denote $z\equiv z_1=z_2$. It will also ensure that $q^+=q^{\prime+}$. We obtain:
\begin{align}
{\cal A}_{\lambda\lambda^{\prime}} & =-2\pi\alpha_{{\rm em}}\sum_{f}q_{f}^{2}\mu^{2-d}g^{2}q^{+}(2\pi)\delta(q^{+}-q^{\prime+})\int_{0}^{1}\!\frac{{\rm d}z}{2\pi}\int\!\frac{{\rm d}^{d}\boldsymbol{k}_1}{(2\pi)^{d}}\frac{{\rm d}^{d}\boldsymbol{k}_2}{(2\pi)^{d}}\nonumber \\
 & \times{\rm tr}[\Psi_{\lambda}(z,\boldsymbol{k}_1-z\boldsymbol{q})\Psi_{\lambda^{\prime}}^{\prime\ast}(z,\boldsymbol{k}_2-z\boldsymbol{q}^{\prime})]\nonumber \\
 & \times\int\!{\rm d}x_{1}^{+}{\rm d}x_{2}^{+}{\rm d}^{d}\boldsymbol{x}_{1}{\rm d}^{d}\boldsymbol{x}_{2}{\rm d}^{d}\boldsymbol{y}_{1}{\rm d}^{d}\boldsymbol{y}_{2}\nonumber \\
 & \times{\rm e}^{-iq^{-}x_{1}^{+}+iq^{\prime-}x_{2}^{+}+i(\boldsymbol{k}_1\cdot\boldsymbol{x}_{1})-i\boldsymbol{k}_2\cdot\boldsymbol{x}_{2}-i\boldsymbol{k}_1-\boldsymbol{q}\cdot\boldsymbol{y}_{1}+i(\boldsymbol{k}_2-\boldsymbol{q}^{\prime})\cdot\boldsymbol{y}_{2}}\label{eq:amp-with-calG}\\
 & \times\left\langle p^{\prime}\left|{\rm tr}\left\{ [A^{-}(x_{1}^{+},\boldsymbol{y}_{1})-A^{-}(x_{1}^{+},\boldsymbol{x}_{1})]{\cal G}_{-\overline{z}q^{+}}(x_{1}^{+},x_{2}^{+};\boldsymbol{y}_{1},\boldsymbol{y}_{2})\right.\right.\right.\nonumber \\
 & \left.\left.\left.\times[A^{-}(x_{2}^{+},\boldsymbol{y}_{2})-A^{-}(x_{2}^{+},\boldsymbol{x}_{2})]{\cal G}_{zq^{+}}(x_{2}^{+},x_{1}^{+};\boldsymbol{x}_{2},\boldsymbol{x}_{1})\right\} \right|p\right\rangle .\nonumber 
\end{align}
The final step towards the most general expression for our amplitude
is to relate the difference of $A^{\mu}$ fields to field strength
tensors: for $A^i=0$, one has
\begin{equation}
A^{-}(x^{+},\boldsymbol{y})-A^{-}(x^{+},\boldsymbol{x})=-(\boldsymbol{y}-\boldsymbol{x})^{i}\int_{0}^{1}{\rm d}sF^{i-}(x^{+},\overline{s}\boldsymbol{x}+s\boldsymbol{y}),\label{eq:AtoF}
\end{equation}
with $\overline{s}\equiv1-s$. The $(\boldsymbol{y}_{1}-\boldsymbol{x}_{1})^{i}$
and $(\boldsymbol{y}_{2}-\boldsymbol{x}_{2})^{j}$ prefactors
can be integrated by parts into derivatives of the wave functions.
Finally, it yields:
\begin{align} 
{\cal A}_{\lambda\lambda^{\prime}} & =-2\pi\alpha_{{\rm em}}\sum_{f}q_{f}^{2}\mu^{2-d}g^{2}q^{+}(2\pi)\delta(q^{+}-q^{\prime+})\int_{0}^{1}\!\frac{{\rm d}z}{2\pi}\int\!\frac{{\rm d}^{d}\boldsymbol{k}_1}{(2\pi)^{d}}\frac{{\rm d}^{d}\boldsymbol{k}_2}{(2\pi)^{d}}\nonumber \\
 & \times{\rm tr}\left[(\partial^{i}\Psi_{\lambda})(z,\boldsymbol{k}_1-z\boldsymbol{q})(\partial^{j}\Psi_{\lambda^{\prime}}^{\prime\ast})(z,\boldsymbol{k}_2-z\boldsymbol{q}^{\prime})\right]\nonumber \\
 & \times\int\!{\rm d}x_{1}^{+}{\rm d}x_{2}^{+}{\rm d}^{d}\boldsymbol{x}_{1}{\rm d}^{d}\boldsymbol{x}_{2}{\rm d}^{d}\boldsymbol{y}_{1}{\rm d}^{d}\boldsymbol{y}_{2}\nonumber \\
 & \times{\rm e}^{-iq^{-}x_{1}^{+}+iq^{\prime-}x_{2}^{+}+i(\boldsymbol{k}_1\cdot\boldsymbol{x}_{1})-i\boldsymbol{k}_2\cdot\boldsymbol{x}_{2}-i\boldsymbol{k}_1-\boldsymbol{q}\cdot\boldsymbol{y}_{1}+i(\boldsymbol{k}_2-\boldsymbol{q}^{\prime})\cdot\boldsymbol{y}_{2}}\label{eq:amp-with-calG-1-and-F}\\
 & \times\int_{0}^{1}\!{\rm d}s\int_{0}^{1}\!{\rm d}s^{\prime}\left\langle p^{\prime}\left|{\rm tr}\left\{ F^{i-}(x_{1}^{+},\overline{s}\boldsymbol{x}_{1}+s\boldsymbol{y}_{1}){\cal G}_{-\overline{z}q^{+}}(x_{1}^{+},x_{2}^{+};\boldsymbol{y}_{1},\boldsymbol{y}_{2})\right.\right.\right.\nonumber \\
 & \left.\left.\left.\times F^{j-}(x_{2}^{+},\overline{s}^{\prime}\boldsymbol{x}_{2}+s^{\prime}\boldsymbol{y}_{2}){\cal G}_{zq^{+}}(x_{2}^{+},x_{1}^{+};\boldsymbol{x}_{2},\boldsymbol{x}_{1})\right\} \right|p\right\rangle .\label{eq:fact-amp-1}\nonumber 
\end{align}
We finally built the most general expression for the $\gamma^{(\ast)} P \rightarrow \gamma^{(\ast)}P$ amplitude with background field methods involving an external $A^-(x^+,0^-,{\boldsymbol x})$ field. Most notably, we were able to factorize the photon wave functions in a similar fashion to the standard high energy factorization (or shock wave approximation), with a key distinction being the enforcement of light-cone time ordering. The latter is crucial to ensure the dependence and convolution in $x$ of our final factorization formula. 

In the next section, we will simplify this expression further in order to retain only necessary powers of the kinematic scales.

\section{Partial Twist Expansion}\label{sec:PTE}

\begin{figure}[ht]
\centering
\includegraphics[width=8cm]{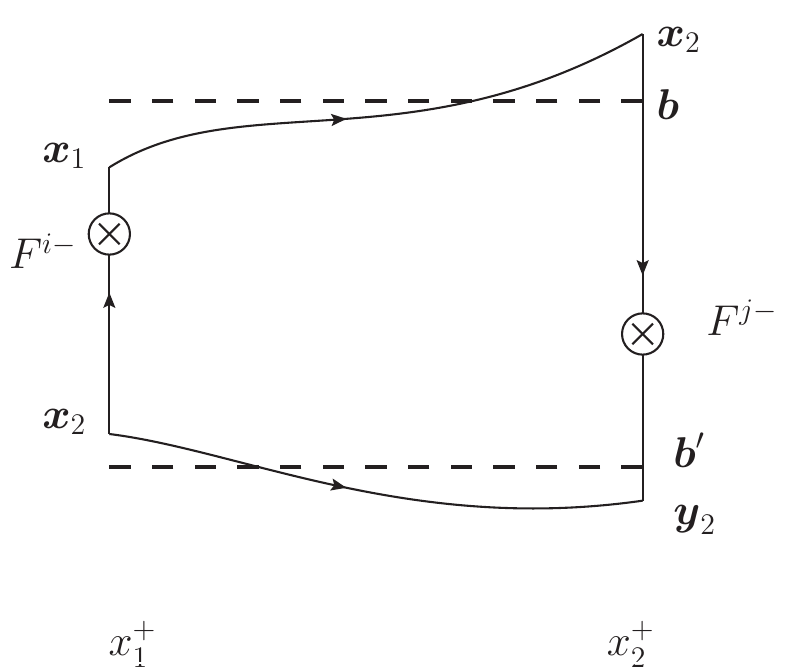}
\caption{Graphical representation of the dipole operator before partial twist expansion (PTE) in Eq. (\ref{eq:amp-with-calG-1-and-F}) and after in Eq. (\ref{eq:amp-with-calG-int-1}) (see text for details). 
\label{fig:figure3} }
\end{figure}

Although we have the wave functions we wanted, the operator in the last 2 lines of Eq.~(\ref{eq:amp-with-calG-1-and-F}) is still unnecessarily complicated: it still contains terms that are suppressed in both Regge and Bjorken limits. We will now perform a Partial Twist Expansion (PTE), which is a procedure that will enable us to get rid of such subleading corrections. 
Let us first examine the leading Gaussian (quantum phase) for a single propagator. For example for $zq^{+}>0$,
\begin{equation}
(\boldsymbol{x}_{2}| {\cal G}_{zq^{+}}^{(0)}(x_{2}^{+},x_{1}^{+}) |\boldsymbol{x}_{1}) \propto{\rm e}^{i\frac{zq^{+}[(\boldsymbol{x}_{2}-\boldsymbol{x}_{1})^{2}+i0]}{2(x_{2}^{+}-x_{1}^{+})}}\theta(x_{2}^{+}-x_{1}^{+}).\label{eq:G0}
\end{equation}
The Gaussian peaks at
\begin{equation}
(\boldsymbol{x}_{2}-\boldsymbol{x}_{1})^{2}\sim2(x_{2}^{+}-x_{1}^{+})/k^{+}.\label{eq:gaussian-arg}
\end{equation}
Parametrically, $zq^{+}\sim q^{+}$. However, we have two longitudinal
variables along the $-$ direction to evaluate the scope of $x_{2}^{+}-x_{1}^{+}$:
$p^{\prime-}-p^{-}$ and $\frac{p^{\prime-}+p^{-}}{2}$. Thankfully,
the correct Fourier conjugation is that differences in positions $x_{2}^{+}-x_{1}^{+}$
are conjugated to the average momentum $\frac{p^{\prime-}+p^{-}}{2}$
and the average position $\frac{x_{1}^{+}+x_{2}^{+}}{2}$ is conjugated
to the momentum transfer $p^{\prime-}-p^{-}$. This means that parametrically
$x_{2}^{+}-x_{1}^{+}\sim1/(p^{\prime-}+p^{-})$ and there is no risk
of enhancement by $1/(p^{\prime-}-p^{-})$ for small values of the
longitudinal momentum transfer. Overall, we get that
\begin{equation}
(\boldsymbol{x}_{2}-\boldsymbol{x}_{1})^{2}\sim\frac{1}{(p^{\prime-}+p^{-})q^{+}},\label{eq:gaussian-arg-approx}
\end{equation}
which is suppressed in both Regge and Bjorken regimes.

\begin{figure}[ht]
\centering
\includegraphics[width=9cm]{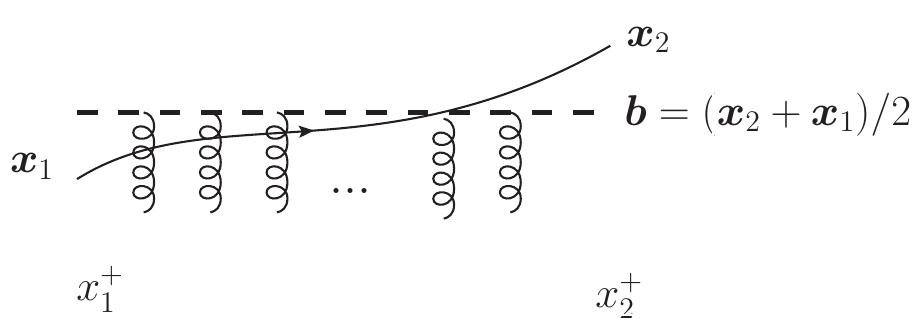}
\caption{Partial Twist Expansion: expansion of the high energy propagator around the straight line trajectory.  The dashed line represents light-like Wilson lines that result from the PTE procedure.\label{fig:figure2} }
\end{figure}

We can thus perform the PTE via a classical approximation~\cite{Boussarie:2020fpb,Boussarie:2021wkn}, comparably to e.g. some steps performed in the earlier works in~\cite{Altinoluk:2014oxa, Altinoluk:2015gia} and write:
\begin{equation}
(\boldsymbol{x}_{2}| {\cal G}_{zq^{+}}(x_{2}^{+},x_{1}^{+}) |\boldsymbol{x}_{1}) \simeq (\boldsymbol{x}_{2}| {\cal G}_{zq^{+}}^{(0)}(x_{2}^{+},x_{1}^{+})|\boldsymbol{x}_{1})[x_{2}^{+},x_{1}^{+}]_{\frac{\boldsymbol{x}_{2}+\boldsymbol{x}_{1}}{2}}\theta(x_{2}^{+}-x_{1}^{+}),.\label{eq:calGq-approx}
\end{equation}
for the quark propagator as depicted  by Figure~\ref{fig:figure2}, and
\begin{equation}
(\boldsymbol{y}_{1}|{\cal G}_{-\overline{z}q^{+}}(x_{1}^{+},x_{2}^{+})|\boldsymbol{y}_{2})\simeq(\boldsymbol{y}_{1}|{\cal G}_{-\overline{z}q^{+}}^{(0)}(x_{1}^{+},x_{2}^{+})|\boldsymbol{y}_{2})[x_{1}^{+},x_{2}^{+}]_{\frac{\boldsymbol{y}_{1}+\boldsymbol{y}_{2}}{2}}\theta(x_{2}^{+}-x_{1}^{+}),\label{eq:calGqbar-approx}
\end{equation}
 for the anti-quark propagator. We may also approximate 
\begin{equation}
F^{i-}(x_{1}^{+},\overline{s}\boldsymbol{x}_{1}+s\boldsymbol{y}_{1})\simeq F^{i-}\left(x_{1}^{+},\overline{s}\frac{\boldsymbol{x}_{2}+\boldsymbol{x}_{1}}{2}+s\frac{\boldsymbol{y}_{2}+\boldsymbol{y}_{1    }}{2}\right)\label{eq:Fi-approx}
\end{equation}
and
\begin{equation}
F^{j-}(x_{2}^{+},\overline{s}^{\prime}\boldsymbol{x}_{2}+s^{\prime}\boldsymbol{y}_{2})\simeq F^{j-}\left(x_{2}^{+},\overline{s}^{\prime}\frac{\boldsymbol{x}_{2}+\boldsymbol{x}_{1}}{2}+s^{\prime}\frac{\boldsymbol{y}_{2}+\boldsymbol{y}_{1}}{2}\right).\label{eq:Fj-approx}
\end{equation}
This allows us to carry out the $\boldsymbol{x}_{2}-\boldsymbol{x}_{1}$ and
$\boldsymbol{y}_{2}-\boldsymbol{y}_{1}$ integrals and keep only the
integrals w.r.t. 
\begin{align} 
\boldsymbol{b}\equiv\frac{\boldsymbol{x}_{2}+\boldsymbol{x}_{1}}{2} \quad \text{and} \quad \boldsymbol{b}^{\prime}\equiv\frac{\boldsymbol{y}_{2}+\boldsymbol{y}_{1}}{2}.
\end{align}
To do so, we simply need the following relations:
\begin{equation}
 \int{\rm d}^{d}(\boldsymbol{x}_{2}-\boldsymbol{x}_{1}){\rm e}^{-i\boldsymbol{K}\cdot(\boldsymbol{x}_{2}-\boldsymbol{x}_{1})}(\boldsymbol{x}_{2}|{\cal G}_{zq^{+}}^{(0)}(x_{2}^{+},x_{1}^{+})|\boldsymbol{x}_{1}) ={\rm e}^{-i\frac{x_{2}^{+}-x_{1}^{+}}{2zq^{+}}(\boldsymbol{K}^{2}-i0)}\theta(x_{2}^{+}-x_{1}^{+})\label{eq:G0q-int}
\end{equation}
and
\begin{equation}
  \int{\rm d}^{d}(\boldsymbol{y}_{2}-\boldsymbol{y}_{1}){\rm e}^{i\boldsymbol{L}\cdot(\boldsymbol{y}_{2}-\boldsymbol{y}_{1})}(\boldsymbol{y}_{1}|{\cal G}_{-\overline{z}q^{+}}^{(0)}(x_{1}^{+},x_{2}^{+})|\boldsymbol{y}_{2})
  ={\rm e}^{-i\frac{x_{2}^{+}-x_{1}^{+}}{2\overline{z}q^{+}}(\boldsymbol{L}^{2}-i0)}\theta(x_{2}^{+}-x_{1}^{+}),\label{eq:G0qbar-int}
\end{equation}
evaluated for 
\begin{equation}
    \boldsymbol{K}=\frac{\boldsymbol{k}_2+\boldsymbol{k}_1}{2} \quad \text{and} \quad \boldsymbol{L}=\frac{\boldsymbol{k}_2+\boldsymbol{k}_1-\boldsymbol{q}-\boldsymbol{q}^\prime}{2}\,,
\end{equation}
to obtain:
\begin{align}
{\cal A}_{\lambda\lambda^{\prime}} & =-2\pi\alpha_{{\rm em}}\sum_{f}q_{f}^{2}\mu^{2-d}g^{2}q^{+}(2\pi)\delta(q^{+}-q^{\prime+})\int\!{\rm d}x_{1}^{+}{\rm d}x_{2}^{+}{\rm d}^{d}\boldsymbol{b}{\rm d}^{d}\boldsymbol{b}^{\prime}\theta(x_{2}^{+}-x_{1}^{+}) \label{eq:amp-with-calG-int-1} \\
 & \times \int_{0}^{1}\!\frac{{\rm d}z}{2\pi}\int\!\frac{{\rm d}^{d}\boldsymbol{k}_1}{(2\pi)^{d}}\frac{{\rm d}^{d}\boldsymbol{k}_2}{(2\pi)^{d}} {\rm e}^{i(q^{\prime-}-q^{-})\frac{x_{2}^{+}+x_{1}^{+}}{2}} {\rm tr}[(\partial^{i}\Psi_{\lambda})(z,\boldsymbol{k}_1-z\boldsymbol{q})(\partial^{j}\Psi_{\lambda^{\prime}}^{\prime\ast})(z,\boldsymbol{k}_2-z\boldsymbol{q}^{\prime})] \nonumber \\
 & \times{\rm e}^{i(\boldsymbol{k}_1-\boldsymbol{k}_2)\cdot\boldsymbol{b}+i(\boldsymbol{k}_2-\boldsymbol{k}_1-\boldsymbol{q}^{\prime}+\boldsymbol{q})\cdot\boldsymbol{b}^{\prime}-i\frac{x_{2}^{+}-x_{1}^{+}}{2z\overline{z}q^{+}}\left[\overline{z}\left(\frac{\boldsymbol{k}_1+\boldsymbol{k}_2}{2}\right)^{2}+z\left(\frac{\boldsymbol{k}_1+\boldsymbol{k}_2}{2}-\frac{\boldsymbol{q}+\boldsymbol{q}^{\prime}}{2}\right)^{2}-z\overline{z}q^{+}(q^{-}+q^{\prime-})-i0\right]}\nonumber \\
 & \times\int_{0}^{1}\!{\rm d}s\int_{0}^{1}\!{\rm d}s^{\prime}\left\langle p^{\prime}\left|{\rm tr}\left\{ F^{i-}(x_{1}^{+},\overline{s}\boldsymbol{b}+s\boldsymbol{b}^{\prime})[x_{1}^{+},x_{2}^{+}]_{\boldsymbol{b}^{\prime}} F^{j-}(x_{2}^{+},\overline{s}^{\prime}\boldsymbol{b}+s^{\prime}\boldsymbol{b}^{\prime})[x_{2}^{+},x_{1}^{+}]_{\boldsymbol{b}}\right\} \right|p\right\rangle .\nonumber 
\end{align}
It is possible to use how the non-forward matrix element changes under translations (i.e. the phase one can extract by translating the operators) 
to integrate w.r.t. $\frac{x_{2}^{+}+x_{1}^{+}}{2}$
and w.r.t. $\frac{\boldsymbol{b}+\boldsymbol{b}^{\prime}}{2}$ and
get momentum conservation relations. With $v^{+}=x_{2}^{+}-x_{1}^{+}$
and $\boldsymbol{v}=\boldsymbol{b}^{\prime}-\boldsymbol{b}$, it follows 
\begin{align}
{\cal A}_{\lambda\lambda^{\prime}} & =-2\pi\alpha_{{\rm em}}\sum_{f}q_{f}^{2}\mu^{2-d}g^{2}(2\pi)^{D}\delta^{D}(q^{\prime}+p^{\prime}-q-p)\nonumber \\
 & \times\int_{0}^{1}\!\frac{{\rm d}z}{2\pi}\int\!\frac{{\rm d}^{d}\boldsymbol{k}_1}{(2\pi)^{d}}\frac{{\rm d}^{d}\boldsymbol{k}_2}{(2\pi)^{d}}{\rm tr}[(\partial^{i}\Psi_{\lambda})(z,\boldsymbol{k}_1-z\boldsymbol{q})(\partial^{j}\Psi_{\lambda^{\prime}}^{\prime\ast})(z,\boldsymbol{k}_2-z\boldsymbol{q}^{\prime})]\nonumber \\
 & \times\int\!{\rm d}v^{+}\int\!{\rm d}^{d}\boldsymbol{v}\theta(v^{+}){\rm e}^{i(\boldsymbol{k}_2-\boldsymbol{k}_1-\frac{\boldsymbol{q}^{\prime}-\boldsymbol{q}}{2})\cdot\boldsymbol{v}}\\
 & \times{\rm e}^{-i\frac{v^{+}}{2z\overline{z}q^{+}}\left[\overline{z}\left(\frac{\boldsymbol{k}_2+\boldsymbol{k}_1}{2}\right)^{2}+z\left(\frac{\boldsymbol{k}_2+\boldsymbol{k}_1}{2}-\frac{\boldsymbol{q}+\boldsymbol{q}^{\prime}}{2}\right)^{2}-z\overline{z}q^{+}(q^{-}+q^{\prime-})-i0\right]}\nonumber \\
 & \times\int_{0}^{1}\!{\rm d}s\int_{0}^{1}\!{\rm d}s^{\prime}\left\langle p^{\prime}\left|{\rm tr}\left\{ F^{i-}(-v^{+}/2,-\boldsymbol{v}/2+s\boldsymbol{v})[-v^{+}/2,v^{+}/2]_{\boldsymbol{v}/2}\right.\right.\right.\nonumber \\
 & \left.\left.\left.\times F^{j-}(v^{+}/2,-\boldsymbol{v}/2+s^{\prime}\boldsymbol{v})[v^{+}/2,-v^{+}/2]_{-\boldsymbol{v}/2}\right\} \right|p\right\rangle ,\nonumber 
\end{align}
where the Gaussian phase can be rewritten into
\begin{equation}
-i\frac{v^{+}}{2z\overline{z}q^{+}}\left[\left(\frac{\boldsymbol{k}_2+\boldsymbol{k}_1}{2}-z\frac{\boldsymbol{q}+\boldsymbol{q}^{\prime}}{2}\right)^{2}-z\overline{z}\left(\frac{q+q^{\prime}}{2}\right)^{2}-i0\right].\label{eq:phase}
\end{equation}
Let us now shift $\boldsymbol{k}_2\rightarrow\boldsymbol{k}_2+z\boldsymbol{q}^{\prime}$
and $\boldsymbol{k}_1\rightarrow\boldsymbol{k}_1+z\boldsymbol{q}$.
As a consequence of transverse boost invariance of the wave functions, we are then left with functions of just $\frac{\boldsymbol{k}_2+\boldsymbol{k}_1}{2}$
and $\boldsymbol{k}_2-\boldsymbol{k}_1$. Renaming $\frac{\boldsymbol{k}_2+\boldsymbol{k}_1}{2}\rightarrow\boldsymbol{\ell}$
and $\boldsymbol{k}_2-\boldsymbol{k}_1\rightarrow\boldsymbol{k}$,
the amplitude becomes
\begin{align}
{\cal A}_{\lambda\lambda^{\prime}} & =-2\pi\alpha_{{\rm em}}\sum_{f}q_{f}^{2}\mu^{2-d}g^{2}q^{+}(2\pi)^{D}\delta^{D}(q^{\prime}+p^{\prime}-q-p)\nonumber \\
 & \times\int_{0}^{1}\!\frac{{\rm d}z}{2\pi}\int\!\frac{{\rm d}^{d}\boldsymbol{\ell}}{(2\pi)^{d}}\frac{{\rm d}^{d}\boldsymbol{k}}{(2\pi)^{d}}{\rm tr}[(\partial^{i}\Psi_{\lambda})(z,\boldsymbol{\ell}-\boldsymbol{k}/2)(\partial^{j}\Psi_{\lambda^{\prime}}^{\prime\ast})(z,\boldsymbol{\ell}+\boldsymbol{k}/2)]\nonumber \\
 & \times\int\!{\rm d}v^{+}\int\!{\rm d}^{d}\boldsymbol{v}\theta(v^{+}){\rm e}^{i(\boldsymbol{k}-\frac{\overline{z}-z}{2}(\boldsymbol{q}^{\prime}-\boldsymbol{q}))\cdot\boldsymbol{v}}\label{eq:amp-almost-final}{\rm e}^{-i\frac{v^{+}}{2z\overline{z}q^{+}}\left[\boldsymbol{\ell}^{2}-z\overline{z}\left(\frac{q+q^{\prime}}{2}\right)^{2}-i0\right]} \\
 & \times\int_{0}^{1}\!{\rm d}s\int_{0}^{1}\!{\rm d}s^{\prime}\left\langle p^{\prime}\left|{\rm tr}\left\{ F^{i-}(-v^{+}/2,-\boldsymbol{v}/2+s\boldsymbol{v})[-v^{+}/2,v^{+}/2]_{\boldsymbol{v}/2}\right.\right.\right.\nonumber \\
 & \left.\left.\left.\times F^{j-}(v^{+}/2,-\boldsymbol{v}/2+s^{\prime}\boldsymbol{v})[v^{+}/2,-v^{+}/2]_{-\boldsymbol{v}/2}\right\} \right|p\right\rangle .\nonumber 
\end{align}
We almost have our final expression. However, the presence of light cone time ordering $\theta(v^+)$ spoils the interpretation of the last 2 lines as a parton distribution. Fortunately, we can rewrite this function without light cone time ordering using the fact that for any function $f$ we have
\begin{align}
\int{\rm d}v^{+}\theta\left(v^{+}\right){\rm e}^{-iv^{+}\left(k^{-}-i0\right)}f\left(v^{+}\right) & =i\int\frac{{\rm d}x}{x-\frac{k^{-}}{P^{-}}+i0}\int\frac{{\rm d}v^{+}}{2\pi}{\rm e}^{-ixP^{-}v^{+}}f\left(v^{+}\right),\label{eq:theta-to-den}
\end{align}
where $P^{-}$ is arbitrary. To recover the standard definition
of GPDs, we will eventually write $P^{-}=\frac{p^{\prime-}+p^{-}}{2}$.
It follows that
\begin{align}
{\cal A}_{\lambda\lambda^{\prime}} & =-2\pi i\alpha_{{\rm em}}\sum_{f}q_{f}^{2}\mu^{2-d}g^{2}(2\pi)^{D}\delta^{D}(q^{\prime}+p^{\prime}-q-p)\nonumber \\
 & \times\int_{0}^{1}\!\frac{{\rm d}z}{2\pi}\int\!\frac{{\rm d}^{d}\boldsymbol{\ell}}{(2\pi)^{d}}\int\!{\rm d}^{d}\boldsymbol{k}{\rm tr}\left[(\partial^{i}\Psi_{\lambda})(z,\boldsymbol{\ell}-\boldsymbol{k}/2)(\partial^{j}\Psi_{\lambda^{\prime}}^{\prime\ast})(z,\boldsymbol{\ell}+\boldsymbol{k}/2)\right]\nonumber \\
 & \times P^{-}q^{+}\int\!\frac{{\rm d}x}{x+\frac{\left(\frac{q+q^{\prime}}{2}\right)^{2}}{2q^{+}P^{-}}-\frac{\boldsymbol{\ell}^{2}}{2z\overline{z}q^{+}P^{-}}+i0}\label{eq:amp-almost-final-1}\\
 & \times\frac{1}{P^{-}}\int\!\frac{{\rm d}v^{+}}{2\pi}\int\frac{\!{\rm d}^{d}\boldsymbol{v}}{(2\pi)^{d}}{\rm e}^{-ixP^{-}v^{+}+i(\boldsymbol{k}-\frac{\overline{z}-z}{2}(\boldsymbol{q}^{\prime}-\boldsymbol{q}))\cdot\boldsymbol{v}}\nonumber \\
 & \times\int_{0}^{1}\!{\rm d}s\int_{0}^{1}\!{\rm d}s^{\prime}\left\langle p^{\prime}\left|{\rm tr}\left\{ F^{i-}(-v^{+}/2,-\boldsymbol{v}/2+s\boldsymbol{v})[-v^{+}/2,v^{+}/2]_{\boldsymbol{v}/2}\right.\right.\right.\nonumber \\
 & \left.\left.\left.\times F^{j-}(v^{+}/2,-\boldsymbol{v}/2+s^{\prime}\boldsymbol{v})[v^{+}/2,-v^{+}/2]_{-\boldsymbol{v}/2}\right\} \right|p\right\rangle .\nonumber 
\end{align}

\begin{figure} 
\centering
\includegraphics[width=8cm]{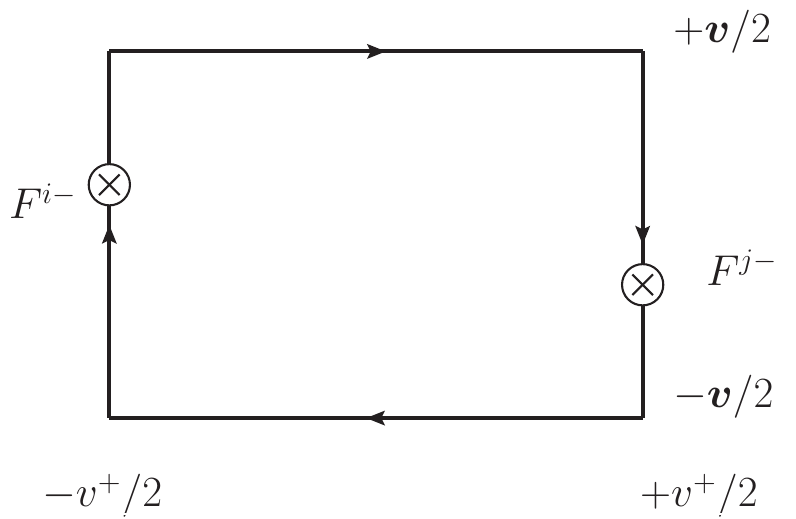}
\caption{Depiction of the 3D gluon operator whose Fourier transform yields the unintegrated gluon distribution. 
\label{fig:figure4}}
\end{figure}
Let us define the standard GPD variables 
\begin{equation}
    \Delta\equiv p^{\prime}-p,\quad P=\frac{p^{\prime}+p}{2},\quad \xi=\frac{p^--p^{\prime -}}{p^-+p^{\prime -}}.
\end{equation}
The correlator in the last 2 lines of Eq.~(\ref{eq:amp-almost-final-1}) defines the unintegrated GPD (cf. Figure~\ref{fig:figure4}):
\begin{align}
{\cal G}^{ij}(x,\xi,\boldsymbol{k},\boldsymbol{\Delta}) & \equiv \frac{1}{P^{-}}\int\!\frac{{\rm d}v^{+}}{2\pi}\int\frac{\!{\rm d}^{d}\boldsymbol{v}}{(2\pi)^{d}}{\rm e}^{-ixP^{-}v^{+}+i(\boldsymbol{k}\cdot\boldsymbol{v})}\label{eq:uGPD-def}\\
 & \times\int_{0}^{1}\!{\rm d}s\int_{0}^{1}\!{\rm d}s^{\prime}\left\langle p^{\prime}\left|{\rm tr}\left\{ F^{i-}(-v^{+}/2,-\boldsymbol{v}/2+s\boldsymbol{v})[-v^{+}/2,v^{+}/2]_{\boldsymbol{v}/2}\right.\right.\right.\nonumber \\
 & \left.\left.\left.\times F^{j-}(v^{+}/2,-\boldsymbol{v}/2+s^{\prime}\boldsymbol{v})[v^{+}/2,-v^{+}/2]_{-\boldsymbol{v}/2}\right\} \right|p\right\rangle .\nonumber 
\end{align}
Up to corrections which are simultaneously twist and energy suppressed where there is non-zero $+$ momentum transfer, ${\cal G}^{ij}$ is a function of only longitudinal and transverse variables, hence the argument in the l.h.s. This distribution is in practice very similar to a Generalized Transverse Momentum Dependent (GTMD) distribution where gluons carry respective momenta $(x-\xi)P^-n_2 + \Delta_\perp/2$ and $(x+\xi)P^-n_2-\Delta_\perp/2$, but with a very peculiar gauge link structure that invalidates simple partonic interpretations. We now have the final expression for the amplitude, valid in both
Regge and Bjorken regimes:
\begin{align}
{\cal A}_{\lambda\lambda^{\prime}} & =-2\pi i\alpha_{{\rm em}}\sum_{f}q_{f}^{2}\mu^{2-d}g^{2}(2\pi)^{D}\delta^{D}(q^{\prime}+p^{\prime}-q-p)\nonumber \\
 & \times P^{-}q^{+}\int\!{\rm d}x\int\!{\rm d}^{d}\boldsymbol{k}{\cal G}^{ij}(x,\xi,\boldsymbol{k}-\frac{z-\overline{z}}{2}\boldsymbol{\Delta},\boldsymbol{\Delta})\label{eq:amp-fin}\\
 & \times\int_{0}^{1}\!\frac{{\rm d}z}{2\pi}\int\!\frac{{\rm d}^{d}\boldsymbol{\ell}}{(2\pi)^{d}}\frac{{\rm tr}[(\partial^{i}\Psi_{\lambda})(z,\boldsymbol{\ell}-\boldsymbol{k}/2)(\partial^{j}\Psi_{\lambda^{\prime}}^{\prime\ast})(z,\boldsymbol{\ell}+\boldsymbol{k}/2)]}{x+\frac{\left(\frac{q+q^{\prime}}{2}\right)^{2}}{2q^{+}P^{-}}-\frac{\boldsymbol{\ell}^{2}}{2z\overline{z}q^{+}P^{-}}+i0}\,,\nonumber 
\end{align}
where ${\boldsymbol \Delta}=\boldsymbol q' - \boldsymbol q $ is the transverse momentum transfer.  

Finally, the ${\cal T}$ matrix given by ${\cal S}=1+i(2\pi)^{D}\delta^{D}(q^{\prime}+p^{\prime}-q-p){\cal T}$
reads,
\begin{align}
{\cal T}_{\lambda\lambda^{\prime}} & =-2\pi\alpha_{{\rm em}}\alpha_{s}\sum_{f}q_{f}^{2}\mu^{2-d}\nonumber \\
 & \times2P^{-}q^{+}\int\!{\rm d}x\int\!{\rm d}^{d}\boldsymbol{k}{\cal G}^{ij}(x,\xi,\boldsymbol{k}-\frac{z-\overline{z}}{2}\boldsymbol{\Delta},\boldsymbol{\Delta})\label{eq:amp-fin-1}\\
 & \times\int_{0}^{1}\!{\rm d}z\int\!\frac{{\rm d}^{d}\boldsymbol{\ell}}{(2\pi)^{d}}\frac{{\rm tr}[(\partial^{i}\Psi_{\lambda})(z,\boldsymbol{\ell}-\boldsymbol{k}/2)(\partial^{j}\Psi_{\lambda^{\prime}}^{\prime\ast})(z,\boldsymbol{\ell}+\boldsymbol{k}/2)]}{x+\frac{\left(\frac{q+q^{\prime}}{2}\right)^{2}}{2q^{+}P^{-}}-\frac{\boldsymbol{\ell}^{2}}{2z\overline{z}q^{+}P^{-}}+i0}.\nonumber 
\end{align}

In the case where $\Delta=0$, $\xi=0$ and $q=q'$ we recover Eq.~(4.21) in \cite{Boussarie:2021wkn}, whose imaginary part yields the inclusive DIS cross section by turning the denominator in the last line into a $\delta$ function relating $x$ to the other variables.

\section{The Bjorken limit: $Q^2 \to +\infty$ and $x$ fixed.}\label{sec:Bjorken}

The one-loop contributions to DVCS, TCS and DDVCS have been computed in the Bjorken regime decades ago \cite{Ji:1997nk, Belitsky:1997rh, Mankiewicz:1997bk, Hoodbhoy:1998vm, Ji:1998xh, Belitsky:1999sg} using the analytic continuation from a non-physical phase space to a physical one, which in practise was tantamount to using a spacelike outgoing photon in intermediate computation steps. This allowed to sweep potential issues for some integrals under the rug, restoring the physical space "by hand" by introducing an imaginary part to $\xi$. The way this imaginary part is introduced, however, depends on the sign of $x_{\rm Bj}$ and can lead to subtleties \cite{Mueller:2012sma}. In the present work, we are performing the honest-to-goodness computation with well defined imaginary parts instead, and find a consistent expression for all values of $x_{\rm Bj}$. Such a computation was done in \cite{Pire:2011st} by a standard computation of Feynman diagrams for one contribution, here we will show all possible cases by evaluating the wave functions which efficiently combine all the Feynman diagrams. We find our results to be compatible with all the known ones.

The Bjorken limit is obtained by taking $\boldsymbol{\ell}^{2}\sim\frac{Q^{2}+Q^{\prime2}}{2}\gg\boldsymbol{k}^{2}$, since $\boldsymbol{k}$ is an intrinsic transverse momentum inside the hadronic target and is thus at most of the order of the saturation scale $Q_s$, which is neglected in the Bjorken regime. In that regime, we also take $|\boldsymbol{\Delta}|\sim \sqrt{-t} \ll Q$. As a result, the  form of the $\boldsymbol{k}$ convolution simplifies, 
\begin{eqnarray}
    \int {\rm d}^d \boldsymbol{k} (\partial^{i}\Psi_{\lambda})(z,\boldsymbol{\ell}-\boldsymbol{k}/2)(\partial^{j}\Psi_{\lambda^{\prime}}^{\prime\ast})(z,\boldsymbol{\ell}+\boldsymbol{k}/2){\cal G}^{ij}(x,\xi,\boldsymbol{k}-\frac{z-\overline{z}}{2}\boldsymbol{\Delta},\boldsymbol{\Delta})   \\ \nonumber \sim (\partial^{i}\Psi_{\lambda})(z,\boldsymbol{\ell})(\partial^{j}\Psi_{\lambda^{\prime}}^{\prime\ast})(z,\boldsymbol{\ell}) \int {\rm d}^d \boldsymbol{k}  {\cal G}^{ij}(x,\xi,\boldsymbol{k},\boldsymbol{\Delta}), 
\end{eqnarray}
and we are left with the integral of the unintegrated GPD.

It is easy to prove that the unintegrated GPD integrates into a typical GPD correlator: writing
\begin{equation}
\int\!{\rm d}^{d}\boldsymbol{k}\, {\cal G}^{ij}(x,\xi,\boldsymbol{k},\boldsymbol{\Delta})\equiv G^{ij}(x,\xi,\boldsymbol{\Delta}),\label{eq:uGPD-to-GPD}
\end{equation}
we find the well known GPD correlator \cite{Diehl:2003ny, Belitsky:2005qn}
\begin{equation}
    G^{ij}(x,\xi,\boldsymbol{\Delta})\equiv \int \frac{{\rm d} v^+}{P^-} {\rm e}^{-ix P^- v^+} \left\langle p^\prime \left| F^{i-}(-v^{+}/2)[-v^{+}/2,v^{+}/2]  F^{j-}(v^{+}/2)[v^{+}/2,-v^{+}/2]  \right|p\right\rangle \label{eq:GPD-def}.
\end{equation}
Then,
\begin{align}
{\cal T}_{\lambda\lambda^{\prime}}^{{\rm Bjorken}} & =-2\pi\alpha_{{\rm em}}\alpha_{s}\sum_{f}q_{f}^{2}2P^{-}q^{+}\int\!{\rm d}xG^{ij}(x,\xi,\boldsymbol{\Delta})\int_{0}^{1}\!{\rm d}z\\
 & \times\mu^{2-d}\int\!\frac{{\rm d}^{d}\boldsymbol{\ell}}{(2\pi)^{d}}\frac{{\rm tr}\left[(\partial^{i}\Psi_{\lambda})(z,\boldsymbol{\ell}-\frac{z-\overline{z}}{4}\boldsymbol{\Delta})(\partial^{j}\Psi_{\lambda^{\prime}}^{\prime\ast})(z,\boldsymbol{\ell}+\frac{z-\overline{z}}{4}\boldsymbol{\Delta})\right]}{x+\frac{\left(\frac{q+q^{\prime}}{2}\right)^{2}}{2q^{+}P^{-}}-\frac{\boldsymbol{\ell}^{2}}{2z\overline{z}q^{+}P^{-}}+i0}.\nonumber
\end{align}
Neglecting $\boldsymbol{\Delta}$ in the wavefunctions, we can actually take the $\boldsymbol{\ell}$ and $z$ integrals to the end.
It is convenient to introduce the following variables:
\begin{equation}
\overline{Q}^{2}\equiv\frac{Q^{2}-Q^{\prime2}}{2},\quad x_{{\rm Bj}}\equiv\frac{\overline{Q}^{2}}{2q^{+}P^{-}}.\label{eq:xB-def}
\end{equation}
Then up to $\left|t\right|/s$ corrections,
\begin{equation}
\frac{\left(\frac{q+q^{\prime}}{2}\right)^{2}}{2q^{+}P^{-}}=-x_{{\rm Bj}},\quad\frac{Q^{2}}{2q^{+}P^{-}}=\xi+x_{{\rm Bj}},\quad\frac{Q^{\prime2}}{2q^{+}P^{-}}=\xi-x_{{\rm Bj}}.\label{eq:xB-rel}
\end{equation}
In terms of these variables, we will now proceed to evaluate our amplitude in the Bjorken regime. We will treat separately the transitions from a longitudinal (L) or transverse (T) photon to an L or T photon.

\subsection{LL transition}

\begin{align}
{\cal T}_{LL}^{{\rm Bjorken}} & =-4\pi\alpha_{s}\alpha_{{\rm em}}\sum_{f}q_{f}^{2} \frac{32QQ^{\prime}}{(2q^{+}P^{-})^{2}}\int\!{\rm d}xG^{ij}(x,\xi,\boldsymbol{\Delta}) \nonumber \\
 & \times\label{eq:amp-fin-2-1} \int_{0}^{1}\!\frac{{\rm d}z}{z\overline{z}}\int\!\mu^{2-d}\frac{{\rm d}^{d}\boldsymbol{\ell}}{(2\pi)^{d}}\frac{\frac{\boldsymbol{\ell}^{i}\boldsymbol{\ell}^{j}}{2z\overline{z}q^{+}P^{-}}}{\frac{\boldsymbol{\ell}^{2}}{2z\overline{z}q^{+}P^{-}}+(x_{{\rm Bj}}-x-i0)} \\
 & \times\frac{1}{\left(\frac{\boldsymbol{\ell}^{2}}{2z\overline{z}q^{+}P^{-}}+(x_{{\rm Bj}}+\xi-i0)\right)^{2}\left(\frac{\boldsymbol{\ell}^{2}}{2z\overline{z}q^{+}P^{-}}+(x_{{\rm Bj}}-\xi-i0)\right)^{2}}.\nonumber 
\end{align}
Owing to the rotational symmetry of the integrand we can make the replacement ${\boldsymbol{\ell}}^i{\boldsymbol{\ell}}^j \to \frac{1}{2}\delta^{ij } {\boldsymbol{\ell}}^2$. 
Then, using the fact that $G^{ii}(x,\xi,\boldsymbol{\Delta})=G^{ii}(-x,\xi,\boldsymbol{\Delta})$\footnote{see e.g. \cite{Diehl:2003ny} or \cite{Belitsky:2005qn}} and using the integrals from Appendix~\ref{sec:intapp}, one gets:
\begin{align}
{\cal T}_{LL}^{{\rm Bjorken}} & =-\alpha_{{\rm em}}\alpha_{s}\sum_{f}q_{f}^{2}2q^{+}P^{-}\frac{16QQ^{\prime}}{(Q^{2}+Q^{\prime2})^{2}}\int\!{\rm d}xG^{ii}(x,\xi,\boldsymbol{\Delta})\label{eq:amp-fin-2-1-1} \\
 & \times\left\{ \frac{1}{x+\xi}\left(\frac{x_{{\rm Bj}}}{\xi}+\frac{x_{{\rm Bj}}+\xi}{x+\xi}\right)\ln\left(\frac{x_{{\rm Bj}}-x-i0}{x_{{\rm Bj}}+\xi-i0}\right)\right.\nonumber \\
 & \left.-\frac{1}{x-\xi}\left(\frac{x_{{\rm Bj}}}{\xi}-\frac{x_{{\rm Bj}}-\xi}{x-\xi}\right)\ln\left(\frac{x_{{\rm Bj}}-x-i0}{x_{{\rm Bj}}-\xi-i0}\right)\right\} .\nonumber 
\end{align}
Note that the $x=\pm\xi$ poles cancel each other. Indeed, we may
have written instead
\begin{align}
{\cal T}_{LL}^{{\rm Bjorken}} & =i\alpha_{{\rm em}}\alpha_{s}\sum_{f}q_{f}^{2}2q^{+}P^{-}\frac{16QQ^{\prime}}{(Q^{2}+Q^{\prime2})^{2}} \int\!{\rm d}x\frac{G^{ii}(x,\xi,\boldsymbol{\Delta})}{x-\xi}\label{eq:ALL-sym} \\
 & \times\left[\left(\frac{x_{{\rm Bj}}}{\xi}-\frac{x_{{\rm Bj}}+\xi}{x-\xi}\right)\ln\left(\frac{x_{{\rm Bj}}+x-i0}{x_{{\rm Bj}}+\xi-i0}\right)+\left(\frac{x_{{\rm Bj}}}{\xi}-\frac{x_{{\rm Bj}}-\xi}{x-\xi}\right)\ln\left(\frac{x_{{\rm Bj}}-x-i0}{x_{{\rm Bj}}-\xi-i0}\right)\right],\nonumber 
\end{align}
where the $x=\xi$ pole cancels explicitly. This ensures continuity at $x=\xi$ and thus prevents end point singularities at this point. 

This result does not contain any collinear divergence. This is expected because the quark contribution to the $LL$ transition amplitude is twist suppressed, hence no counterterm from the evolution of the quark GPD would regularize the $LL$ transition from gluons. Our expression is compatible with the one found in~\cite{Mankiewicz:1997bk}.

\subsection{LT and TL transitions}

In the considered limit, the $\left[(\partial^{i}\Psi_{L})(z,\boldsymbol{\ell})(\partial^{j}\Psi_{h^{\prime}}^{\prime\ast})(z,\boldsymbol{\ell})\right]$
and $\left[(\partial^{i}\Psi_{h})(z,\boldsymbol{\ell})(\partial^{j}\Psi_{L}^{\prime\ast})(z,\boldsymbol{\ell})\right]$
overlaps are odd under $\boldsymbol{\ell}\leftrightarrow-\boldsymbol{\ell}$,
thus the LT and TL transition amplitudes cancel upon $\boldsymbol{\ell}$
integration.

\subsection{TT transition}

The GPD operator from Eq.~(\ref{eq:GPD-def}) has two open transverse indices $i$ and $j$, it can thus be decomposed into an unpolarized (U) term with a projection on $\delta^{ij}$, a polarized (P) term with a projection on $\epsilon^{ij}$ and transversity (T) terms with a projector that is symmetric and traceless for two index pairs $\tau^{ij,pq}=\delta^{ip}\delta^{jq}+\delta^{iq}\delta^{jp}-\frac{2}{d}\delta^{ij}\delta^{pq}$. Let us write 
\begin{equation}
G^{ij}(x,\xi,\boldsymbol{\Delta})\equiv\frac{\delta^{ij}}{d}G(x,\xi,t)+\frac{\epsilon^{ij}}{d}\widetilde{G}(x,\xi,t)+\frac{\tau^{ij,pq}}{d}G_{T}^{pq}(x,\xi,t).\label{eq:GPD-decomp}
\end{equation}
This is allowed if the smallness of $\Delta_\perp \sim \sqrt{|t|}$ allows to neglect the possible $\boldsymbol{\Delta}^i \boldsymbol{\Delta}^j$ tensor. In this decomposition, the r.h.s. then involves the well known unpolarized ($G$), polarized ($\widetilde{G}$) and transversity ($G_T$) gluon GPDs. Let us define the projected amplitudes:
\begin{equation}
{\cal T}_{hh^{\prime};u}^{{\rm Bjorken}}\equiv\left.{\cal T}_{hh^{\prime}}^{{\rm Bjorken}}\right|_{G^{ij}\rightarrow\frac{\delta^{ij}}{d}G},\quad{\cal T}_{hh^{\prime};p}^{{\rm Bjorken}}\equiv\left.{\cal T}_{hh^{\prime}}^{{\rm Bjorken}}\right|_{G^{ij}\rightarrow\frac{\epsilon^{ij}}{d}\widetilde{G}},\quad{\cal T}_{hh^{\prime};T}^{{\rm Bjorken}}\equiv\left.{\cal T}_{hh^{\prime}}^{{\rm Bjorken}}\right|_{G^{ij}\rightarrow\frac{\tau^{ij,pq}}{d}G_{T}^{pq}}\label{eq:A-decomp}
\end{equation}
Using the integrals from Appendix~\ref{sec:intapp} along with the symmetry
(resp. antisymmetry) of $G,G_{T}$ (resp. $\widetilde{G}$)\footnote{see\cite{Diehl:2003ny} or \cite{Belitsky:2005qn}} we get:
\begin{align}
{\cal T}_{hh^{\prime};u}^{{\rm Bjorken}} & =-2\alpha_{{\rm em}}\alpha_{s}\sum_{f}q_{f}^{2}\delta^{mn}\boldsymbol{e}_{h}^{m}\boldsymbol{e}_{h^{\prime}}^{\prime n\ast}\int\!{\rm d}x\frac{G(x,\xi,\boldsymbol{\Delta})}{(x+\xi-i0x_{{\rm Bj}})(x-\xi+i0x_{{\rm Bj}})}\nonumber \\
 & \times\frac{1}{\epsilon}(1-2\epsilon)\left(\frac{{\rm e}^{\gamma_{E}}}{4\pi}\frac{Q^{2}+Q^{\prime2}}{2\mu^{2}}\right)^{\epsilon}\label{eq:ATTu}\\
 & \times\left\{ \frac{x^{2}-2x_{{\rm Bj}}x+2x_{{\rm Bj}}^{2}-\xi^{2}}{(x+\xi-i0x_{{\rm Bj}})(x-\xi+i0x_{{\rm Bj}})}\ln(x_{{\rm Bj}}-x-i0)\right.\nonumber \\
 & +\frac{x^{2}+2x_{{\rm Bj}}x+2x_{{\rm Bj}}^{2}-\xi^{2}}{(x+\xi-i0x_{{\rm Bj}})(x-\xi+i0x_{{\rm Bj}})}\ln(x_{{\rm Bj}}+x-i0)\nonumber \\
 & +\frac{x_{{\rm Bj}}-\xi}{\xi}\frac{x^{2}-2x_{{\rm Bj}}\xi-\xi^{2}}{(x+\xi-i0x_{{\rm Bj}})(x-\xi+i0x_{{\rm Bj}})}\ln(x_{{\rm Bj}}-\xi-i0)\nonumber \\
 & -\frac{x_{{\rm Bj}}+\xi}{\xi}\frac{x^{2}+2x_{{\rm Bj}}\xi-\xi^{2}}{(x+\xi-i0x_{{\rm Bj}})(x-\xi+i0x_{{\rm Bj}})}\ln(x_{{\rm Bj}}+\xi-i0)\nonumber \\
 & +\epsilon\Big[\ln(x_{{\rm Bj}}-x-i0)+\ln(x_{{\rm Bj}}+x-i0)\nonumber \\
 & -\frac{x_{{\rm Bj}}+\xi}{\xi}\ln(x_{{\rm Bj}}+\xi-i0)+\frac{x_{{\rm Bj}}-\xi}{\xi}\ln(x_{{\rm Bj}}-\xi-i0)\nonumber \\
 & +\frac{1}{2}\frac{(x^{2}-2x_{{\rm Bj}}x+2x_{{\rm Bj}}^{2}-\xi^{2})}{(x+\xi-i0x_{{\rm Bj}})(x-\xi+i0x_{{\rm Bj}})}\ln^{2}\left(\frac{x_{{\rm Bj}}-x-i0}{\xi}\right)\nonumber \\
 & +\frac{1}{2}\frac{(x^{2}+2x_{{\rm Bj}}x+2x_{{\rm Bj}}^{2}-\xi^{2})}{(x+\xi-i0x_{{\rm Bj}})(x-\xi+i0x_{{\rm Bj}})}\ln^{2}\left(\frac{x_{{\rm Bj}}+x-i0}{\xi}\right)\nonumber \\
 & +\frac{1}{2}\frac{x_{{\rm Bj}}-\xi}{\xi}\frac{x^{2}-2x_{{\rm Bj}}\xi-\xi^{2}}{(x+\xi-i0x_{{\rm Bj}})(x-\xi+i0x_{{\rm Bj}})}\ln^{2}\left(\frac{x_{{\rm Bj}}-\xi-i0}{\xi}\right)\nonumber \\
 &  \left.\left.-\frac{1}{2}\frac{x_{{\rm Bj}}+\xi}{\xi}\frac{x^{2}+2x_{{\rm Bj}}\xi-\xi^{2}}{(x+\xi-i0x_{{\rm Bj}})(x-\xi+i0x_{{\rm Bj}})}\ln^{2}\left(\frac{x_{{\rm Bj}}+\xi-i0}{\xi}\right)\right] \right\} ,\nonumber 
\end{align}
for the unpolarized contribution, 
\begin{align}\label{eq:Tpol}
{\cal T}_{hh^{\prime};p}^{{\rm Bjorken}} & =2\alpha_{{\rm em}}\alpha_{s}\sum_{f}q_{f}^{2}\epsilon^{mn}\boldsymbol{e}_{h}^{m}\boldsymbol{e}_{h^{\prime}}^{\prime n\ast}\frac{1}{\epsilon}\left(\frac{{\rm e}^{\gamma_{E}}}{4\pi}\frac{Q^{2}+Q^{\prime2}}{2\mu^{2}}\right)^{\epsilon}\nonumber\\
 & \times\int\!{\rm d}x\frac{\widetilde{G}(x,\xi,\boldsymbol{\Delta})}{(x+\xi-i0x_{{\rm Bj}})^{2}(x-\xi+i0x_{{\rm Bj}})^{2}}\\
 & \times\left\{ (x^{2}-2xx_{{\rm Bj}}+\xi^{2})\ln(x_{{\rm Bj}}-x-i0)\right. -(x^{2}+2xx_{{\rm Bj}}+\xi^{2})\ln(x_{{\rm Bj}}+x-i0)\nonumber\\
 & +2x(x_{{\rm Bj}}+\xi)\ln(x_{{\rm Bj}}+\xi-i0) +2x(x_{{\rm Bj}}-\xi)\ln(x_{{\rm Bj}}-\xi-i0)\nonumber\\
 & +\epsilon \Big[-2(2x^{2}-3xx_{{\rm Bj}}+\xi^{2})\ln(x_{{\rm Bj}}-x-i0) +2(2x^{2}+3xx_{{\rm Bj}}+\xi^{2})\ln(x_{{\rm Bj}}+x-i0) \nonumber\\
 & -6x(x_{{\rm Bj}}+\xi)\ln(x_{{\rm Bj}}+\xi-i0) -6\epsilon x(x_{{\rm Bj}}-\xi)\ln(x_{{\rm Bj}}-\xi-i0)\nonumber\\
 & +\frac{1}{2}(x^{2}-2xx_{{\rm Bj}}+\xi^{2})\ln^{2}\left(\frac{x_{{\rm Bj}}-x-i0}{\xi}\right) -\frac{1}{2}(x^{2}+2xx_{{\rm Bj}}+\xi^{2})\ln^{2}\left(\frac{x_{{\rm Bj}}+x-i0}{\xi}\right)\nonumber\\
 & \left. \left. + x(x_{{\rm Bj}}+\xi)\ln^{2}\left(\frac{x_{{\rm Bj}}+\xi-i0}{\xi}\right)+x(x_{{\rm Bj}}-\xi)\ln^{2}\left(\frac{x_{{\rm Bj}}-\xi-i0}{\xi}\right)\right] \right\} ,\centering \nonumber
\end{align}
for the polarized contribution, and
\begin{align}
{\cal T}_{hh^{\prime};T}^{{\rm Bjorken}} & =2\alpha_{{\rm em}}\alpha_{s}\sum_{f}q_{f}^{2}\tau^{mn,ij}\boldsymbol{e}_{h}^{m}\boldsymbol{e}_{h^{\prime}}^{\prime n\ast}\int\!{\rm d}x\frac{G_{T}^{ij}(x,\xi,\boldsymbol{\Delta})}{(x-\xi+i0x_{{\rm Bj}})^{2}(x+\xi-i0x_{{\rm Bj}})^{2}}\nonumber \\
 & \times\left[(x^{2}-\xi^{2})+(x_{{\rm Bj}}^{2}-\xi^{2})\ln\frac{(x_{{\rm Bj}}-x-i0)(x_{{\rm Bj}}+x-i0)}{(x_{{\rm Bj}}-\xi-i0)(x_{{\rm Bj}}+\xi-i0)}\right]\label{eq:ATT-T}
\end{align}
for the transversity contribution.

Using the symmetrized expression from Eq.~(\ref{eq:ATTu}), it is
easy to check that our results are compatible with \cite{Pire:2011st} after a misprint in Eq. (10) from that reference has been corrected. Indeed, the denominator in that equation is only correct in the DVCS limit. It should read instead $(x-\xi+i0x_{\rm Bj})(x+\xi-i0x_{\rm Bj})$. Also, the DVCS limit $x_{\rm Bj} \rightarrow \xi$ of Eq.~(\ref{eq:ATT-T}) is compatible with~\cite{Hoodbhoy:1998vm}.

\subsection{Cancelling the divergence}

\begin{figure} 
\label{fig:figure5}
\begin{subfigure}[b]{0.5\textwidth}
         \centering
         \includegraphics[width=7cm]{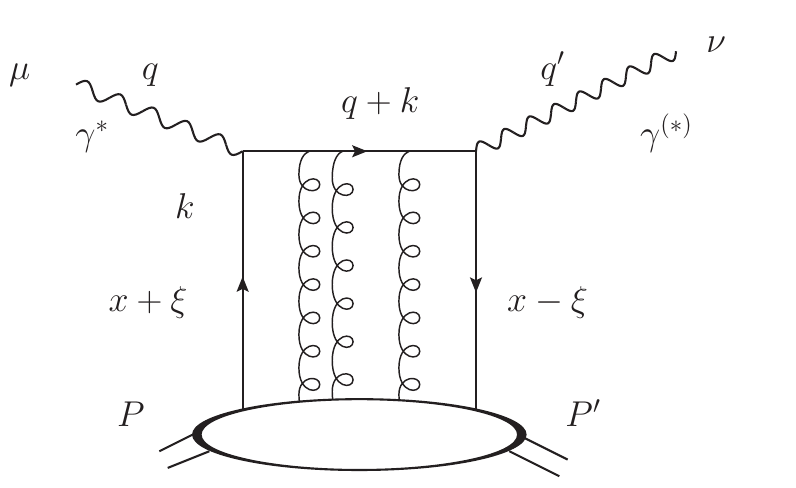}
         \caption{LO contribution}
         \label{fig:figure5-a}
     \end{subfigure}
     \qquad
     \begin{subfigure}[b]{0.5\textwidth}
         \centering
         \includegraphics[width=7cm]{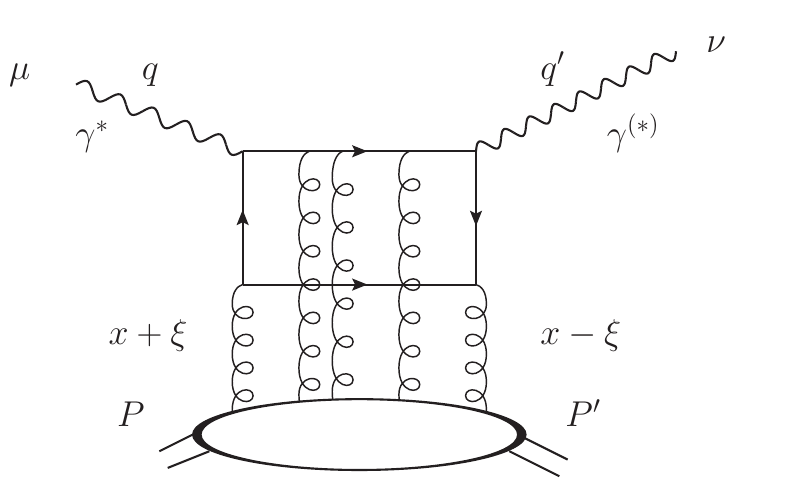}
         \caption{Gluon contribution to the NLO. }
         \label{fig:figure5-b}
     \end{subfigure}

\caption{Diagrams contributing to the $\gamma^{\ast} P\rightarrow\gamma^{\ast}P$ amplitude at LO and NLO in the Bjorken limit.    }
\end{figure}

Some of the transitions presented above have divergences for $\epsilon \rightarrow 0$. Indeed, the gluon contribution we are computing is in fact a loop correction to the leading order handbag diagram (cf. Figure~\ref{fig:figure5-a})  and such a divergence is expected. It must be canceled via the renormalization of the GPD operators, yielding their well known evolution equations. The convolution of the leading order handbag amplitude with the evolution equation then provides a counterterm to our divergence, and the purpose of this section is to prove the fact that the counterterm and the divergence cancel each other.

\subsubsection{Leading order (LO) quark amplitude}
The handbag amplitude is straightforward to compute. Factorizing in spinor space using the Fierz identity with the base of Fierz matrices $\Gamma^\lambda$, it reads:
\begin{align}
{\cal A}^{q} & =-\frac{ie_{q}^{2}}{4}(2\pi)^{D}\delta^{D}(q^{\prime}-q+p^{\prime}-p)\int{\rm d}^{D}z\langle p^{\prime}|\overline{\psi}(-\frac{z}{2})\Gamma^{\lambda}\psi(\frac{z}{2})|p\rangle\\
 & \times\int\frac{{\rm d}^{D}k}{(2\pi)^{D}}\frac{1}{k^{2}+i0}{\rm tr}\left[\slashed{\varepsilon}_{q^{\prime}}^{\ast}\slashed{k}\slashed{\varepsilon}_{q}\Gamma_{\lambda}{\rm e}^{i\left(k-\frac{q^{\prime}+q}{2}\right)\cdot z}+\slashed{\varepsilon}_{q}\slashed{k}\slashed{\varepsilon}_{q^{\prime}}^{\ast}\Gamma_{\lambda}{\rm e}^{i\left(k+\frac{q^{\prime}+q}{2}\right)\cdot z}\right]. \nonumber
\end{align}
Only vector $\Gamma^\lambda = \gamma^\lambda$ and axial-vector $\Gamma^\lambda = \gamma^\lambda\gamma_5$ projections contribute. Up to twist corrections the GPDs can also be taken to depend only on $z^{-}$, and their Lorentz index $\lambda$ can be taken to be $-$.  Neglecting
target mass corrections, in forward kinematics and in a frame where
the incoming photon has no transverse momentum, we also have $\boldsymbol{q}^{\prime}=\boldsymbol{q}=\boldsymbol{0},$
$q^{+}=q^{\prime+},$ $q^{-}=\frac{-Q^{2}}{2q^{+}},$ $q^{\prime-}=\frac{Q^{\prime2}}{2q^{+}}$.
The $LT$ and $TL$ transitions are identically $0$ for $\boldsymbol{\Delta}=\boldsymbol{0}$,
and the $LL$ transition is also suppressed at leading twist so we
will only compute the $TT$ transition amplitude. The photon polarizations
are transverse in the sense of Minkowski space, which simplifies the
traces. All in all, one finds:
\begin{align}
{\cal A}^{q} & =-\frac{ie_{q}^{2}}{4}(2\pi)^{D}\delta^{D}(q^{\prime}-q+p^{\prime}-p)\frac{1}{2}\int{\rm d}x\, {\rm e}^{ixP^{-}z^{+}} \nonumber \\
 & \times\int\frac{{\rm d}z^{+}}{2\pi}\langle p^{\prime}|\overline{\psi}(-\frac{1}{2}z^{+})\Gamma^{-}\psi(\frac{1}{2}z^{+})|p\rangle\\
 & \times{\rm tr}\left[\frac{\slashed{\varepsilon}_{q^{\prime}\perp}^{\ast}\gamma^{-}\slashed{\varepsilon}_{q\perp}\Gamma^{+}}{x+\frac{q^{\prime-}+q^{-}}{2P^{-}}+i0}+\frac{\slashed{\varepsilon}_{q\perp}\gamma^{-}\slashed{\varepsilon}_{q^{\prime}\perp}^{\ast}\Gamma^{+}}{x-\frac{q^{\prime-}+q^{-}}{2P^{-}}-i0}\right].\nonumber
\end{align}

 Defining the leading twist quark GPD operators with the conventions of~\cite{Diehl:2003ny}
\begin{align}
F^{q}(x) & =\frac{1}{2}\int\frac{{\rm d}z^{+}}{2\pi}{\rm e}^{ixP^{-}z^{+}}\left\langle p^{\prime}\left|\overline{\psi}(-\frac{1}{2}z^{+})\gamma^{-}\psi(\frac{1}{2}z^{+})\right|p\right\rangle \\
\widetilde{F}^{q}(x) & =\frac{1}{2}\int\frac{{\rm d}z^{+}}{2\pi}{\rm e}^{ixP^{-}z^{+}}\left\langle p^{\prime}\left|\overline{\psi}(-\frac{1}{2}z^{+})\gamma^{-}\gamma_{5}\psi(\frac{1}{2}z^{+})\right|p\right\rangle 
\end{align}
we get the standard result:
\begin{align}
{\cal A}^{q} & =-ie_{q}^{2}(2\pi)^{D}\delta^{D}(q^{\prime}-q+p^{\prime}-p)\,\boldsymbol{\varepsilon}_{q}^{i}\boldsymbol{\varepsilon}_{q^{\prime}}^{j\ast}\int{\rm d}x\nonumber\\
 & \times\left[\delta^{ij}F^{q}(x)\left(\frac{1}{x-x_{{\rm Bj}}+i0}+\frac{1}{x+x_{{\rm Bj}}-i0}\right)\right.\\
 & \left.-i\boldsymbol{\epsilon}^{ij}\widetilde{F}^{q}(x)\left(\frac{1}{x-x_{{\rm Bj}}+i0}-\frac{1}{x+x_{{\rm Bj}}-i0}\right)\right],\nonumber
\end{align}
where we used ${\rm tr}\left[\slashed{\varepsilon}_{q^{\prime}\perp}^{\ast}\slashed{\varepsilon}_{q\perp}\gamma^{-}\gamma^{+}\gamma_{5}\right]=-4i\boldsymbol{\epsilon}^{ij}\boldsymbol{\varepsilon}_{q}^{i}\boldsymbol{\varepsilon}_{q^{\prime}}^{j\ast}.$

\subsubsection{Evolution equations and counterterms}

In this article, we only considered gluon contributions to our observables. Nevertheless, what we computed also accounts for an indirect quark contribution that emerges from the
collinearly divergent part in the form of the $K^{qg}$ splitting
functions. This contribution has to be subtracted using the evolution
equation for the quark GPDs.
Let us study the unpolarized term in detail first, as the polarized counterterms will be obtained in a similar fashion. If $F^{q}$ is the unpolarized quark GPD operator as defined earlier,
the connection between its bare $F_{q}^{q}$ and renormalized $F_{R}^{q}$
forms in the $\overline{MS}$ scheme reads:
\begin{equation}
\left.F_{0}^{q}(y)\right|_{g}=F_{R}^{q}(y)-\frac{2}{\epsilon}\left(\frac{{\rm e}^{\gamma_{E}}\mu_{F}^{2}}{4\pi\mu^{2}}\right)^{\epsilon}\int_{-1}^{1}{\rm d}xK^{qg}(y,x)G(x,\xi,\mu^{2}), \label{eq:F0R}
\end{equation}
where we did not write the quark contribution from $K^{qq}$. Note the factor $2$
here, which is absent from the usual formulation for the evolution.
It is due to our definition for $G$ where we used the trace of gluon
field strength tensors so we have a $T_{F}=1/2$ factor compared to the usual
definition. The $qg$ kernel
reads \cite{Ji:1996nm}:
\begin{align}
K^{qg}\left(y,x\right) & =\frac{\alpha_{s}}{2\pi}T_{F}\frac{y^{2}+(x-y)^{2}-\xi^{2}}{(x^{2}-\xi^{2})^{2}}\left[\theta(y-\xi)\theta(x-y)-\theta(-\xi-y)\theta(y-x)\right]\nonumber \\
 & +\frac{\alpha_{s}}{2\pi}T_{F}\frac{(y+\xi)(x-2y+\xi)}{2\xi(x+\xi)(x^{2}-\xi^{2})}\theta(x-y)\theta(\xi-y)\theta(y+\xi) \\
 & +\frac{\alpha_{s}}{2\pi}T_{F}\frac{(y-\xi)(x-2y-\xi)}{2\xi(x-\xi)(x^{2}-\xi^{2})}\theta(y-x)\theta(\xi-y)\theta(y+\xi). \nonumber
\end{align}
At the level of the kernel, we need to distinguish between the $(1>x>\xi)$, $(\xi>x>-\xi)$ and $(-\xi>x>-1)$ regions. That said, all regions can be shown to contribute equally to the counterterm, see Appendix~\ref{sec:regions}.
If we write the amplitude for the quark contribution as 
\begin{equation}
{\cal A}^{q}(\mu^{2})=\int_{-1}^{1}{\rm d}yC^{q}(y)F_{R}^{q}(y),
\end{equation}
we obtain a counterterm to the gluon contribution in the following form:
\begin{equation}
{\cal A}_{{\rm div};u}=-\frac{2}{\epsilon}\sum_{f}\left(\frac{{\rm e}^{\gamma_{E}}\mu_{F}^{2}}{4\pi\mu^{2}}\right)^{\epsilon}\int_{-1}^{1}{\rm d}x\int_{-1}^{1}{\rm d}yC^{q}(y)K^{qg}(y,x)G(x,\xi,\mu^{2}).
\end{equation}
Explicitly,
\begin{align}
{\cal A}_{{\rm div};u} & =4\pi i\alpha_{{\rm em}}\sum_{f}q_{f}^{2}(2\pi)^{D}\delta^{D}(q^{\prime}-q+p^{\prime}-p)(\boldsymbol{\varepsilon}_{q}\cdot\boldsymbol{\varepsilon}_{q^{\prime}}^{\ast})\nonumber\\
 & \times\frac{2}{\epsilon}\left(\frac{{\rm e}^{\gamma_{E}}\mu_{F}^{2}}{4\pi\mu^{2}}\right)^{\epsilon}\int_{-1}^{1}{\rm d}xG(x,\xi,\mu^{2})\label{eq:counterterm-in}\\
 & \times\int_{-1}^{1}{\rm d}yK^{qg}(y,x)\left(\frac{1}{y-x_{{\rm Bj}}+i0}+\frac{1}{y+x_{{\rm Bj}}-i0}\right).\nonumber
\end{align}
We thus need the following integral:
\begin{align}
I^{qg} & =\int_{-1}^{1}{\rm d}yK^{qg}(y,x)\left(\frac{1}{y-x_{{\rm Bj}}+i0}+\frac{1}{y+x_{{\rm Bj}}-i0}\right)\,,
\end{align}
which is straightforward to compute, and reads:
\begin{align}
I^{qg} & =\frac{\alpha_{s}}{2\pi}\frac{T_{F}}{(x^{2}-\xi^{2})^{2}}\\
 & \times\left[(x^{2}-2x_{{\rm Bj}}x+2x_{{\rm Bj}}^{2}-\xi^{2})\ln(x_{{\rm Bj}}-x-i0)\right. +(x^{2}+2x_{{\rm Bj}}x+2x_{{\rm Bj}}^{2}-\xi^{2})\ln(x_{{\rm Bj}}+x-i0)\nonumber \\
 & -\frac{x_{{\rm Bj}}+\xi}{\xi}(x^{2}+2\xi x_{{\rm Bj}}-\xi^{2})\ln(x_{{\rm Bj}}+\xi-i0) \left.+\frac{x_{{\rm Bj}}-\xi}{\xi}(x^{2}-2x_{{\rm Bj}}\xi-\xi^{2})\ln(x_{{\rm Bj}}-\xi-i0)\right]. \nonumber
\end{align}
Finally summing over all quark flavors, the total counterterm to the
unpolarized ${\cal T}$ matrix from the renormalization of the unpolarized quark GPD operators
becomes:
\begin{align}
{\cal T}_{{\rm div};u} & =\alpha_{{\rm em}}\alpha_{s}\sum_{f}q_{f}^{2}(\boldsymbol{\varepsilon}_{q}\cdot\boldsymbol{\varepsilon}_{q^{\prime}}^{\ast}) \frac{2}{\epsilon}\left(\frac{{\rm e}^{\gamma_{E}}\mu_{F}^{2}}{4\pi\mu^{2}}\right)^{\epsilon}\int_{-1}^{1}{\rm d}x\frac{G(x,\xi,\mu^{2})}{(x^{2}-\xi^{2})^{2}}\label{eq:counterterm-u}\\
 & \times\left[(x^{2}-2x_{{\rm Bj}}x+2x_{{\rm Bj}}^{2}-\xi^{2})\ln(x_{{\rm Bj}}-x-i0)\right. +(x^{2}+2x_{{\rm Bj}}x+2x_{{\rm Bj}}^{2}-\xi^{2})\ln(x_{{\rm Bj}}+x-i0)\nonumber\\
 & -\frac{x_{{\rm Bj}}+\xi}{\xi}(x^{2}+2\xi x_{{\rm Bj}}-\xi^{2})\ln(x_{{\rm Bj}}+\xi-i0)\left.+\frac{x_{{\rm Bj}}-\xi}{\xi}(x^{2}-2x_{{\rm Bj}}\xi-\xi^{2})\ln(x_{{\rm Bj}}-\xi-i0)\right].\nonumber
\end{align}
Note that the double poles for $x=\pm\xi$ are not a cause for concern: the numerator is quadratic in $(x\mp\xi)$ for $x\rightarrow\pm\xi$.

The polarized case is very similar. The gluonic part of the associated kernel reads
\begin{align}
\Delta K^{qg}(y,x) & =i\frac{\alpha_{s}}{2\pi}T_{F}\frac{y^{2}-(x-y)^{2}-\xi^{2}}{(x^{2}-\xi^{2})^{2}}\left[\theta(x-y)\theta(y-\xi)-\theta(y-x)\theta(-\xi-y)\right]\nonumber\\
 & -i\frac{\alpha_{s}}{2\pi}\frac{T_{F}}{2\xi}\frac{(y+\xi)(x-\xi)}{(x+\xi)(x^{2}-\xi^{2})}\theta(x-y)\theta(\xi-y)\theta(y+\xi)\\
 & -i\frac{\alpha_{s}}{2\pi}\frac{T_{F}}{2\xi}\frac{(y-\xi)(x+\xi)}{(x-\xi)(x^{2}-\xi^{2})}\theta(y-x)\theta(\xi-y)\theta(y+\xi).\nonumber
\end{align}
This time, the integral for the counterterm reads as follows:
\begin{align}
 & \int_{-1}^{1}{\rm d}y\left(\frac{1}{y-x_{{\rm Bj}}+i0}-\frac{1}{y+x_{{\rm Bj}}-i0}\right)\Delta K^{qg}(y,x)\\
 & =-i\frac{\alpha_{s}}{2\pi}T_{F}\frac{1}{(x^{2}-\xi^{2})^{2}}\left[ 2x(x_{{\rm Bj}}+\xi)\ln(x_{{\rm Bj}}+\xi-i0) +2x(x_{{\rm Bj}}-\xi)\ln(x_{{\rm Bj}}-\xi-i0) \right. \nonumber\\
 & \left. +(x^{2}-2xx_{{\rm Bj}}+\xi^{2})\ln(x_{{\rm Bj}}-x-i0) -(x^{2}+2xx_{{\rm Bj}}+\xi^{2})\ln(x_{{\rm Bj}}+x-i0) \right]\nonumber
\end{align}
and the counterterm to the ${\cal T}$ matrix becomes:
\begin{align}
\widetilde{{\cal T}}_{{\rm div};p} & =\alpha_{{\rm em}}\alpha_{s}\sum_{f}q_{f}^{2}\boldsymbol{\epsilon}^{ij}\boldsymbol{\varepsilon}_{q}^{i}\boldsymbol{\varepsilon}_{q^{\prime}}^{j\ast} \frac{-2}{\epsilon}\left(\frac{{\rm e}^{\gamma_{E}}\mu_{F}^{2}}{4\pi\mu^{2}}\right)^{\epsilon}\int_{-1}^{1}{\rm d}x\frac{\widetilde{G}(x,\xi,\mu^{2})}{(x^{2}-\xi^{2})^{2}}\label{eq:counterterm-p}\\
 & \times\left[(x^{2}-2xx_{{\rm Bj}}+\xi^{2})\ln(x_{{\rm Bj}}-x-i0) -(x^{2}+2xx_{{\rm Bj}}+\xi^{2})\ln(x_{{\rm Bj}}+x-i0)\right.\nonumber\\
 & \left. +2x(x_{{\rm Bj}}+\xi)\ln(x_{{\rm Bj}}+\xi-i0) +2x(x_{{\rm Bj}}-\xi)\ln(x_{{\rm Bj}}-\xi-i0)\right]\nonumber
\end{align}
We thus found the counterterms to Eqs.~(\ref{eq:ATTu}, \ref{eq:Tpol}), whose purpose will be to cancel the $1/\epsilon$ poles therein. We will now proceed to prove that cancellation.

\subsubsection{Combining counterterms and unrenormalized terms}

It is natural to choose a factorization scale of the order of $Q^2+Q^{\prime 2}$, so let us take the evolution equation for $\mu_{F}^{2}=\alpha\frac{Q^{2}+Q^{\prime2}}{2}$ and add the resulting counterterm to our divergent one loop amplitudes. The unpolarized combination is now finite:
\begin{align}
{\cal T}_{hh^{\prime};u}^{{\rm Bjorken}}&+{\cal T}_{{\rm div};u} =2\alpha_{{\rm em}}\alpha_{s}\sum_{f}q_{f}^{2}\int\!{\rm d}x\frac{(\boldsymbol{\varepsilon}_{q}\cdot\boldsymbol{\varepsilon}_{q^{\prime}}^{\ast})G(x,\xi,\boldsymbol{\Delta})}{(x+\xi-i0x_{{\rm Bj}})^{2}(x-\xi+i0x_{{\rm Bj}})^{2}}\nonumber \\
 & \times\frac{1}{2\xi}\left\{ \left[(x_{{\rm Bj}}+\xi)(x^{2}-\xi^{2}+4\xi x_{{\rm Bj}}+4\xi x)\ln\left(\frac{x_{{\rm Bj}}+x-i0}{x_{{\rm Bj}}+\xi-i0}\right)\right.\right.\label{eq:TBu-fin} \\
 & -\frac{x_{{\rm Bj}}+\xi}{2}(x^{2}-\xi^{2}+2\xi x_{{\rm Bj}}+2\xi x)\left[\ln^{2}\left(\frac{x_{{\rm Bj}}+x-i0}{\alpha\xi}\right)-\ln^{2}\left(\frac{x_{{\rm Bj}}+\xi-i0}{\alpha\xi}\right)\right]\nonumber\\
 & +\frac{x_{{\rm Bj}}-\xi}{2}(x^{2}-\xi^{2}-2\xi x_{{\rm Bj}}-2\xi x)\left[\ln^{2}\left(\frac{x_{{\rm Bj}}+x-i0}{\alpha\xi}\right)-\ln^{2}\left(\frac{x_{{\rm Bj}}-\xi-i0}{\alpha\xi}\right)\right]\nonumber \\
 & \left.\left.-(x_{{\rm Bj}}-\xi)(x^{2}-\xi^{2}-4\xi x_{{\rm Bj}}+4\xi)\ln\left(\frac{x_{{\rm Bj}}+x-i0}{x_{{\rm Bj}}-\xi-i0}\right)\right]+(x\rightarrow-x)\right\} \nonumber 
\end{align}
The choice of $\alpha=1$ or $2$ or another "constant" number seems most natural to us, but previous computations have been using $\alpha=\frac{Q^{2}-Q^{\prime2}}{Q^{2}+Q^{\prime2}}=\frac{x_{{\rm Bj}}}{\xi}$. With that choice, we find an exact match between our unpolarized amplitude and results from \cite{Mankiewicz:1997bk}. However, that choice makes it artificially look like the amplitude may be ill-defined when $Q^2 = Q^{\prime 2}$ although it is finite so we suggest to stick to $\alpha=1$ or $2$ instead.
Similarly the polarized expression becomes:
\begin{align}
{\cal T}_{hh^{\prime};p}^{{\rm Bjorken}} & + {\cal T}_{{\rm div};p} =2\alpha_{{\rm em}}\alpha_{s}\sum_{f}q_{f}^{2}\int\!{\rm d}x\frac{\epsilon^{mn}\boldsymbol{e}_{h}^{m}\boldsymbol{e}_{h^{\prime}}^{\prime n\ast}\widetilde{G}(x,\xi,\boldsymbol{\Delta})}{(x+\xi-i0x_{{\rm Bj}})^{2}(x-\xi+i0x_{{\rm Bj}})^{2}}\nonumber \\
 & \times\left\{ \left[2(2x^{2}+\xi^{2})\ln\left(\frac{x_{{\rm Bj}}+x-i0}{\xi}\right)\right.\right.\label{eq:TBp-fin}\\
 & +3x(x_{{\rm Bj}}+\xi)\ln\left(\frac{x_{{\rm Bj}}+x-i0}{x_{{\rm Bj}}+\xi-i0}\right)+3x(x_{{\rm Bj}}-\xi)\ln\left(\frac{x_{{\rm Bj}}+x-i0}{x_{{\rm Bj}}-\xi-i0}\right)\nonumber \\
 & -\frac{1}{2}x(x_{{\rm Bj}}+\xi)\left[\ln^{2}\left(\frac{x_{{\rm Bj}}+x-i0}{\alpha\xi}\right)-\ln^{2}\left(\frac{x_{{\rm Bj}}+\xi-i0}{\alpha\xi}\right)\right]\nonumber \\
 & -\frac{1}{2}x(x_{{\rm Bj}}-\xi)\left[\ln^{2}\left(\frac{x_{{\rm Bj}}+x-i0}{\alpha\xi}\right)-\ln^{2}\left(\frac{x_{{\rm Bj}}-\xi-i0}{\alpha\xi}\right)\right]\nonumber \\
 & \left.\left.-\frac{1}{2}(x^{2}+\xi^{2})\ln^{2}\left(\frac{x_{{\rm Bj}}+x-i0}{\alpha\xi}\right)\right]-\left(x\rightarrow-x\right)\right\} \nonumber 
\end{align}
which is also compatible with \cite{Mankiewicz:1997bk}.

\section{Regge limit}\label{sec:Regge}

In the previous sections, we obtained finite results for the Bjorken limit of our amplitude and confirmed their compatibility with known results. Here, we will now investigate the Regge limit. Naively\footnote{We will elaborate on the more rigorous Regge limit in Section~\ref{sec:Double Limit}}, it is obtained by picking the cut contribution from the
denominator in Eq.~(\ref{eq:amp-fin}), neglecting all $1/(2q^{+}P^{-})$
corrections, that is, 
\begin{equation}
\frac{1}{x+\frac{\left(\frac{q+q^{\prime}}{2}\right)^{2}}{2q^{+}P^{-}}-\frac{\boldsymbol{\ell}^{2}}{2z\overline{z}q^{+}P^{-}}+i0}\rightarrow-i\pi\delta(x).\label{eq:Regge-cut}
\end{equation}
Also going to coordinate space as follows:
\begin{align}
 & {\rm tr}[(\partial^{i}\Psi_{\lambda})(z,\boldsymbol{\ell}-\boldsymbol{k}/2)(\partial^{j}\Psi_{\lambda^{\prime}}^{\prime\ast})(z,\boldsymbol{\ell}+\boldsymbol{k}/2)]\label{eq:PsiToPhi}\\
 & =\int{\rm d}^{d}\boldsymbol{x}_{1}{\rm d}^{d}\boldsymbol{x}_{2}\boldsymbol{x}_{1}^{i}\boldsymbol{x}_{2}^{j}{\rm e}^{-i(\boldsymbol{\ell}-\frac{\boldsymbol{k}}{2})\cdot\boldsymbol{x}_{1}+i(\boldsymbol{\ell}+\frac{\boldsymbol{k}}{2})\cdot\boldsymbol{x}_{2}}{\rm tr}[\Phi_{\lambda}(z,\boldsymbol{x}_{1})\Phi_{\lambda^{\prime}}^{\prime\ast}(z,\boldsymbol{x}_{2})],\nonumber 
\end{align}
the previous step allows for full integration w.r.t. $\boldsymbol{\ell}$
into a $\delta$ function:
\begin{align}
 & \int\!\frac{{\rm d}^{d}\boldsymbol{\ell}}{(2\pi)^{d}}\frac{{\rm tr}[(\partial^{i}\Psi_{\lambda})(z,\boldsymbol{\ell}-\boldsymbol{k}/2)(\partial^{j}\Psi_{\lambda^{\prime}}^{\prime\ast})(z,\boldsymbol{\ell}+\boldsymbol{k}/2)]}{x+\frac{\left(\frac{q+q^{\prime}}{2}\right)^{2}}{2q^{+}P^{-}}-\frac{\boldsymbol{\ell}^{2}}{2z\overline{z}q^{+}P^{-}}+i0}\nonumber \\
 & \rightarrow(-i\pi)\delta(x)\int\!\frac{{\rm d}^{d}\boldsymbol{\ell}}{(2\pi)^{d}}{\rm tr}[(\partial^{i}\Psi_{\lambda})(z,\boldsymbol{\ell}-\boldsymbol{k}/2)(\partial^{j}\Psi_{\lambda^{\prime}}^{\prime\ast})(z,\boldsymbol{\ell}+\boldsymbol{k}/2)]\nonumber \\
 & =(-i\pi)\delta(x)\int{\rm d}^{d}\boldsymbol{r}{\rm e}^{i(\boldsymbol{k}\cdot\boldsymbol{r})}\boldsymbol{r}^{i}\boldsymbol{r}^{j}{\rm tr}[\Phi_{\lambda}(z,\boldsymbol{r})\Phi_{\lambda^{\prime}}^{\prime\ast}(z,\boldsymbol{r})].\label{eq:ell-int-Regge}
\end{align}
As a crucial consequence of neglecting the $\boldsymbol \ell$ dependence in the delta function, both photon wave functions are evaluated a the same transverse coordinate $\boldsymbol r$ that is associated with the size of the quark dipole that interacts instantaneously with the target. Then,
\begin{align}
{\cal T}_{\lambda\lambda^{\prime}}^{{\rm Regge}} & =2i\pi^{2}\alpha_{s}\alpha_{{\rm em}}\sum_{f}q_{f}^{2}\mu^{2-d}\nonumber \\
 & \times2P^{-}q^{+}\int_{0}^{1}\!{\rm d}z\int{\rm d}^{d}\boldsymbol{r}{\rm tr}[\Phi_{\lambda}(z,\boldsymbol{r})\Phi_{\lambda^{\prime}}^{\prime\ast}(z,\boldsymbol{r})]\label{eq:amp-regge-int}\\
 & \times\int\!{\rm d}^{d}\boldsymbol{k}{\rm e}^{i(\boldsymbol{k}\cdot\boldsymbol{r})}\boldsymbol{r}^{i}\boldsymbol{r}^{j}{\cal G}^{ij}(x=0,\xi,\boldsymbol{k}-\frac{z-\overline{z}}{2}\boldsymbol{\Delta},\boldsymbol{\Delta}).\nonumber 
\end{align}
We now need to examine the $x=0$ limit of the uGPD in order to reconstruct the dipole operator which follows from the semi-classical description in the shock wave approximation. 
First, note that
\begin{align}
 & \int\!{\rm d}^{d}\boldsymbol{k}{\rm e}^{i(\boldsymbol{k}\cdot\boldsymbol{r})}\boldsymbol{r}^{i}\boldsymbol{r}^{j}{\cal G}^{ij}(x=0,\xi,\boldsymbol{k}-\frac{z-\overline{z}}{2}\boldsymbol{\Delta},\boldsymbol{\Delta})\nonumber \\
 & =\frac{1}{P^{-}}\int\!\frac{{\rm d}v^{+}}{2\pi}{\rm e}^{-i\frac{z-\overline{z}}{2}(\boldsymbol{\Delta}\cdot\boldsymbol{r})}\label{eq:uGPD-regge}\\
 & \times\left\langle p^{\prime}\left|{\rm tr}\left\{ [A^{-}(-v^{+}/2,-\boldsymbol{r}/2)-A^{-}(-v^{+}/2,\boldsymbol{r}/2)][-v^{+}/2,v^{+}/2]_{-\boldsymbol{r}/2}\right.\right.\right.\nonumber \\
 & \left.\left.\left.\times[A^{-}(v^{+}/2,-\boldsymbol{r}/2)-A^{-}(v^{+}/2,\boldsymbol{r}/2)][v^{+}/2,-v^{+}/2]_{\boldsymbol{r}/2}\right\} \right|p\right\rangle .\nonumber 
\end{align}
The Regge limit also implies $\xi\simeq0$. Let us isolate the involved operator
\begin{align}
{\cal O}\left(-\frac{v^{+}}{2},-\frac{\boldsymbol{r}}{2};\frac{v^{+}}{2},\frac{\boldsymbol{r}}{2}\right) & \equiv{\rm tr}\left\{ [A^{-}(-v^{+}/2,-\boldsymbol{r}/2)-A^{-}(-v^{+}/2,\boldsymbol{r}/2)][-v^{+}/2,v^{+}/2]_{-\boldsymbol{r}/2}\right.\nonumber \\
 & \left.\times[A^{-}(v^{+}/2,-\boldsymbol{r}/2)-A^{-}(v^{+}/2,\boldsymbol{r}/2)][v^{+}/2,-v^{+}/2]_{\boldsymbol{r}/2}\right\} .\label{eq:ope}
\end{align}
For target states normalized as $\left\langle p^\prime|p\right\rangle =2p^{-}(2\pi)^{d+1}\delta(p^{\prime-}-p^-)\delta^{d}(\boldsymbol{p}^\prime-\boldsymbol{p})$,
one has
\begin{align}
 & \int\!\frac{{\rm d}v^{+}}{2\pi}\left\langle p^{\prime}\left|{\cal O}\left(-\frac{v^{+}}{2},\frac{\boldsymbol{r}}{2};\frac{v^{+}}{2},-\frac{\boldsymbol{r}}{2}\right)\right|p\right\rangle \nonumber \\
 & =2p^{-}\int\!\frac{{\rm d}v^{+}}{2\pi}\int{\rm d}v_{1}^{+}{\rm d}^{d}\boldsymbol{b}\frac{\left\langle p^{\prime}\left|{\cal O}\left(-\frac{v^{+}}{2},-\frac{\boldsymbol{r}}{2};\frac{v^{+}}{2},\frac{\boldsymbol{r}}{2}\right)\right|p\right\rangle }{\left\langle p'|p\right\rangle}\label{eq:ope-manip-1}\\
 & =2p^{-}\int\!\frac{{\rm d}v^{+}}{2\pi}\int{\rm d}v_{1}^{+}{\rm d}^{d}\boldsymbol{b}\, {\rm e}^{i(p^{\prime-}-p^{-})v_{1}^{+}-i(\boldsymbol{p}^{\prime}-\boldsymbol{p})\cdot\boldsymbol{b}}\nonumber \\
 & \times\frac{\left\langle p^{\prime}\left|{\cal O}\left(v_{1}^{+}-\frac{v^{+}}{2},\boldsymbol{b}-\frac{\boldsymbol{r}}{2};v_{1}^{+}+\frac{v^{+}}{2},\boldsymbol{b}+\frac{\boldsymbol{r}}{2}\right)\right|p\right\rangle }{\left\langle p'|p\right\rangle}\label{eq:ope-manip-2}\\
 & =\frac{p^{-}}{\pi}\int\!{\rm d}x_{1}^{+}\int{\rm d}x_{2}^{+}{\rm e}^{i(p^{\prime-}-p^{-})\frac{x_{1}^{+}+x_{2}^{+}}{2}}\int{\rm d}^{d}\boldsymbol{b}\, {\rm e}^{-i(\boldsymbol{p}^{\prime}-\boldsymbol{p})\cdot\boldsymbol{b}}\nonumber \\
 & \times\frac{\left\langle p^{\prime}\left|{\cal O}\left(x_{1}^{+},\boldsymbol{b}-\frac{\boldsymbol{r}}{2};x_{2}^{+},\boldsymbol{b}+\frac{\boldsymbol{r}}{2}\right)\right|p\right\rangle }{\left\langle p'|p\right\rangle}.\label{eq:ope-manip-3}
\end{align}
In the Regge limit, we must neglect the $\left(p^{\prime-}-p^{-}\right)$
phase. Also note $P^{-}\simeq p^{-}$. 
A crucial step is to use the following identity 
\begin{align}
{\cal O}\left(x_{1}^{+},\boldsymbol{b}-\frac{\boldsymbol{r}}{2};x_{2}^{+},\boldsymbol{b}+\frac{\boldsymbol{r}}{2}\right) & =\frac{1}{g^{2}}\frac{\partial}{\partial x_{1}^{+}}\frac{\partial}{\partial x_{2}^{+}}{\rm tr}\left\{ [x_{1}^{+},x_{2}^{+}]_{\boldsymbol{b}-\boldsymbol{r}/2}[x_{2}^{+},x_{1}^{+}]_{\boldsymbol{b}+\boldsymbol{r}/2}\right\} .\label{eq:ope-manip-fin}
\end{align}
 We can then write
\begin{align}
 & \int\!{\rm d}^{d}\boldsymbol{k}\,{\rm e}^{i(\boldsymbol{k}\cdot\boldsymbol{r})}\boldsymbol{r}^{i}\boldsymbol{r}^{j}{\cal G}^{ij}(x=0,\xi,\boldsymbol{k}-\frac{z-\overline{z}}{2}\boldsymbol{\Delta},\boldsymbol{\Delta})\nonumber \\
 & =\frac{1}{\pi g^{2}}\int{\rm d}^{d}\boldsymbol{b}\,{\rm e}^{-i(\boldsymbol{\Delta}\cdot\boldsymbol{b})}\,{\rm e}^{-i\frac{z-\overline{z}}{2}(\boldsymbol{\Delta}\cdot\boldsymbol{r})}\label{eq:uGPD-as-der}\\
 & \times\int\!{\rm d}x_{1}^{+}\int{\rm d}x_{2}^{+}\frac{\partial}{\partial x_{1}^{+}}\frac{\partial}{\partial x_{2}^{+}}\frac{\left\langle p^{\prime}\left|{\rm tr}\left\{ [x_{1}^{+},x_{2}^{+}]_{\boldsymbol{b}-\boldsymbol{r}/2}[x_{2}^{+},x_{1}^{+}]_{\boldsymbol{b}+\boldsymbol{r}/2}\right\} \right|p\right\rangle }{\left\langle p'|p\right\rangle}.\nonumber 
\end{align}
Introducing the infinite Wilson line operators
\begin{equation}
V_{\boldsymbol{x}}\equiv[\infty,-\infty]_{\boldsymbol{x}},\quad V_{\boldsymbol{x}}^{\dagger}\equiv[-\infty,\infty]_{\boldsymbol{x}}\label{eq:V-def}
\end{equation}
we finally find, by taking the trivial integral in the last line,
\begin{align}
 & \int\!{\rm d}^{d}\boldsymbol{k}\,{\rm e}^{i(\boldsymbol{k}\cdot\boldsymbol{r})}\boldsymbol{r}^{i}\boldsymbol{r}^{j}{\cal G}^{ij}(x=0,\xi,\boldsymbol{k}-\frac{z-\overline{z}}{2}\boldsymbol{\Delta},\boldsymbol{\Delta})\nonumber \\
 & =\frac{2}{\pi g^{2}}\int{\rm d}^{d}\boldsymbol{b}\,{\rm e}^{-i(\boldsymbol{\Delta}\cdot\boldsymbol{b})}\,{\rm e}^{-i\frac{z-\overline{z}}{2}(\boldsymbol{\Delta}\cdot\boldsymbol{r})}\label{eq:uGPD-as-dip}\frac{\left\langle p^{\prime}\left|{\rm tr}\left(1-{\rm Re}(V_{\boldsymbol{b}+\boldsymbol{r}/2}V_{\boldsymbol{b}-\boldsymbol{r}/2}^{\dagger})\right)\right|p\right\rangle }{\left\langle p'|p\right\rangle}. 
\end{align}
Hence,
\begin{align}
{\cal T}_{\lambda\lambda^{\prime}}^{{\rm Regge}} & =i\alpha_{{\rm em}}\sum_{f}q_{f}^{2}\mu^{2-d}\int_{0}^{1}\!{\rm d}z\int{\rm d}^{d}\boldsymbol{r}\,{\rm e}^{-i\frac{z-\overline{z}}{2}(\boldsymbol{\Delta}\cdot\boldsymbol{r})}{\rm tr}[\Phi_{\lambda}(z,\boldsymbol{r})\Phi_{\lambda^{\prime}}^{\prime\ast}(z,\boldsymbol{r})]\label{eq:amp-regge-dip}\\
 & \times2P^{-}q^{+}\int{\rm d}^{d}\boldsymbol{b}\,{\rm e}^{-i(\boldsymbol{\Delta}\cdot\boldsymbol{b})}\frac{\left\langle p^{\prime}\left|{\rm tr}\left(1-{\rm Re}(V_{\boldsymbol{b}+\boldsymbol{r}/2}V_{\boldsymbol{b}-\boldsymbol{r}/2}^{\dagger})\right)\right|p\right\rangle }{\left\langle p'|p\right\rangle}.\nonumber 
\end{align}
Shifting $\boldsymbol{b}\rightarrow\boldsymbol{b}+\overline{z}\boldsymbol{r}-\boldsymbol{r}/2$
to get more standard variables yields our final result:
\begin{align}
{\cal T}_{\lambda\lambda^{\prime}}^{{\rm Regge}} & =2iP^{-}q^{+}\alpha_{{\rm em}}\sum_{f}q_{f}^{2}\int_{0}^{1}\!{\rm d}z\int{\rm d}^{d}\boldsymbol{r}\,{\rm tr}[\Phi_{\lambda}(z,\boldsymbol{r})\Phi_{\lambda^{\prime}}^{\prime\ast}(z,\boldsymbol{r})]\nonumber \\
 & \times\int{\rm d}^{d}\boldsymbol{b}\,{\rm e}^{-i(\boldsymbol{\Delta}\cdot\boldsymbol{b})}\frac{\left\langle p^{\prime}\left|{\rm tr}\left(1-{\rm Re}(V_{\boldsymbol{b}+\overline{z}\boldsymbol{r}}V_{\boldsymbol{b}-z\boldsymbol{r}}^{\dagger})\right)\right|p\right\rangle }{\left\langle p'|p\right\rangle}.\label{eq:amp-regge-dip-shifted}
\end{align}
This has the exact form one would expect when computing the amplitude in the context of a shock wave approximated low $x_{\rm Bj}$ formalism such as the Color Glass Condensate or the dipole framework, see for example \cite{Hatta:2017cte}.
The wave function overlaps are given below for the sake of completeness:
\begin{align}
{\rm tr}[\Phi_{L}(z,\boldsymbol{r})\Phi_{L}^{\prime\ast}(z,\boldsymbol{r})]= & -\frac{16z^{2}\overline{z}^{2}QQ^{\prime}}{(2\pi)^{d}(\boldsymbol{r}^{2})^{d-2}}(\sqrt{z\overline{z}Q^{2}\boldsymbol{r}^{2}})^{\frac{d}{2}-1}K_{\frac{d}{2}-1}(\sqrt{z\overline{z}Q^{2}\boldsymbol{r}^{2}})\nonumber \\
 & \times(-i\sqrt{z\overline{z}Q^{\prime2}\boldsymbol{r}^{2}})^{\frac{d}{2}-1}K_{\frac{d}{2}-1}(-i\sqrt{z\overline{z}Q^{\prime2}\boldsymbol{r}^{2}})\label{eq:PhiLPhiL}
\end{align}
for the LL transition,
\begin{align}
{\rm tr}[\Phi_{L}(z,\boldsymbol{r})\Phi_{h^{\prime}}^{\prime\ast}(z,\boldsymbol{r})] & =-\frac{8iz\overline{z}(z-\overline{z})Q(\boldsymbol{e}_{h^{\prime}}^{\prime\ast}\cdot\boldsymbol{r})}{(2\pi)^{d}(\boldsymbol{r}^{2})^{d-1}}(\sqrt{z\overline{z}Q^{2}\boldsymbol{r}^{2}})^{\frac{d}{2}-1}K_{\frac{d}{2}-1}(\sqrt{z\overline{z}Q^{2}\boldsymbol{r}^{2}})\nonumber \\
 & \times(-i\sqrt{z\overline{z}Q^{\prime2}\boldsymbol{r}^{2}})^{\frac{d}{2}}K_{\frac{d}{2}}(-i\sqrt{z\overline{z}Q^{\prime2}\boldsymbol{r}^{2}})\label{eq:PhiLPhiT}
\end{align}
for the LT transition,
\begin{align}
{\rm tr}[\Phi_{h}(z,\boldsymbol{r})\Phi_{L}^{\prime\ast}(z,\boldsymbol{r})] & =\frac{8iz\overline{z}(z-\overline{z})Q^{\prime}(\boldsymbol{e}_{h}\cdot\boldsymbol{r})}{(2\pi)^{d}(\boldsymbol{r}^{2})^{d-1}}(\sqrt{z\overline{z}Q^{2}\boldsymbol{r}^{2}})^{\frac{d}{2}}K_{\frac{d}{2}}(\sqrt{z\overline{z}Q^{2}\boldsymbol{r}^{2}})\nonumber \\
 & \times(-i\sqrt{z\overline{z}Q^{\prime2}\boldsymbol{r}^{2}})^{\frac{d}{2}-1}K_{\frac{d}{2}-1}(-i\sqrt{z\overline{z}Q^{\prime2}\boldsymbol{r}^{2}})\label{eq:PhiTPhiL}
\end{align}
for the TL transition, and finally
\begin{align}
{\rm tr}[\Phi_{h}(z,\boldsymbol{r})\Phi_{h^{\prime}}^{\prime\ast}(z,\boldsymbol{r})] & =\frac{4\boldsymbol{e}_{h}^{i}\boldsymbol{e}_{h^{\prime}}^{\prime j\ast}(4z\overline{z}\boldsymbol{r}^{i}\boldsymbol{r}^{j}-\boldsymbol{r}^{2}\delta^{ij})}{(2\pi)^{d}(\boldsymbol{r}^{2})^{d}}(\sqrt{z\overline{z}Q^{2}\boldsymbol{r}^{2}})^{\frac{d}{2}}K_{\frac{d}{2}}(\sqrt{z\overline{z}Q^{2}\boldsymbol{r}^{2}})\nonumber \\
 & \times(-i\sqrt{z\overline{z}Q^{\prime2}\boldsymbol{r}^{2}})^{\frac{d}{2}}K_{\frac{d}{2}}(-i\sqrt{z\overline{z}Q^{\prime2}\boldsymbol{r}^{2}})\label{eq:PhiTPhiT}
\end{align}
for the TT transition.

\section{On the commutativity of the limits $s \to \infty$ and $Q^2\to \infty$}\label{sec:Double Limit}

In this Section, we will finally discuss how the leading twist limit of the naive Regge limit and the small $\xi,x_{\rm Bj}$ limit of the Bjorken limit compare to one another. In particular, we will show the more rigorous way to obtain the Regge limit and a crucial hypothesis it implies and requires.

\subsection{Leading twist limit of the Regge limit}\label{sec:LT of Regge}
Let us now compute the $Q^2+Q^{\prime 2}\rightarrow\infty$ limit of the Regge result. It is most convenient to go back to the momentum space expression for Eq.~(\ref{eq:amp-regge-int}), then to write (see Section~\ref{sec:Bjorken}) $|\boldsymbol{k}|,|\boldsymbol{\Delta}|\ll|\boldsymbol{\ell}|$ to obtain
\begin{align}
\left.{\cal T}_{\lambda\lambda^{\prime}}^{{\rm Regge}}\right|_{Q^{2}+Q^{\prime2}\rightarrow\infty} & =2i\pi^{2}\alpha_{s}\alpha_{{\rm em}}\sum_{f}q_{f}^{2}\mu^{2-d}2P^{-}q^{+}\int\!{\rm d}^{d}\boldsymbol{k}{\cal G}^{ij}(x=0,\xi,\boldsymbol{k},\boldsymbol{\Delta}) \nonumber\\
 & \times\int_{0}^{1}\!{\rm d}z\int\!\frac{{\rm d}^{d}\boldsymbol{\ell}}{(2\pi)^{d}}{\rm tr}[(\partial^{i}\Psi_{\lambda})(z,\boldsymbol{\ell})(\partial^{j}\Psi_{\lambda^{\prime}}^{\prime\ast})(z,\boldsymbol{\ell})].\label{eq:}  
\end{align}
Given the simple relation between the integral of the unintegrated GPD and the GPD correlator from Eq.~(\ref{eq:uGPD-to-GPD}), we get:
\begin{align}
\left.{\cal T}_{\lambda\lambda^{\prime}}^{{\rm Regge}}\right|_{Q^{2}+Q^{\prime2}\rightarrow\infty} & =2i\pi^{2}\alpha_{s}\alpha_{{\rm em}}\sum_{f}q_{f}^{2}\mu^{2-d}\nonumber 2P^{-}q^{+}G^{ij}(x=0,\xi,\boldsymbol{\Delta})\nonumber \\
 & \times\int_{0}^{1}\!{\rm d}z\int\!\frac{{\rm d}^{d}\boldsymbol{\ell}}{(2\pi)^{d}}{\rm tr}[(\partial^{i}\Psi_{\lambda})(z,\boldsymbol{\ell})(\partial^{j}\Psi_{\lambda^{\prime}}^{\prime\ast})(z,\boldsymbol{\ell})]. 
\end{align}
Because of the $|\boldsymbol{\Delta}|\sim 0$ approximation yielding oddness for $\boldsymbol{\ell}\rightarrow-\boldsymbol{\ell}$, the LT and TL overlaps cancel after taking the integral in $\boldsymbol{\ell}$. We are left with
\begin{equation}
{\rm tr}[\partial^{i}\Psi_{L}(z,\boldsymbol{\ell})\partial^{j}\Psi_{L}^{\prime\ast}(z,\boldsymbol{\ell})]=\frac{-64z^{2}\overline{z}^{2}QQ^{\prime}\boldsymbol{\ell}^{i}\boldsymbol{\ell}^{j}}{\left(\boldsymbol{\ell}^{2}+z\overline{z}Q^{2}-i0\right)^{2}\left(\boldsymbol{\ell}^{2}-z\overline{z}Q^{\prime2}-i0\right)^{2}}
\end{equation}
and
\begin{align}
{\rm tr}[\partial^{i}\Psi_{h}(z,\boldsymbol{\ell})\partial^{j}\Psi_{h^{\prime}}^{\prime\ast}(z,\boldsymbol{\ell})] & =4\boldsymbol{e}_{h}^{m}\boldsymbol{e}_{h^{\prime}}^{\prime\ast n}\left[(1-4z\overline{z})\delta^{km}\delta^{ln}-\delta^{kn}\delta^{lm}+\delta^{kl}\delta^{mn}\right]\nonumber \\
 & \times\left\{\frac{\delta^{ik}}{\boldsymbol{\ell}^{2}+z\overline{z}Q^{2}-i0}-\frac{2\boldsymbol{\ell}^{i}\boldsymbol{\ell}^{k}}{\left(\boldsymbol{\ell}^{2}+z\overline{z}Q^{2}-i0\right)^{2}}\right\}\\
 & \times\left\{\frac{\delta^{jl}}{\boldsymbol{\ell}^{2}-z\overline{z}Q^{\prime2}-i0}-\frac{2\boldsymbol{\ell}^{j}\boldsymbol{\ell}^{l}}{\left(\boldsymbol{\ell}^{2}-z\overline{z}Q^{\prime2}-i0\right)^{2}}\right\}.\nonumber 
\end{align}
There is no subtlety about the integrals of these overlaps over $z$ and $\boldsymbol{\ell}$. Standard algebra yields:
\begin{align}
\left.{\cal T}_{LL}^{{\rm Regge}}\right|_{Q^{2}+Q^{\prime2}\rightarrow\infty} & =\frac{2i\pi}{\xi}\alpha_{s}\alpha_{{\rm em}}\sum_{f}q_{f}^{2}G^{ii}(x=0,\xi,\boldsymbol{\Delta})\nonumber \\
 & \times 4\frac{QQ^{\prime}}{Q^{2}+Q^{\prime2}}\left[2-\frac{Q^{2}-Q^{\prime2}}{Q^{2}+Q^{\prime2}}\ln\left(\frac{Q^2}{-Q^{\prime 2}-i0}\right)\right],
\end{align}
and
\begin{align}
\left.{\cal T}_{hh^{\prime}}^{{\rm Regge}}\right|_{Q^{2}+Q^{\prime2}\rightarrow\infty} & =4\frac{i\pi}{\xi}\alpha_{s}\alpha_{{\rm em}}\sum_{f}q_{f}^{2} G^{ij}(x=0,\xi,\boldsymbol{\Delta})\nonumber \\
 &\times \boldsymbol{e}_{h}^{m}\boldsymbol{e}_{h^{\prime}}^{\prime\ast n}\left(\frac{{\rm e}^{\gamma_{E}}}{4\pi}\right)^{\epsilon}\left[\left(\frac{1}{\epsilon}-2\right)\delta^{ij}\delta^{mn}-\frac{1}{2}\tau^{ij,mn}\right]\nonumber \\
 & \times \left\{ \frac{Q^{2}-Q^{\prime2}}{Q^{2}+Q^{\prime2}}+\frac{2Q^{2}Q^{\prime2}}{(Q^{2}+Q^{\prime2})^{2}}\ln\left(\frac{Q^{2}}{-Q^{\prime2}-i0}\right)\right.\label{eq:TT-RtoB}\\
 & +\epsilon\left[\frac{Q^{2}}{Q^{2}+Q^{\prime2}}\ln\left(\frac{Q^{2}}{\mu^{2}}\right)-\frac{Q^{\prime2}}{Q^{2}+Q^{\prime2}}\ln\left(-\frac{Q^{\prime2}}{\mu^{2}}-i0\right)\right]\nonumber \\
 & \left.+\epsilon\frac{Q^{2}Q^{\prime2}}{(Q^{2}+Q^{\prime2})^{2}}\left[\ln^{2}\left(\frac{Q^{2}}{\mu^{2}}\right)-\ln^{2}\left(-\frac{Q^{\prime2}}{\mu^{2}}-i0\right)\right]\right\}.\nonumber
\end{align}
The TT transition term is divergent and requires counterterms from the shock wave limit of the GPD evolution equation. We will discuss the proper way of computing such counterterms below. 

\subsection{Shock wave limit of the Bjorken results}

How to properly take the shock wave limit of the $x$-dependent results in order to match the leading twist limit of the Regge results is the most delicate step of the present analysis. Indeed, the amplitude has a dependence on three longitudinal variables $(x,x_{\rm Bj},\xi)$ out of which two $(x_{\rm Bj},\xi)$ are suppressed by definition of this limit. It is not obvious, upon seeing cumbersome, fully integrated expressions such as Eq.~(\ref{eq:TBu-fin}), what one should do with the remaining $x$ variable. The notion of taking a cut in $x$, when facing the complexity of the amplitude after all integrations have been taken, and given that several terms could contribute to the imaginary part of the amplitude, does not make sense anymore. In this section, we will argue that if the amplitude has the form of the convolution
\begin{equation}
    \int_{-1}^1 {\rm d}x G^{ij}(x,\xi,t) {\cal H}^{ij}(x,x_{\rm Bj}, \xi, t),
\end{equation}
then to match the shock wave limit one has to approximate it by
\begin{equation}
    G^{ij}(x=0,\xi,t) \lim_{x_{\rm Bj},\xi \ll 1} \int_{-1}^1 {\rm d}x  {\cal H}^{ij}(x,x_{\rm Bj}, \xi, t),
\end{equation}
i.e. one has to use the fact that the $x$ integral peaks around $x=0$ for small values of $\xi$ and $x_{\rm Bj}$ by assuming that the GPD is constant around $x=0$ and taking it out of the integral.
In Appendix~\ref{sec:intintapp}, we calculate the $x_{\rm Bj},\xi \rightarrow 0$ limit of the $x$ integrals of all possible hard parts ${\cal H}^{ij}(x,x_{\rm Bj}, \xi, t)$ involved in our observable. Rewriting the result of these integrals Eqs.~(\ref{eq:ILL-int-fin}),(\ref{eq:ITT-int-almost}) in terms of $Q$ and $Q^\prime$ instead of $x_{\rm Bj}$ and $\xi$, we find:
\begin{equation}
    {\cal T}_{LL}^{\rm Bjorken} \rightarrow \frac{i\pi}{\xi} \alpha_s\alpha_{\rm em}\sum_f q_f^2 \frac{8QQ^\prime}{Q^2+Q^{\prime 2}}G^{ii}(0,\xi,\Delta) 
 \left[ 2 -\frac{Q^2-Q^{\prime 2}}{Q^2+Q^{\prime 2}}\ln \left( \frac{Q^2}{-Q^{\prime 2}-i0} \right)  \right],  \label{eq:LL-Bj-then-R}
\end{equation}
and
\begin{align}
{\cal T}_{hh^{\prime}}^{{\rm Bjorken}} & \rightarrow\frac{4i\pi}{\xi}\alpha_{{\rm em}}\alpha_{s}\sum_{f}q_{f}^{2}G^{ij}(0,\xi,t)\label{eq:TT-Bj-then-R}\\
 & \times\boldsymbol{e}_{h}^{m}\boldsymbol{e}_{h^{\prime}}^{\prime n\ast}\left(\frac{{\rm e}^{\gamma_{E}}}{4\pi}\right)^{\epsilon}\left[\left(\frac{1}{\epsilon}-2\right)\delta^{ij}\delta^{mn}-\frac{\tau^{ij;mn}}{2}\right]\nonumber \\
 & \times\left\{ \frac{Q^{2}-Q^{\prime2}}{Q^{2}+Q^{\prime2}}+2\frac{Q^{2}Q^{\prime2}}{(Q^{2}+Q^{\prime2})^{2}}\ln\left(\frac{Q^{2}}{-Q^{\prime2}-i0}\right)\right.\nonumber \\
 & +\epsilon\left[\frac{Q^{2}}{Q^{2}+Q^{\prime2}}\ln\left(\frac{Q^{2}}{\mu^{2}}\right)-\frac{Q^{\prime2}}{Q^{2}+Q^{\prime2}}\ln\left(-\frac{Q^{\prime2}}{\mu^{2}}-i0\right)\right]\nonumber \\
 & \left.+\epsilon\frac{Q^{2}Q^{\prime2}}{(Q^{2}+Q^{\prime2})^{2}}\left[\ln^{2}\left(\frac{Q^{2}}{\mu^{2}}\right)-\ln^{2}\left(\frac{-Q^{\prime2}-i0}{\mu^{2}}\right)\right]\right\} .\nonumber 
\end{align}
These match the amplitudes obtained by taking the leading twist limit of the Regge results. \\
The TT transition amplitude contains a collinear divergence, which has to be cancelled using the counterterms in Eqs.~(\ref{eq:counterterm-u}, \ref{eq:counterterm-p}). First of all, note that polarized GPDs $\tilde{G}$ cancel at $x=0$, so we do not need the polarized counterterm: the polarized amplitude cancels. With the same procedure as described above, the counterterm becomes:
\begin{align}
{\cal T}_{{\rm div}} & \rightarrow-4\frac{i\pi}{\xi}\alpha_{{\rm em}}\alpha_{s}\sum_{f}q_{f}^{2}(\boldsymbol{\varepsilon}_{q}\cdot\boldsymbol{\varepsilon}_{q^{\prime}}^{\ast})\frac{1}{\epsilon}\left(\frac{{\rm e}^{\gamma_{E}}\mu_{F}^{2}}{4\pi\mu^{2}}\right)^{\epsilon}G(0,\xi,\mu^{2})\nonumber \\
 & \times\left[\frac{Q^{2}-Q^{\prime2}}{Q^{2}+Q^{\prime2}}+2\frac{Q^{2}Q^{\prime2}}{(Q^{2}+Q^{\prime2})^{2}}\ln\left(\frac{Q^{2}-i0}{-Q^{\prime2}-i0}\right)\right],\label{eq:counterterms-x0}
\end{align}
which cancels the divergence. One is left with the following result for the double limit:
\begin{align}
{\cal T}_{hh^{\prime}}^{{\rm double}} & = 4\frac{i\pi}{\xi}\alpha_{s}\alpha_{{\rm em}}\sum_{f}q_{f}^{2}\boldsymbol{e}_{h}^{m}\boldsymbol{e}_{h^{\prime}}^{\prime\ast n}G^{ij}(0,\xi,\boldsymbol{\Delta})\label{eq:T-double-fin}\\
 & \times\left\{ \left(-2\delta^{ij}\delta^{mn}-\frac{1}{2}\tau^{ij,mn}\right)\left[\frac{Q^{2}-Q^{\prime2}}{Q^{2}+Q^{\prime2}}+\frac{2Q^{2}Q^{\prime2}}{(Q^{2}+Q^{\prime2})^{2}}\ln\left(\frac{Q^{2}}{-Q^{\prime2}-i0}\right)\right]\right.\nonumber \\
 & +\delta^{ij}\delta^{mn}\left[\frac{Q^{2}}{Q^{2}+Q^{\prime2}}\ln\left(\frac{Q^{2}}{\mu_{F}^{2}}\right)-\frac{Q^{\prime2}}{Q^{2}+Q^{\prime2}}\ln\left(-\frac{Q^{\prime2}}{\mu_{F}^{2}}-i0\right)\right]\nonumber \\
 & \left.+\delta^{ij}\delta^{mn}\frac{Q^{2}Q^{\prime2}}{(Q^{2}+Q^{\prime2})^{2}}\left[\ln^{2}\left(\frac{Q^{2}}{\mu_{F}^{2}}\right)-\ln^{2}\left(\frac{-Q^{\prime2}-i0}{\mu_{F}^{2}}\right)\right]\right\} \nonumber 
\end{align}
Note that this is the first time the collinear limit of small $x$ exclusive Compton Scattering is computed beyond the simple DVCS case.
The way the counterterm is obtained is key to our argument about the proper way of taking the Regge limit: the correct way of finding the counterterms to the divergence is to take $x=0$ in the distribution and thus integrate the hard part wrt $x$. There is no argument like taking a cut in $x$ to set the correct expression for the counterterms. In the next section, we will make an argument as to why the limits seem to commute, and the underlying assumption.

\subsection{'Naive' vs actual shock wave limit}

We finally found a way to match the leading twist limit of the Regge result with the shock wave limit of the leading twist limit. However, taking the latter required a different procedure to that described in Eq.~(\ref{eq:Regge-cut}). One can thus wonder why there seems to be a match between the 'naive' shock wave limit, taken by picking a specific cut in the $q^+P^-\rightarrow\infty$ limit, and the 'actual' shock wave limit, taken by assuming the distribution is constant in $x$ and then taking the $q^+P^-\rightarrow\infty$ limit. \\
A quick glance at the generic formula in Eq.~(\ref{eq:amp-fin-1}) yields the answer to that question. Indeed, the wave functions do not depend on $x$ so the dependence of the hard part on this variable is very simple. The convolution in $x$ has the form
\begin{equation}
\int_{-1}^{1}{\rm d}x\frac{{\cal G}^{ij}(x,\xi,\boldsymbol{k}-\frac{z-\bar{z}}{2}\boldsymbol{\Delta},\boldsymbol{\Delta})}{x+\frac{\left(\frac{q+q^{\prime}}{2}\right)^{2}}{2q^{+}P^{-}}-\frac{\boldsymbol{\ell}^{2}}{2z\bar{z}q^{+}P^{-}}+i0}\label{eq:conv-x}.
\end{equation}
Taking the cut as in Eq.~(\ref{eq:Regge-cut}) turns this into 
\begin{equation}
-i\pi{\cal G}^{ij}(0,\xi,\boldsymbol{k}-\frac{z-\bar{z}}{2}\boldsymbol{\Delta},\boldsymbol{\Delta})\label{eq:conv-x-1}.
\end{equation}
On the other hand, assuming ${\cal G}^{ij}$ is a constant in the $x\to 0$ limit,  yields
\begin{equation}
{\cal G}^{ij}(0,\xi,\boldsymbol{k}-\frac{z-\bar{z}}{2}\boldsymbol{\Delta},\boldsymbol{\Delta})\ln\left(\frac{1+\frac{\left(\frac{q+q^{\prime}}{2}\right)^{2}}{2q^{+}P^{-}}-\frac{\boldsymbol{\ell}^{2}}{2z\bar{z}q^{+}P^{-}}+i0}{-1+\frac{\left(\frac{q+q^{\prime}}{2}\right)^{2}}{2q^{+}P^{-}}-\frac{\boldsymbol{\ell}^{2}}{2z\bar{z}q^{+}P^{-}}+i0}\right)\label{eq:conv-x-const}.
\end{equation}
In the $q^+ P^-\rightarrow \infty$ limit, the logarithm becomes $\ln(\frac{1}{-1+i0})=-i\pi$ and the result finally matches Eq.~(\ref{eq:conv-x-1}). In other words, taking the distribution to be a constant naturally implies the naive cut one takes in order to get the Regge result.\\
It is however crucial to note that the procedures are not always entirely equivalent: the cut is merely a consequence of the other. We can see this difference play out when studying the collinear divergence. Indeed, the counterterm in Eq.~(\ref{eq:counterterm-u}) does not have an $i0$ in its poles in $x$ since the brackets completely cancel the pole parts. This means there is no actual cut to take. An alternative approach involving cuts would be to get back to Eq.~(\ref{eq:counterterm-in}) where the $x=x_{\rm  Bj}$ poles are present in the leading order quark hard part, and to take said poles. However, this approach does not yield the full counterterm from Eq.~(\ref{eq:counterterms-x0}) required to cancel the collinear divergence. Instead, it yields:
\begin{align}
{\cal T}_{{\rm div}} & \rightarrow-4\frac{i\pi}{\xi}\alpha_{{\rm em}}\alpha_{s}\sum_{f}q_{f}^{2}(\boldsymbol{\varepsilon}_{q}\cdot\boldsymbol{\varepsilon}_{q^{\prime}}^{\ast})\frac{1}{\epsilon}\left(\frac{{\rm e}^{\gamma_{E}}\mu_{F}^{2}}{4\pi\mu^{2}}\right)^{\epsilon}G(0,\xi,\mu^{2})\nonumber \\
 & \times\left[\frac{Q^{2}-Q^{\prime2}}{Q^{2}+Q^{\prime2}}+2\frac{Q^{2}Q^{\prime2}}{(Q^{2}+Q^{\prime2})^{2}}\ln\left(\frac{Q^{2}}{Q^{\prime2}}\right)\right]\label{eq:counterterms-x0}.
\end{align}
This expression is missing the real part which comes from the $\ln(-Q^{\prime 2}-i0)$ logarithm. Incidentally, this real part cancels exactly in the DVCS limit, which is the reason why the cute procedure, when applied to DVCS \cite{Hatta:2017cte}, led to the correct result. In full generality taking the cut is the wrong approach since it misses the real part, while the correct procedure is the one which relies solely on the hypothesis that the GPD is independent on $x$ for small values of $x$.

\section{Conclusion}

Using the consistent, first-principle scheme developped in~\cite{Boussarie:2021wkn}, we provided a unified expression for exclusive Compton scattering processes $\gamma^{(\ast)}P \rightarrow \gamma^{(\ast)}P$. Our result constitutes an interpolation between the leading power of $\sqrt{s}$ in the Regge ($s \gg Q^2 \gg \Lambda_{\rm QCD}^2$) limit and the leading power of $Q$ in the Bjorken ($s \sim Q^2 \gg \Lambda_{\rm QCD}^2$) limit. We have explicitly checked that both limits are recovered in their entirety, which means among other notable properties that all powers of all three longitudinal variables $x$, $x_{\rm Bj}$ and $\xi$ are fully restored. Remarkably, although our scheme is inspired by semi-classical descriptions of low $x$ observables, the polarized GPD contribution to the amplitude is obtained by the scheme as well, despite it being strictly null in the leading power of the low $x$ regime.

We thus found an efficient way to compute observables in the Bjorken limit, as well as a consistent way to include explicit dependence on longitudinal variables in the Regge limit. Based on the interpolating formula, we confirmed once more~(see \cite{Boussarie:2021wkn}) that for the leading twist limit of semi-classical low $x$ observables and the eikonal limit of the same observables as written in leading twist QCD factorizaction to be compatible with each other, it is necessary for the parton distributions to be constant at $x=0$. 

Our interpolating result is written as a 3-dimensional convolution between a hard subamplitude and an unintegrated GPD correlator with dependence on the Feynman $x$ variable, an intrinsic transverse momentum $k_\perp$, and momentum transfer $\Delta$. This distribution spans both the non-diagonal matrix element of the dipole operator one encounters in semi-classical small $x$ schemes such as the Color Glass Condensate, and the Generalized Parton Distribution one finds in QCD factorization. Because our scheme does not rely on assumptions on light cone time separations, and thus it is consistent with orderings in longitudinal variables and with momentum conservation, the evolution equation(s) for this distribution will naturally provide long sought collinear corrections to the low $x$ evolution kernel.

\acknowledgments
Y. M.-T.'s work has been supported by the U.S. Department of Energy under Contract No. DE-SC0012704. This work is supported in part by Laboratory Directed Research and Development (LDRD) funds from Brookhaven Science Associates.

\appendix

\section{Bjorken limit integrals}\label{sec:intapp}

We want to compute the following integrals:
\begin{align}
I_{LL}^{ij} & \equiv\mu^{2-d}\int_{0}^{1}\!\frac{{\rm d}z}{z\overline{z}}\int\!\frac{{\rm d}^{d}\boldsymbol{\ell}}{(2\pi)^{d}}\frac{\frac{\boldsymbol{\ell}^{i}\boldsymbol{\ell}^{j}}{2z\overline{z}q^{+}P^{-}}}{\frac{\boldsymbol{\ell}^{2}}{2z\overline{z}q^{+}P^{-}}+(x_{{\rm Bj}}-x-i0)}\label{eq:ILL-def}\\
 & \times\frac{1}{\left[\frac{\boldsymbol{\ell}^{2}}{2z\overline{z}q^{+}P^{-}}+(x_{{\rm Bj}}+\xi-i0)\right]^{2}\left[\frac{\boldsymbol{\ell}^{2}}{2z\overline{z}q^{+}P^{-}}+(x_{{\rm Bj}}-\xi-i0)\right]^{2}},\nonumber 
\end{align}
and
\begin{align}
I_{TT}^{ijmn} & \equiv\mu^{2-d}\int_{0}^{1}\!\frac{{\rm d}z}{z^{2}\overline{z}^{2}}\int\!\frac{{\rm d}^{d}\boldsymbol{\ell}}{(2\pi)^{d}}\frac{(1-4z\overline{z})\delta^{km}\delta^{ln}+\delta^{kl}\delta^{mn}-\delta^{lm}\delta^{kn}}{\frac{\boldsymbol{\ell}^{2}}{2z\overline{z}q^{+}P^{-}}+(x_{{\rm Bj}}-x-i0)}\label{eq:ITT-def}\\
 & \times\left\{\frac{\delta^{ik}}{\frac{\boldsymbol{\ell}^{2}}{2z\overline{z}q^{+}P^{-}}+(x_{{\rm Bj}}+\xi-i0)}-2\frac{\frac{\boldsymbol{\ell}^{i}\boldsymbol{\ell}^{k}}{2z\overline{z}q^{+}P^{-}}}{\left[\frac{\boldsymbol{\ell}^{2}}{2z\overline{z}q^{+}P^{-}}+(x_{{\rm Bj}}+\xi-i0)\right]^{2}}\right\}\nonumber \\
 & \times\left\{\frac{\delta^{jl}}{\frac{\boldsymbol{\ell}^{2}}{2z\overline{z}q^{+}P^{-}}+(x_{{\rm Bj}}-\xi-i0)}-2\frac{\frac{\boldsymbol{\ell}^{j}\boldsymbol{\ell}^{l}}{2z\overline{z}q^{+}P^{-}}}{\left[\frac{\boldsymbol{\ell}^{2}}{2z\overline{z}q^{+}P^{-}}+(x_{{\rm Bj}}-\xi-i0)\right]^{2}}\right\}.\nonumber 
\end{align}
Let us detail how $I_{LL}^{ij}$ only is derived. Similar steps are
necessary for $I_{TT}^{ijkl}$. The first step is to rescale $\boldsymbol{\ell}$ by $2z\overline{z}q^+P^-$,
and to integrate $z$ using the well known integral
\begin{equation}
\int_{0}^{1}{\rm d}zz^{p}\overline{z}^{q}=\frac{\Gamma(p+1)\Gamma(q+1)}{\Gamma(p+q+2)}.\label{eq:beta-int}
\end{equation}
It yields:
\begin{align}
I_{LL}^{ij} & =\mu^{2-d}(2q^{+}P^{-})^{\frac{d}{2}}\frac{\Gamma(\frac{d}{2})\Gamma(\frac{d}{2})}{\Gamma(d)}\int\!\frac{{\rm d}^{d}\boldsymbol{\ell}}{(2\pi)^{d}}\frac{\boldsymbol{\ell}^{i}\boldsymbol{\ell}^{j}}{\boldsymbol{\ell}^{2}+(x_{{\rm Bj}}-x-i0)}\nonumber \\
 & \times\frac{1}{\left[ 
 \boldsymbol{\ell}^{2}+(x_{{\rm Bj}}+\xi-i0)\right]^{2}\left[\boldsymbol{\ell}^{2}+(x_{{\rm Bj}}-\xi-i0)\right]^{2}}.\label{eq:ILL-zint}
\end{align}
We can perform $\boldsymbol{\ell}^{i}\boldsymbol{\ell}^{j}\rightarrow\frac{1}{d}\delta^{ij}\boldsymbol{\ell}^{2}$
and decompose $I_{LL}^{ij}$ into simpler base integrals as follows:
\begin{align}
I_{LL}^{ij} & =\frac{1}{4\xi^{2}}(2q^{+}P^{-})^{\frac{d}{2}}\frac{\Gamma(\frac{d}{2})\Gamma(\frac{d}{2})}{d\Gamma(d)}\delta^{ij}\label{eq:ILL-dec}\\
 & \times\left[\frac{x_{{\rm Bj}}}{\xi}(I_{-}-I_{+})-(x_{{\rm Bj}}+\xi)J_{+}-(x_{{\rm Bj}}-\xi)J_{-}\right],\nonumber 
\end{align}
with 
\begin{align}
I_{\pm} & \equiv\mu^{2-d}\int\!\frac{{\rm d}^{d}\boldsymbol{\ell}}{(2\pi)^{d}}\frac{1}{\left[\boldsymbol{\ell}^{2}+(x_{{\rm Bj}}-x-i0)\right]\left[\boldsymbol{\ell}^{2}+(x_{{\rm Bj}}\pm\xi-i0)\right]},\label{eq:Ipm-def}
\end{align}
and
\begin{align}
J_{\pm} & \equiv\mu^{2-d}\int\!\frac{{\rm d}^{d}\boldsymbol{\ell}}{(2\pi)^{d}}\frac{1}{\left[\boldsymbol{\ell}^{2}+(x_{{\rm Bj}}-x-i0)\right]\left[\boldsymbol{\ell}^{2}+(x_{{\rm Bj}}\pm\xi-i0)\right]^{2}}.\label{eq:Jpm-def}
\end{align}
These integrals are fairly standard, with a few extra subtleties.
Let us detail $I_{\pm}$ a bit as an example. Feynman's parametrization yields:
\begin{align}
I_{\pm} & =\mu^{2-d}\int_{0}^{1}{\rm d}\alpha\int\!\frac{{\rm d}^{d}\boldsymbol{\ell}}{(2\pi)^{d}}\frac{1}{\left[\boldsymbol{\ell}^{2}+\alpha(x_{{\rm Bj}}-x)+(1-\alpha)(x_{{\rm Bj}}\pm\xi)-i0\right]^{2}}.\label{eq:Ipm-feynman}
\end{align}
Using
\begin{eqnarray}
\int\frac{{\rm d}^{d}\boldsymbol{\ell}}{(\boldsymbol{\ell}^{2}+M^{2}-i0)^{n}} & = & \frac{\pi^{\frac{d}{2}}\Gamma(n-\frac{d}{2})}{\Gamma(n)(M^{2}-i0)^{n-\frac{d}{2}}},\label{eq:base-int}
\end{eqnarray}
we get
\begin{align}
I_{\pm} & =\frac{\Gamma(2-\frac{d}{2})\mu^{2-d}}{(4\pi)^{\frac{d}{2}}}\int_{0}^{1}{\rm d}\alpha\frac{1}{\left[-\alpha(x\pm\xi)+(x_{{\rm Bj}}\pm\xi-i0)\right]^{2-\frac{d}{2}}}.\label{eq:Ipm-feynman-1}
\end{align}
We now need to detail all possible cases for the sign or cancellation
of $x\pm\xi$ and $x_{{\rm Bj}}\pm\xi$, since the standard computation require steps which are not always mathematically sound depending on the case. For $x_{{\rm Bj}}\pm\xi\neq0$,
it is actually possible to assume $x\pm\xi>0$ and to proceed, but
for the sake of completeness let us compute this integral in all 3
possible cases. For $x\pm\xi>0$, we have:
\begin{align}
\left.I_{\pm}\right|_{x\pm\xi>0}^{x_{{\rm Bj}}\pm\xi\neq0} & =\frac{\Gamma(2-\frac{d}{2})\mu^{2-d}}{(4\pi)^{\frac{d}{2}}(x\pm\xi)^{2-\frac{d}{2}}}\int_{0}^{1}{\rm d}\alpha\frac{1}{\left[-\alpha+\left(\frac{x_{{\rm Bj}}\pm\xi-i0}{x\pm\xi}\right)\right]^{2-\frac{d}{2}}}\nonumber \\
 & =\frac{\Gamma(2-\frac{d}{2})\mu^{2-d}}{(4\pi)^{\frac{d}{2}}(x\pm\xi)^{2-\frac{d}{2}}}\frac{\left(\frac{x_{{\rm Bj}}-x-i0}{x\pm\xi}\right)^{\frac{d}{2}-1}-\left(\frac{x_{{\rm Bj}}\pm\xi-i0}{x\pm\xi}\right)^{\frac{d}{2}-1}}{1-\frac{d}{2}}\label{eq:Ipm-pos}\\
 & =\frac{\Gamma(2-\frac{d}{2})\mu^{2-d}}{(4\pi)^{\frac{d}{2}}(x\pm\xi)}\frac{(x_{{\rm Bj}}-x-i0)^{\frac{d}{2}-1}-(x_{{\rm Bj}}\pm\xi-i0)^{\frac{d}{2}-1}}{1-\frac{d}{2}}.\nonumber 
\end{align}
This is continuous for $x \rightarrow \mp \xi$, but for consistency we will compute that case separately instead. For $x\pm\xi=0$, Eq.(\ref{eq:Ipm-feynman-1}) leads directly to:
\begin{align}
\left.I_{\pm}\right|_{x\pm\xi=0}^{x_{{\rm Bj}}\pm\xi\neq0} & =\frac{\Gamma(2-\frac{d}{2})}{(4\pi)^{\frac{d}{2}}}\frac{\mu^{2-d}}{(x_{{\rm Bj}}\pm\xi-i0)^{2-\frac{d}{2}}},\label{eq:Ipm-eq}
\end{align}
Finally, the $x\pm\xi<0$ case yields:
\begin{align}
\left.I_{\pm}\right|_{x\pm\xi<0}^{x_{{\rm Bj}}\pm\xi\neq0} & =\frac{\Gamma(2-\frac{d}{2})\mu^{2-d}}{(4\pi)^{\frac{d}{2}}\left[-(x\pm\xi)\right]^{2-\frac{d}{2}}}\int_{0}^{1}{\rm d}\alpha\frac{1}{\left[\alpha-\left(\frac{x_{{\rm Bj}}\pm\xi-i0}{x\pm\xi}\right)\right]^{2-\frac{d}{2}}}\label{eq:Ipm-neg}\\
 & =\frac{\Gamma(2-\frac{d}{2})\mu^{2-d}}{(4\pi)^{\frac{d}{2}}\left[-(x\pm\xi)\right]^{2-\frac{d}{2}}}\frac{-\left(-\frac{x_{{\rm Bj}}-x-i0}{x\pm\xi}\right)^{\frac{d}{2}-1}+\left(-\frac{x_{{\rm Bj}}\pm\xi-i0}{x\pm\xi}\right)^{\frac{d}{2}-1}}{1-\frac{d}{2}}\nonumber \\
 & =\frac{\Gamma(2-\frac{d}{2})\mu^{2-d}}{(4\pi)^{\frac{d}{2}}(x\pm\xi)}\frac{(x_{{\rm Bj}}-x-i0)^{\frac{d}{2}-1}-(x_{{\rm Bj}}\pm\xi-i0)^{\frac{d}{2}-1}}{1-\frac{d}{2}},\nonumber 
\end{align}
which is perfectly consistent with the $x\pm\xi>0$ case.
Overall, we get the unified expression
\begin{align}
\left.I_{\pm}\right|^{x_{{\rm Bj}}\pm\xi\neq0} & =\frac{\Gamma(2-\frac{d}{2})\mu^{2-d}}{(4\pi)^{\frac{d}{2}}(x\pm\xi)}\frac{(x_{{\rm Bj}}-x-i0)^{\frac{d}{2}-1}-(x_{{\rm Bj}}\pm\xi-i0)^{\frac{d}{2}-1}}{1-\frac{d}{2}}.\label{eq:Ipm-res}
\end{align}
in dimensional regularization:
\begin{align}
\left.I_{\pm}\right|^{x_{{\rm Bj}}\pm\xi=0} & =\mu^{2-d}\int\!\frac{{\rm d}^{d}\boldsymbol{\ell}}{(2\pi)^{d}}\frac{1}{\boldsymbol{\ell}^{2}\left[\boldsymbol{\ell}^{2}+(x_{{\rm Bj}}-x-i0)\right]}\label{eq:Ipm-0}\\
 & =-\frac{\Gamma(1-\frac{d}{2})\mu^{2-d}}{(4\pi)^{\frac{d}{2}}}(x_{{\rm Bj}}-x-i0)^{\frac{d}{2}-2}.\nonumber \\
 & =\frac{\Gamma(1-\frac{d}{2})\mu^{2-d}}{(4\pi)^{\frac{d}{2}}(x\pm\xi+i0)}(x_{{\rm Bj}}-x-i0)^{\frac{d}{2}-1}.\nonumber 
\end{align}
In the last line, we used that $x_{\rm Bj} = \mp \xi$ in order to cast the result in a similar form to the previous results.
We can finally write the fully unified expression in the following fashion. First note that since the r.h.s. of Eq.~(\ref{eq:Ipm-res}) is finite for $x=\mp\xi$, we can add a $\pm i0$ regulator with any sign with no consequence if $x_{{\rm Bj}}\pm\xi\neq0$. If $x_{{\rm Bj}}\pm\xi=0$, we must add $i0$ to $x\pm\xi$ in the denominator. In order to have an expression which is valid for
both $I_{+}$ and $I_{-}$ and in both DVCS and TCS cases,
we must write $x\pm\xi-i0x_{{\rm Bj}}$ as the unified
regulator. Indeed, this term regularizes $x=\xi$ for $x_{\rm Bj}=\xi$ and $x=-\xi$ for $x_{\rm Bj}=-\xi$, and if $\xi\neq \pm x_{\rm Bj}$ the regulator does not contribute and can thus be added for free. Finally, we can write
in general:
\begin{align}
I_{\pm} & =\frac{\Gamma(2-\frac{d}{2})\mu^{2-d}}{(4\pi)^{\frac{d}{2}}(x\pm\xi-i0x_{{\rm Bj}})}\frac{(x_{{\rm Bj}}-x-i0)^{\frac{d}{2}-1}-(x_{{\rm Bj}}\pm\xi-i0)^{\frac{d}{2}-1}}{1-\frac{d}{2}}.\label{eq:Ipm-res-fin}
\end{align}
 This integral is finite for $x_{{\rm Bj}}\pm\xi=0$ if one takes
this limit before expanding in the dim reg parameter, which is the correct order to proceed. In practice,
we will find the amplitude to also be finite if the limit is taken
after the expansion.

The other integral is obtained in a similar way. We get:
\begin{align}
J_{\pm} & =\frac{\Gamma(2-\frac{d}{2})\mu^{2-d}}{(4\pi)^{\frac{d}{2}}(x\pm\xi-i0x_{{\rm Bj}})}\left[\frac{(x_{{\rm Bj}}-x-i0)^{\frac{d}{2}-1}-(x_{{\rm Bj}}\pm\xi-i0)^{\frac{d}{2}-1}}{(1-\frac{d}{2})(x\pm\xi-i0x_{{\rm Bj}})}-(x_{{\rm Bj}}\pm\xi-i0)^{\frac{d}{2}-2}\right].\label{eq:Jpm-res}
\end{align}
 Finally,
\begin{align}
I_{LL}^{ij} & =\frac{\delta^{ij}}{4\xi^{2}}\mu^{2-d}(2q^{+}P^{-})^{\frac{d}{2}}\frac{\Gamma(\frac{d}{2})\Gamma(\frac{d}{2})}{d\Gamma(d)}\frac{\Gamma(2-\frac{d}{2})}{(4\pi)^{\frac{d}{2}}}\\
 & \times\left\{ \left(\frac{x_{{\rm Bj}}}{\xi}+\frac{x_{{\rm Bj}}+\xi}{x+\xi-i0x_{{\rm Bj}}}\right)\frac{(x_{{\rm Bj}}-x-i0)^{\frac{d}{2}-1}-(x_{{\rm Bj}}+\xi-i0)^{\frac{d}{2}-1}}{(\frac{d}{2}-1)(x+\xi-i0x_{{\rm Bj}})}\right.\nonumber\\
 & -\left(\frac{x_{{\rm Bj}}}{\xi}-\frac{x_{{\rm Bj}}-\xi}{x-\xi+i0x_{{\rm Bj}}}\right)\frac{(x_{{\rm Bj}}-x-i0)^{\frac{d}{2}-1}-(x_{{\rm Bj}}-\xi-i0)^{\frac{d}{2}-1}}{(\frac{d}{2}-1)(x-\xi+i0x_{{\rm Bj}})}\nonumber\\
 & \left.+\frac{(x_{{\rm Bj}}+\xi-i0)^{\frac{d}{2}-1}}{(x+\xi-i0x_{{\rm Bj}})}+\frac{(x_{{\rm Bj}}-\xi-i0)^{\frac{d}{2}-1}}{(x-\xi+i0x_{{\rm Bj}})}\right\} \nonumber.
\end{align}
 It is finite in 4 dimensions:
\begin{align}
I_{LL}^{ij} & =\frac{\delta^{ij}}{8\xi^{2}}\frac{2q^{+}P^{-}}{4\pi}\left\{ \frac{1}{x+\xi}\left(\frac{x_{{\rm Bj}}}{\xi}+\frac{x_{{\rm Bj}}+\xi}{x+\xi}\right)\ln\left(\frac{x_{{\rm Bj}}-x-i0}{x_{{\rm Bj}}+\xi-i0}\right)\right.\nonumber\\
 & \left.-\frac{1}{x-\xi}\left(\frac{x_{{\rm Bj}}}{\xi}-\frac{x_{{\rm Bj}}-\xi}{x-\xi}\right)\ln\left(\frac{x_{{\rm Bj}}-x-i0}{x_{{\rm Bj}}-\xi-i0}\right)+\frac{2x}{x^{2}-\xi^{2}}\right\} .
\end{align}
Note that $i0x_{{\rm Bj}}$ in denominators were removed here, since
DVCS and TCS do not involve an LL transition so the $x_{{\rm Bj}}\pm\xi=0$ case which required this special care is absent from the LL contribution.

Similarly steps lead to:
\begin{align}
I_{TT}^{ijmn} & =2q^{+}P^{-}\frac{2}{\xi}\frac{\Gamma(1-\epsilon)}{4\pi(1+\epsilon)}\frac{\Gamma(1+\epsilon)\Gamma(1+\epsilon)}{\Gamma(1+2\epsilon)}\left(\frac{Q^{2}+Q^{\prime2}}{4\pi\mu^{2}}\right)^{\epsilon}\label{eq:ITT-before-exp}\\
 & \times\left[\frac{1}{2\epsilon}(\delta^{km}\delta^{ln}+\delta^{kl}\delta^{mn}-\delta^{lm}\delta^{kn})-\frac{1}{1+2\epsilon}\delta^{km}\delta^{ln}\right]\nonumber \\
 & \times\left\{ \left(\epsilon\delta^{ik}\delta^{jl}-\frac{x_{{\rm Bj}}^{2}-\xi^{2}}{\xi^{2}}D^{ijkl}\right)\frac{\left(\frac{x_{{\rm Bj}}-x-i0}{2\xi}\right)^{\epsilon}-\left(\frac{x_{{\rm Bj}}+\xi-i0}{2\xi}\right)^{\epsilon}}{\epsilon(x+\xi-i0x_{{\rm Bj}})}\right.\nonumber \\
 & -\left(\epsilon\delta^{ik}\delta^{jl}-\frac{x_{{\rm Bj}}^{2}-\xi^{2}}{\xi^{2}}D^{ijkl}\right)\frac{\left(\frac{x_{{\rm Bj}}-x-i0}{2\xi}\right)^{\epsilon}-\left(\frac{x_{{\rm Bj}}-\xi-i0}{2\xi}\right)^{\epsilon}}{\epsilon(x-\xi+i0x_{{\rm Bj}})}\nonumber \\
 & +(x_{{\rm Bj}}+\xi)\left(\delta^{ik}\delta^{jl}-\frac{x_{{\rm Bj}}+\xi}{\xi}D^{ijkl}\right)\frac{\left(\frac{x_{{\rm Bj}}-x-i0}{2\xi}\right)^{\epsilon}-\left(\frac{x_{{\rm Bj}}+\xi-i0}{2\xi}\right)^{\epsilon}}{\epsilon(x+\xi-i0x_{{\rm Bj}})^{2}}\nonumber \\
 & -(x_{{\rm Bj}}-\xi)\left(\delta^{ik}\delta^{jl}+\frac{x_{{\rm Bj}}-\xi}{\xi}D^{ijkl}\right)\frac{\left(\frac{x_{{\rm Bj}}-x-i0}{2\xi}\right)^{\epsilon}-\left(\frac{x_{{\rm Bj}}-\xi-i0}{2\xi}\right)^{\epsilon}}{\epsilon(x-\xi+i0x_{{\rm Bj}})^{2}}\nonumber \\
 & -\left(\delta^{ik}\delta^{jl}+\frac{x_{{\rm Bj}}-\xi}{\xi}D^{ijkl}\right)\frac{1}{x-\xi+i0x_{{\rm Bj}}}\left(\frac{x_{{\rm Bj}}-\xi-i0}{2\xi}\right)^{\epsilon}\nonumber \\
 & \left.+\left(\delta^{ik}\delta^{jl}-\frac{x_{{\rm Bj}}+\xi}{\xi}D^{ijkl}\right)\frac{1}{x+\xi-i0x_{{\rm Bj}}}\left(\frac{x_{{\rm Bj}}+\xi-i0}{2\xi}\right)^{\epsilon}\right\} ,\nonumber 
\end{align}
where we defined $D^{ijkl}\equiv\frac{\delta^{ij}\delta^{kl}+\delta^{ik}\delta^{jl}+\delta^{il}\delta^{jk}}{d+2}$.
Let us go to $4+2\epsilon$ dimensions, and distinguish all possible
projections for the $(ij)$ pair of indices: unpolarized $(\delta^{ij})$,
polarized $(\epsilon^{ij})$ and transversity $(\tau^{ij,pq})$. We find:
\begin{align}
\frac{1}{d}\delta^{ij}I_{TT}^{ijmn} & =\frac{2q^{+}P^{-}}{\xi}\frac{\delta^{mn}}{4\pi}\left(\frac{1}{\epsilon}-2\right)\left(\frac{{\rm e}^{\gamma_{E}}}{4\pi}\frac{Q^{2}+Q^{\prime2}}{\mu^{2}}\right)^{\epsilon}\\
 & \times\left\{ -\frac{x_{{\rm Bj}}}{\xi}\left(\frac{1}{x+\xi-i0x_{{\rm Bj}}}+\frac{1}{x-\xi+i0x_{{\rm Bj}}}\right)\right.\nonumber\\
 & -\frac{x_{{\rm Bj}}^{2}-\xi^{2}}{\xi^{2}}\left(\frac{\ln\left(\frac{x_{{\rm Bj}}-x-i0}{x_{{\rm Bj}}+\xi-i0}\right)}{x+\xi-i0x_{{\rm Bj}}}-\frac{\ln\left(\frac{x_{{\rm Bj}}-x-i0}{x_{{\rm Bj}}-\xi-i0}\right)}{x-\xi+i0x_{{\rm Bj}}}\right)\nonumber\\
 & -\frac{x_{{\rm Bj}}}{\xi}\left(\frac{(x_{{\rm Bj}}+\xi)\ln\left(\frac{x_{{\rm Bj}}-x-i0}{x_{{\rm Bj}}+\xi-i0}\right)}{(x+\xi-i0x_{{\rm Bj}})^{2}}+\frac{(x_{{\rm Bj}}-\xi)\ln\left(\frac{x_{{\rm Bj}}-x-i0}{x_{{\rm Bj}}-\xi-i0}\right)}{(x-\xi+i0x_{{\rm Bj}})^{2}}\right)\nonumber\\
 & -\epsilon\frac{1}{\xi}\left(\frac{(x_{{\rm Bj}}-\xi)\ln\left(\frac{x_{{\rm Bj}}-\xi-i0}{2\xi}\right)}{x-\xi+i0x_{{\rm Bj}}}+\frac{(x_{{\rm Bj}}+\xi)\ln\left(\frac{x_{{\rm Bj}}+\xi-i0}{2\xi}\right)}{x+\xi-i0x_{{\rm Bj}}}\right)\nonumber\\
 & +\epsilon\left(\frac{1}{x+\xi-i0x_{{\rm Bj}}}-\frac{1}{x-\xi+i0x_{{\rm Bj}}}\right)\ln\left(\frac{x_{{\rm Bj}}-x-i0}{2\xi}\right)\nonumber\\
 & +\frac{\epsilon}{2}\frac{x_{{\rm Bj}}^{2}-\xi^{2}}{\xi^{2}}\frac{\ln^{2}\left(\frac{x_{{\rm Bj}}-x-i0}{2\xi}\right)-\ln^{2}\left(\frac{x_{{\rm Bj}}-\xi-i0}{2\xi}\right)}{(x-\xi+i0x_{{\rm Bj}})}\nonumber\\
 & -\frac{\epsilon}{2}\frac{x_{{\rm Bj}}^{2}-\xi^{2}}{\xi^{2}}\frac{\ln^{2}\left(\frac{x_{{\rm Bj}}-x-i0}{2\xi}\right)-\ln^{2}\left(\frac{x_{{\rm Bj}}+\xi-i0}{2\xi}\right)}{(x+\xi-i0x_{{\rm Bj}})}\nonumber\\
 & -\frac{\epsilon}{2}(x_{{\rm Bj}}+\xi)\frac{x_{{\rm Bj}}}{\xi}\frac{\ln^{2}\left(\frac{x_{{\rm Bj}}-x-i0}{2\xi}\right)-\ln^{2}\left(\frac{x_{{\rm Bj}}+\xi-i0}{2\xi}\right)}{(x+\xi-i0x_{{\rm Bj}})^{2}}\nonumber\\
 & \left.-\frac{\epsilon}{2}(x_{{\rm Bj}}-\xi)\frac{x_{{\rm Bj}}}{\xi}\frac{\ln^{2}\left(\frac{x_{{\rm Bj}}-x-i0}{2\xi}\right)-\ln^{2}\left(\frac{x_{{\rm Bj}}-\xi-i0}{2\xi}\right)}{(x-\xi+i0x_{{\rm Bj}})^{2}}\right\} ,\nonumber
\end{align}
for the unpolarized term,
\begin{align}
\frac{1}{d}\epsilon^{ij}I_{TT}^{ijmn} & =\epsilon^{mn}\frac{2q^{+}P^{-}}{\xi}\frac{\Gamma(1-\epsilon)}{(4\pi)^{1+\epsilon}}\left(\frac{Q^{2}+Q^{\prime2}}{2\mu^{2}}\right)^{\epsilon}\\
 & \times\frac{1}{\epsilon}(1-3\epsilon)\left\{ \frac{1}{x+\xi-i0x_{{\rm Bj}}}-\frac{1}{x-\xi+i0x_{{\rm Bj}}}\right.\nonumber\\
 & +\frac{x_{{\rm Bj}}+\xi}{(x+\xi-i0x_{{\rm Bj}})^{2}}\ln\left(\frac{x_{{\rm Bj}}-x-i0}{x_{{\rm Bj}}+\xi-i0}\right) -\frac{x_{{\rm Bj}}-\xi}{(x-\xi+i0x_{{\rm Bj}})^{2}}\ln\left(\frac{x_{{\rm Bj}}-x-i0}{x_{{\rm Bj}}-\xi-i0}\right)\nonumber\\
 & +\epsilon\left(\frac{1}{x+\xi-i0x_{{\rm Bj}}}-\frac{1}{x-\xi+i0x_{{\rm Bj}}}\right)\ln\left(\frac{x_{{\rm Bj}}-x-i0}{\xi}\right)\nonumber\\
 & +\frac{\epsilon}{2}\frac{x_{{\rm Bj}}+\xi}{(x+\xi-i0x_{{\rm Bj}})^{2}}\left[\ln^{2}\left(\frac{x_{{\rm Bj}}-x-i0}{\xi}\right)-\ln^{2}\left(\frac{x_{{\rm Bj}}+\xi-i0}{\xi}\right)\right]\nonumber\\
 & \left.-\frac{\epsilon}{2}\frac{x_{{\rm Bj}}-\xi}{(x-\xi+i0x_{{\rm Bj}})^{2}}\left[\ln^{2}\left(\frac{x_{{\rm Bj}}-x-i0}{\xi}\right)-\ln^{2}\left(\frac{x_{{\rm Bj}}-\xi-i0}{\xi}\right)\right]\right\} \nonumber
\end{align}
for the polarized term, and 
\begin{align}
\frac{1}{d}\tau^{ij,pq}I_{TT}^{ijmn} & =\frac{q^{+}P^{-}}{\xi^{2}}\frac{\tau^{mn,pq}}{4\pi}\left\{ \frac{x_{{\rm Bj}}+\xi}{x-\xi+i0x_{{\rm Bj}}}+\frac{x_{{\rm Bj}}-\xi}{x+\xi-i0x_{{\rm Bj}}}\right.\\
 & +\frac{x_{{\rm Bj}}^{2}-\xi^{2}}{x+\xi-i0x_{{\rm Bj}}}\left(\frac{1}{\xi}+\frac{1}{x+\xi-i0x_{{\rm Bj}}}\right)\ln\left(\frac{x_{{\rm Bj}}-x-i0}{x_{{\rm Bj}}+\xi-i0}\right)\nonumber\\
 & \left.-\frac{x_{{\rm Bj}}^{2}-\xi^{2}}{x-\xi+i0x_{{\rm Bj}}}\left(\frac{1}{\xi}-\frac{1}{x-\xi+i0x_{{\rm Bj}}}\right)\ln\left(\frac{x_{{\rm Bj}}-x-i0}{x_{{\rm Bj}}-\xi-i0}\right)\right\} \nonumber
\end{align}
for the transversity term.
\section{Integrals of the integrals}\label{sec:intintapp}
To study the consistency of the double log limit, we need the integrals over $x$ of the Bjorken amplitudes. This means we need to integrate
the integrals from the previous appendix: $\int_{-1}^{1}{\rm d}xI_{LL}^{ij}$
and $\int_{-1}^{1}{\rm d}xI_{TT}^{ij}$. Most terms can easily be rewritten as total derivatives of logarithms and dilogarithms:
\begin{align}
I_{LL}^{ij} & =\frac{\delta^{ij}}{8\xi^{2}}\frac{2q^{+}P^{-}}{4\pi}\frac{\partial}{\partial x}\left\{ \frac{x-x_{{\rm Bj}}}{x+\xi}\ln\left(\frac{x_{{\rm Bj}}-x-i0}{x_{{\rm Bj}}+\xi-i0}\right)+\frac{x-x_{{\rm Bj}}}{x-\xi}\ln\left(\frac{x_{{\rm Bj}}-x-i0}{x_{{\rm Bj}}-\xi-i0}\right)\right.\nonumber \\
 & \left.+\frac{x_{{\rm Bj}}}{\xi}{\rm Li}_{2}\left(\frac{x-\xi}{x_{{\rm Bj}}-\xi-i0}\right)-\frac{x_{{\rm Bj}}}{\xi}{\rm Li}_{2}\left(\frac{x+\xi}{x_{{\rm Bj}}+\xi-i0}\right)\right\} ,\label{eq:ILL-preint}
\end{align}
and
\begin{align}
I_{TT}^{ijmn} & =\frac{4}{\xi^{2}}\mu^{2-d}\frac{\Gamma(2-\frac{d}{2})}{(4\pi)^{\frac{d}{2}}}\frac{\Gamma(\frac{d}{2})\Gamma(\frac{d}{2})}{d\Gamma(d)}\left(\frac{Q^{2}+Q^{\prime2}}{2}\right)^{\frac{d}{2}}\nonumber \\
 & \times\left[\left(\frac{d-1}{d-2}\right)(\delta^{km}\delta^{ln}+\delta^{kl}\delta^{mn}-\delta^{lm}\delta^{kn})-\delta^{km}\delta^{ln}\right]\nonumber \\
 & \times\left\{ -2\frac{x_{{\rm Bj}}}{\xi}D^{ijkl}\left[\frac{\partial}{\partial x}\frac{\left(\frac{x_{{\rm Bj}}-x-i0}{\xi}\right)^{\frac{d}{2}-1}}{(\frac{d}{2}-1)}\right]\right.\label{eq:ITT-rewritten}\\
 & +\left[(\frac{d}{2}-1)+\frac{x_{{\rm Bj}}+\xi}{\xi}\right]\left(\frac{x_{{\rm Bj}}-\xi-i0}{\xi}\right)^{\frac{d}{2}}D^{ijkl}\frac{1}{x-\xi}\left[\frac{\left(\frac{x_{{\rm Bj}}-x-i0}{x_{{\rm Bj}}-\xi-i0}\right)^{\frac{d}{2}-1}-1}{(\frac{d}{2}-1)}\right]\nonumber \\
 & +\left[(\frac{d}{2}-1)-\frac{x_{{\rm Bj}}-\xi}{\xi}\right]\left(\frac{x_{{\rm Bj}}+\xi-i0}{\xi}\right)^{\frac{d}{2}}D^{ijkl}\frac{1}{x+\xi}\left[\frac{\left(\frac{x_{{\rm Bj}}-x-i0}{x_{{\rm Bj}}+\xi-i0}\right)^{\frac{d}{2}-1}-1}{(\frac{d}{2}-1)}\right]\nonumber \\
 & +\xi\left(\delta^{ik}\delta^{jl}+\frac{x_{{\rm Bj}}-\xi}{\xi}D^{ijkl}\right)\left(\frac{x_{{\rm Bj}}-\xi-i0}{\xi}\right)^{\frac{d}{2}}\left[\frac{\partial}{\partial x}\frac{\left(\frac{x_{{\rm Bj}}-x-i0}{x_{{\rm Bj}}-\xi-i0}\right)^{\frac{d}{2}-1}-1}{(\frac{d}{2}-1)(x-\xi)}\right]\nonumber \\
 & \left.-\xi\left(\delta^{ik}\delta^{jl}-\frac{x_{{\rm Bj}}+\xi}{\xi}D^{ijkl}\right)\left(\frac{x_{{\rm Bj}}+\xi-i0}{\xi}\right)^{\frac{d}{2}}\left[\frac{\partial}{\partial x}\frac{\left(\frac{x_{{\rm Bj}}-x-i0}{x_{{\rm Bj}}+\xi-i0}\right)^{\frac{d}{2}-1}-1}{(\frac{d}{2}-1)(x+\xi)}\right]\right\} \nonumber 
\end{align}
 Note that the last 2 lines are suppressed when $\xi\rightarrow0$.
For this reason, the Regge limit only contains contributions from
the $D^{ijkl}$ tensor, and thus from unpolarized and transversity
GPDs. Once these two lines have been neglected, $D^{ijkl}$ contracted,
and after dim reg expansion, we get:
\begin{align}
I_{TT}^{ijmn} & =\frac{2q^{+}P^{-}}{\xi}\frac{1}{4\pi}\left(\frac{{\rm e}^{\gamma_{E}}}{4\pi}\frac{Q^{2}+Q^{\prime2}}{2\mu^{2}}\right)^{\epsilon}\left((\frac{1}{\epsilon}-2)\delta^{ij}\delta^{mn}-\frac{\tau^{ij;mn}}{2}\right)\nonumber \\
 & \times\left\{ -2\frac{x_{{\rm Bj}}}{\xi}\frac{\partial}{\partial x}\left[\ln\left(\frac{x_{{\rm Bj}}-x-i0}{\xi}\right)+\frac{\epsilon}{2}\ln^{2}\left(\frac{x_{{\rm Bj}}-x-i0}{\xi}\right)\right]\right.\label{eq:ITT-reduced-expanded}\\
 & +\left(\frac{x_{{\rm Bj}}-\xi}{\xi}\epsilon+\frac{x_{{\rm Bj}}^{2}-\xi^{2}}{\xi^{2}}\right)\left(\frac{x_{{\rm Bj}}-\xi-i0}{\xi}\right)^{\epsilon}\frac{1}{x-\xi}\ln\left(\frac{x_{{\rm Bj}}-x-i0}{x_{{\rm Bj}}-\xi-i0}\right)\nonumber \\
 & +\left(\frac{x_{{\rm Bj}}+\xi}{\xi}\epsilon-\frac{x_{{\rm Bj}}^{2}-\xi^{2}}{\xi^{2}}\right)\left(\frac{x_{{\rm Bj}}+\xi-i0}{\xi}\right)^{\epsilon}\frac{1}{x+\xi}\ln\left(\frac{x_{{\rm Bj}}-x-i0}{x_{{\rm Bj}}+\xi-i0}\right)\nonumber \\
 & \left.+\frac{\epsilon}{2}\left(\frac{x_{{\rm Bj}}^{2}-\xi^{2}}{\xi^{2}}\right)\frac{1}{x-\xi}\ln^{2}\left(\frac{x_{{\rm Bj}}-x-i0}{x_{{\rm Bj}}-\xi-i0}\right)-\frac{\epsilon}{2}\left(\frac{x_{{\rm Bj}}^{2}-\xi^{2}}{\xi^{2}}\right)\frac{1}{x+\xi}\ln^{2}\left(\frac{x_{{\rm Bj}}-x-i0}{x_{{\rm Bj}}+\xi-i0}\right)\right\} ,\nonumber 
\end{align}
with $\tau^{ij;mn}=\delta^{im}\delta^{jn}+\delta^{in}\delta^{jm}-\delta^{ij}\delta^{mn}$
the transversity projector. For both $LL$ and $TT$ cases, we will
need to study the asymptotics of polylogarithms. Indeed, terms like $\frac{1}{x-\xi}\ln\left(\frac{x_{{\rm Bj}}-x-i0}{x_{{\rm Bj}}-\xi-i0}\right)$ and $\frac{1}{x-\xi}\ln^2\left(\frac{x_{{\rm Bj}}-x-i0}{x_{{\rm Bj}}-\xi-i0}\right)$ integrate into di- and tri-logarithms. In practise, we will
need the following:
\begin{equation}
{\rm Li}_{2}\left(\frac{1}{X-i0}\right)\sim_{X\rightarrow0}-\frac{1}{2}\ln^{2}(X-i0)-i\pi\ln(X-i0)+\frac{\pi^{2}}{3}\label{eq:dilog-1}
\end{equation}
and
\begin{equation}
{\rm Li}_{2}\left(\frac{-1}{X-i0}\right)\sim_{X\rightarrow0}-\frac{1}{2}\ln^{2}(X-i0)-\frac{\pi^{2}}{6}\label{eq:dilog-(-1)}
\end{equation}
for dilogarithms, as well as
\begin{equation}
{\rm Li}_{3}\left(\frac{1}{X-i0}\right)\sim_{X\rightarrow0}\frac{1}{6}\ln^{3}(X-i0)+\frac{1}{2}i\pi\ln^{2}(X-i0)-\frac{\pi^{2}}{3}\ln(X-i0)\label{eq:trilog-1}
\end{equation}
and
\begin{equation}
{\rm Li}_{3}\left(\frac{-1}{X-i0}\right)\sim_{X\rightarrow0}\frac{1}{6}\ln^{3}(X-i0)+\frac{\pi^{2}}{6}\ln(X-i0)\label{eq:trilog-(-1)}
\end{equation}
for trilogarithms. After some algebra, we then get:
\begin{align}
 & \lim_{x_{{\rm Bj}},\xi\rightarrow0}\int_{-1}^{1}{\rm d}xI_{LL}^{ij}\label{eq:ILL-int-fin}\\
 & =-2iq^{+}P^{-}\frac{\delta^{ij}}{32\xi^{2}}\left[2-\frac{Q^{2}-Q^{\prime2}}{Q^{2}+Q^{\prime2}}\ln\left(\frac{Q^{2}}{-Q^{\prime2}-i0}\right)\right],\nonumber 
\end{align}
and
\begin{align}
\lim_{x_{{\rm Bj}},\xi\rightarrow0}\int_{-1}^{1}{\rm d}x&I_{TT}^{ijmn} =i\pi\frac{2q^{+}P^{-}}{\xi}\frac{1}{4\pi}\left(\frac{{\rm e}^{\gamma_{E}}}{4\pi}\right)^{\epsilon}\left[\left(\frac{1}{\epsilon}-2\right)\delta^{ij}\delta^{mn}-\frac{\tau^{ij;mn}}{2}\right]\nonumber \\
 & \times\left\{ 2\frac{x_{{\rm Bj}}}{\xi}-\frac{x_{{\rm Bj}}^{2}-\xi^{2}}{\xi^{2}}\ln\left(\frac{x_{{\rm Bj}}+\xi-i0}{x_{{\rm Bj}}-\xi-i0}\right)\right.\label{eq:ITT-int-almost}\\
 & +\left. \epsilon \frac{x_{{\rm Bj}}-\xi}{\xi}\ln\left(\frac{2q^{+}P^{-}}{\mu^{2}}(x_{{\rm Bj}}-\xi-i0)\right) \nonumber \right. \\  & \left.  +\epsilon \frac{x_{{\rm Bj}}+\xi}{\xi}\ln\left(\frac{2q^{+}P^{-}}{\mu^{2}}(x_{{\rm Bj}}+\xi-i0)\right)\nonumber \right. \\
 & \left.-\frac{\epsilon}{2}\frac{x_{{\rm Bj}}^{2}-\xi^{2}}{\xi^{2}} \ln^{2}\left(\frac{2q^{+}P^{-}}{\mu^{2}}(x_{{\rm Bj}}+\xi-i0)\right) \nonumber \right. \\ & \left.+ \frac{\epsilon}{2}\frac{x_{{\rm Bj}}^{2}-\xi^{2}}{\xi^{2}}\ln^{2}\left(\frac{2q^{+}P^{-}}{\mu^{2}}(x_{{\rm Bj}}-\xi-i0)\right) \right\} .\nonumber 
\end{align}

\section{A remark on the evolution kernel}\label{sec:regions}

It is worth showing that the ERBL ($\xi>x>-\xi)$ and DGLAP ($x>\xi$
and $-\xi>x$) regions of the kernel will contribute identically to Eq.~(\ref{eq:F0R}) as
far as the $y$ integral is concerned. Indeed, distinguishing all
3 regions we can rewrite the kernel into:
\begin{align}
K^{qg}(y,x) & =\frac{\alpha_{s}}{2\pi}T_{F}\theta(x-\xi)\frac{y^{2}+(x-y)^{2}-\xi^{2}}{(x^{2}-\xi^{2})^{2}}\theta(x-y)\theta(y-\xi)\nonumber\\
 & +\frac{\alpha_{s}}{2\pi}T_{F}\theta(x-\xi)\frac{(y+\xi)(x-2y+\xi)}{2\xi(x+\xi)(x^{2}-\xi^{2})}\theta(\xi-y)\theta(y+\xi)\nonumber\\
 & -\frac{\alpha_{s}}{2\pi}T_{F}\theta(-\xi-x)\frac{y^{2}+(x-y)^{2}-\xi^{2}}{(x^{2}-\xi^{2})^{2}}\theta(-\xi-y)\theta(y-x)\\
 & +\frac{\alpha_{s}}{2\pi}T_{F}\theta(-\xi-x)\frac{(y-\xi)(x-2y-\xi)}{2\xi(x-\xi)(x^{2}-\xi^{2})}\theta(y+\xi)\theta(\xi-y)\nonumber\\
 & +\frac{\alpha_{s}}{2\pi}T_{F}\theta(\xi-x)\theta(x+\xi)\frac{(y+\xi)(x-2y+\xi)}{2\xi(x+\xi)(x^{2}-\xi^{2})}\theta(x-y)\theta(y+\xi)\nonumber\\
 & +\frac{\alpha_{s}}{2\pi}T_{F}\theta(\xi-x)\theta(x+\xi)\frac{(y-\xi)(x-2y-\xi)}{2\xi(x-\xi)(x^{2}-\xi^{2})}\theta(y-x)\theta(\xi-y).\nonumber
\end{align}
For a test function $\phi(y),$ we then have
\begin{align}
\int_{-1}^{1}{\rm d}y\phi(y)K^{qg}(y,x) & =\frac{\alpha_{s}}{2\pi}T_{F}\theta(x-\xi)\int_{\xi}^{x}{\rm d}y\phi(y)\frac{y^{2}+(x-y)^{2}-\xi^{2}}{(x^{2}-\xi^{2})^{2}}\nonumber\\
 & +\frac{\alpha_{s}}{2\pi}T_{F}\theta(x-\xi)\int_{-\xi}^{\xi}{\rm d}y\phi(y)\frac{(y+\xi)(x-2y+\xi)}{2\xi(x+\xi)(x^{2}-\xi^{2})}\nonumber\\
 & -\frac{\alpha_{s}}{2\pi}T_{F}\theta(-\xi-x)\int_{x}^{-\xi}{\rm d}y\phi(y)\frac{y^{2}+(x-y)^{2}-\xi^{2}}{(x^{2}-\xi^{2})^{2}}\\
 & +\frac{\alpha_{s}}{2\pi}T_{F}\theta(-\xi-x)\int_{-\xi}^{\xi}{\rm d}y\phi(y)\frac{(y-\xi)(x-2y-\xi)}{2\xi(x-\xi)(x^{2}-\xi^{2})}\nonumber\\
 & +\frac{\alpha_{s}}{2\pi}T_{F}\theta(\xi-x)\theta(x+\xi)\int_{-\xi}^{x}{\rm d}y\phi(y)\frac{(y+\xi)(x-2y+\xi)}{2\xi(x+\xi)(x^{2}-\xi^{2})}\nonumber\\
 & +\frac{\alpha_{s}}{2\pi}T_{F}\theta(\xi-x)\theta(x+\xi)\int_{x}^{\xi}{\rm d}y\phi(y)\frac{(y-\xi)(x-2y-\xi)}{2\xi(x-\xi)(x^{2}-\xi^{2})}.\nonumber
\end{align}
As long as $\phi$ is integrable on $[-1,1]$, we can simplify this
further by playing with the integration boundaries:
\begin{align}
\int_{-1}^{1}{\rm d}y\phi(y)K^{qg}(y,x) & =\frac{\alpha_{s}}{2\pi}T_{F}\theta(x-\xi)\int_{-\xi}^{x}{\rm d}y\phi(y)\frac{y^{2}+(x-y)^{2}-\xi^{2}}{(x^{2}-\xi^{2})^{2}}\nonumber\\
 & +\frac{\alpha_{s}}{2\pi}T_{F}\theta(x-\xi)\int_{-\xi}^{\xi}{\rm d}y\phi(y)\frac{(y-\xi)(x-2y-\xi)}{2\xi(x-\xi)(x^{2}-\xi^{2})}\nonumber\\
 & +\frac{\alpha_{s}}{2\pi}T_{F}\theta(-\xi-x)\int_{-\xi}^{x}{\rm d}y\phi(y)\frac{y^{2}+(x-y)^{2}-\xi^{2}}{(x^{2}-\xi^{2})^{2}}\nonumber\\
 & +\frac{\alpha_{s}}{2\pi}T_{F}\theta(-\xi-x)\int_{-\xi}^{\xi}{\rm d}y\phi(y)\frac{(y-\xi)(x-2y-\xi)}{2\xi(x-\xi)(x^{2}-\xi^{2})}\\
 & +\frac{\alpha_{s}}{2\pi}T_{F}\theta(\xi-x)\theta(x+\xi)\int_{-\xi}^{x}{\rm d}y\phi(y)\frac{y^{2}+(x-y)^{2}-\xi^{2}}{(x^{2}-\xi^{2})^{2}}\nonumber\\
 & +\frac{\alpha_{s}}{2\pi}T_{F}\theta(\xi-x)\theta(x+\xi)\int_{-\xi}^{\xi}{\rm d}y\phi(y)\frac{(y-\xi)(x-2y-\xi)}{2\xi(x-\xi)(x^{2}-\xi^{2})}
\nonumber\end{align}
All 3 regions contribute the same, and we can thus use:
\begin{align}
\int_{-1}^{1}{\rm d}y\phi(y)K^{qg}(y,x) & =\frac{\alpha_{s}}{2\pi}\frac{T_{F}}{(x-\xi)(x^{2}-\xi^{2})}\left[\int_{-\xi}^{x}{\rm d}y\phi(y)\frac{y^{2}+(x-y)^{2}-\xi^{2}}{x+\xi}\right.\nonumber\\
 & \left.+\int_{-\xi}^{\xi}{\rm d}y\phi(y)\frac{(y-\xi)(x-2y-\xi)}{2\xi}\right].
\end{align}


\begin{thebibliography}{99}

\bibitem{Collins:1998be}
J.~C.~Collins and A.~Freund,
Phys. Rev. D \textbf{59}, 074009 (1999)
doi:10.1103/PhysRevD.59.074009
[arXiv:hep-ph/9801262 [hep-ph]].

\bibitem{Muller:1994ses}
D.~M\"uller, D.~Robaschik, B.~Geyer, F.~M.~Dittes and J.~Ho\v{r}ej\v{s}i,
Fortsch. Phys. \textbf{42}, 101-141 (1994)
doi:10.1002/prop.2190420202
[arXiv:hep-ph/9812448 [hep-ph]].

\bibitem{Ji:1996nm}
X.~D.~Ji,
Phys. Rev. D \textbf{55}, 7114-7125 (1997)
doi:10.1103/PhysRevD.55.7114
[arXiv:hep-ph/9609381 [hep-ph]].

\bibitem{Radyushkin:1997ki}
A.~V.~Radyushkin,
Phys. Rev. D \textbf{56}, 5524-5557 (1997)
doi:10.1103/PhysRevD.56.5524
[arXiv:hep-ph/9704207 [hep-ph]].

\bibitem{Diehl:2003ny}
M.~Diehl,
Phys. Rept. \textbf{388}, 41-277 (2003)
doi:10.1016/j.physrep.2003.08.002
[arXiv:hep-ph/0307382 [hep-ph]].

\bibitem{Belitsky:2005qn}
A.~V.~Belitsky and A.~V.~Radyushkin,
Phys. Rept. \textbf{418}, 1-387 (2005)
doi:10.1016/j.physrep.2005.06.002
[arXiv:hep-ph/0504030 [hep-ph]].

\bibitem{McLerran:1993ni}
L.~D.~McLerran and R.~Venugopalan,
Phys. Rev. D \textbf{49}, 2233-2241 (1994)
doi:10.1103/PhysRevD.49.2233
[arXiv:hep-ph/9309289 [hep-ph]].

\bibitem{McLerran:1993ka}
L.~D.~McLerran and R.~Venugopalan,
Phys. Rev. D \textbf{49}, 3352-3355 (1994)
doi:10.1103/PhysRevD.49.3352
[arXiv:hep-ph/9311205 [hep-ph]].

\bibitem{McLerran:1994vd}
L.~D.~McLerran and R.~Venugopalan,
Phys. Rev. D \textbf{50}, 2225-2233 (1994)
doi:10.1103/PhysRevD.50.2225
[arXiv:hep-ph/9402335 [hep-ph]].

\bibitem{Balitsky:1995ub}
I.~Balitsky,
Nucl. Phys. B \textbf{463}, 99-160 (1996)
doi:10.1016/0550-3213(95)00638-9
[arXiv:hep-ph/9509348 [hep-ph]].

\bibitem{Iancu:2003xm}
E.~Iancu and R.~Venugopalan,
doi:10.1142/9789812795533\_0005
[arXiv:hep-ph/0303204 [hep-ph]].

\bibitem{Gelis:2010nm}
F.~Gelis, E.~Iancu, J.~Jalilian-Marian and R.~Venugopalan,
Ann. Rev. Nucl. Part. Sci. \textbf{60}, 463-489 (2010)
doi:10.1146/annurev.nucl.010909.083629
[arXiv:1002.0333 [hep-ph]].

\bibitem{Altinoluk:2019fui}
T.~Altinoluk, R.~Boussarie and P.~Kotko,
JHEP \textbf{05}, 156 (2019)
doi:10.1007/JHEP05(2019)156
[arXiv:1901.01175 [hep-ph]].

\bibitem{Altinoluk:2019wyu}
T.~Altinoluk and R.~Boussarie,
JHEP \textbf{10}, 208 (2019)
doi:10.1007/JHEP10(2019)208
[arXiv:1902.07930 [hep-ph]].

\bibitem{Boussarie:2020vzf}
R.~Boussarie and Y.~Mehtar-Tani,
Phys. Rev. D \textbf{103}, no.9, 094012 (2021)
doi:10.1103/PhysRevD.103.094012
[arXiv:2001.06449 [hep-ph]].

\bibitem{Hatta:2017cte}
Y.~Hatta, B.~W.~Xiao and F.~Yuan,
Phys. Rev. D \textbf{95}, no.11, 114026 (2017)
doi:10.1103/PhysRevD.95.114026
[arXiv:1703.02085 [hep-ph]].

\bibitem{Boussarie:2020fpb}
R.~Boussarie and Y.~Mehtar-Tani,
Phys. Lett. B \textbf{831}, 137125 (2022)
doi:10.1016/j.physletb.2022.137125
[arXiv:2006.14569 [hep-ph]].

\bibitem{Boussarie:2021wkn}
R.~Boussarie and Y.~Mehtar-Tani,
JHEP \textbf{07}, 080 (2022)
doi:10.1007/JHEP07(2022)080
[arXiv:2112.01412 [hep-ph]].

\bibitem{Bialas:2000xs}
A.~Bialas, H.~Navelet and R.~B.~Peschanski,
Nucl. Phys. B \textbf{593}, 438-450 (2001)
doi:10.1016/S0550-3213(00)00640-4
[arXiv:hep-ph/0009248 [hep-ph]].

\bibitem{Kutak:2004ym}
K.~Kutak and A.~M.~Stasto,
Eur. Phys. J. C \textbf{41}, 343-351 (2005)
doi:10.1140/epjc/s2005-02223-0
[arXiv:hep-ph/0408117 [hep-ph]].

\bibitem{Altinoluk:2014oxa}
T.~Altinoluk, N.~Armesto, G.~Beuf, M.~Mart\'\i{}nez and C.~A.~Salgado,
JHEP \textbf{07}, 068 (2014)
doi:10.1007/JHEP07(2014)068
[arXiv:1404.2219 [hep-ph]].

\bibitem{Altinoluk:2015gia}
T.~Altinoluk, N.~Armesto, G.~Beuf and A.~Moscoso,
JHEP \textbf{01}, 114 (2016)
doi:10.1007/JHEP01(2016)114
[arXiv:1505.01400 [hep-ph]].

\bibitem{Ji:1997nk}
X.~D.~Ji and J.~Osborne,
Phys. Rev. D \textbf{57}, 1337-1340 (1998)
doi:10.1103/PhysRevD.57.R1337
[arXiv:hep-ph/9707254 [hep-ph]].

\bibitem{Belitsky:1997rh}
A.~V.~Belitsky and D.~Mueller,
Phys. Lett. B \textbf{417}, 129-140 (1998)
doi:10.1016/S0370-2693(97)01390-7
[arXiv:hep-ph/9709379 [hep-ph]].

\bibitem{Mankiewicz:1997bk}
L.~Mankiewicz, G.~Piller, E.~Stein, M.~Vanttinen and T.~Weigl,
Phys. Lett. B \textbf{425}, 186-192 (1998)
[erratum: Phys. Lett. B \textbf{461}, 423-423 (1999)]
doi:10.1016/S0370-2693(98)00190-7
[arXiv:hep-ph/9712251 [hep-ph]].

\bibitem{Hoodbhoy:1998vm}
P.~Hoodbhoy and X.~D.~Ji,
Phys. Rev. D \textbf{58}, 054006 (1998)
doi:10.1103/PhysRevD.58.054006
[arXiv:hep-ph/9801369 [hep-ph]].

\bibitem{Ji:1998xh}
X.~D.~Ji and J.~Osborne,
Phys. Rev. D \textbf{58}, 094018 (1998)
doi:10.1103/PhysRevD.58.094018
[arXiv:hep-ph/9801260 [hep-ph]].

\bibitem{Belitsky:1999sg}
A.~V.~Belitsky, D.~Mueller, L.~Niedermeier and A.~Schafer,
Phys. Lett. B \textbf{474}, 163-169 (2000)
doi:10.1016/S0370-2693(99)01283-6
[arXiv:hep-ph/9908337 [hep-ph]].
\bibitem{Blaizot:2015lma}
J.~P.~Blaizot and Y.~Mehtar-Tani,
Int. J. Mod. Phys. E \textbf{24}, no.11, 1530012 (2015)
doi:10.1142/S021830131530012X
[arXiv:1503.05958 [hep-ph]].
\bibitem{Blaizot:2012fh}
J.~P.~Blaizot, F.~Dominguez, E.~Iancu and Y.~Mehtar-Tani,
JHEP \textbf{01}, 143 (2013)
doi:10.1007/JHEP01(2013)143
[arXiv:1209.4585 [hep-ph]].

\bibitem{Mueller:2012sma}
D.~Mueller, B.~Pire, L.~Szymanowski and J.~Wagner,
Phys. Rev. D \textbf{86}, 031502 (2012)
doi:10.1103/PhysRevD.86.031502
[arXiv:1203.4392 [hep-ph]].

\bibitem{Pire:2011st}
B.~Pire, L.~Szymanowski and J.~Wagner,
Phys. Rev. D \textbf{83}, 034009 (2011)
doi:10.1103/PhysRevD.83.034009
[arXiv:1101.0555 [hep-ph]].

\end{thebibliography}
\end{document}